\documentclass[5p]{elsarticle}          % look and feel of the final two-column format

\journal{ISPRS Journal of Photogrammetry and Remote Sensing}

\usepackage{graphicx}
\usepackage[caption=false,font=footnotesize]{subfig}
\usepackage{amsmath}
\usepackage{amssymb}
\usepackage{booktabs}
\usepackage[all]{nowidow}
\usepackage{stfloats}
\usepackage{url}
\usepackage{makecell}
\usepackage{array, makecell}
\usepackage{multirow}
\usepackage{enumitem}
\usepackage{adjustbox}
\usepackage{xspace}
\usepackage[dvipsnames]{xcolor}
\usepackage{arydshln}
\usepackage[super]{nth}
\usepackage[pagebackref=false,breaklinks=true,colorlinks,bookmarks=false]{hyperref}
\usepackage{xcolor}
\usepackage{dirtytalk}

\usepackage[normalem]{ulem}  % for strike-through text

\usepackage{ulem}

\xspaceaddexceptions{]\}}

\newcommand{\resdepth}{\textsc{ResDepth}\xspace}

\newcommand{\zurichOne}{\textsc{ZUR1}\xspace}
\newcommand{\zurichTwo}{\textsc{ZUR2}\xspace}
\newcommand{\zurichThree}{\textsc{ZUR3}\xspace}
\newcommand{\zurich}{\textsc{ZUR}\xspace}

\newcommand{\berlinOne}{\textsc{BER1}\xspace}
\newcommand{\berlinTwo}{\textsc{BER2}\xspace}
\newcommand{\berlin}{\textsc{BER}\xspace}

\begin{document}

\begin{frontmatter}

\title{\resdepth: A Deep Residual Prior For 3D Reconstruction From High-resolution Satellite Images}

\author{Corinne Stucker\corref{mycorrespondingauthor}}
\cortext[mycorrespondingauthor]{Corresponding author}
\ead{corinne.stucker@geod.baug.ethz.ch}

\author{Konrad Schindler}
\ead{schindler@ethz.ch}

\address{Photogrammetry and Remote Sensing, ETH Zurich, 8093 Z\"urich, Switzerland}

\begin{abstract}
    Modern optical satellite sensors enable high-resolution stereo reconstruction from space. But the challenging imaging conditions when observing the Earth from space push stereo matching to its limits. In practice, the resulting digital surface models (DSMs) are fairly noisy and often do not attain the accuracy needed for high-resolution applications such as 3D city modeling.
    Arguably, stereo correspondence based on low-level image similarity is insufficient and should be complemented with a-priori knowledge about the expected surface geometry beyond basic local smoothness.
    To that end, we introduce \resdepth, a convolutional neural network that learns such an expressive geometric prior from example data.
    \resdepth refines an initial, raw stereo DSM while conditioning the refinement on the images. I.e., it acts as a smart, learned post-processing filter and can seamlessly complement any stereo matching pipeline.
    In a series of experiments, we find that the proposed method consistently improves stereo DSMs both quantitatively and qualitatively. We show that the prior encoded in the network weights captures meaningful geometric characteristics of urban design, which also generalize across different districts and even from one city to another. Moreover, we demonstrate that, by training on a variety of stereo pairs, \resdepth can acquire a sufficient degree of invariance against variations in imaging conditions and acquisition geometry.
\end{abstract}


\end{frontmatter}

\section{Introduction}
Deriving \textit{digital surface models} (DSMs) from remotely sensed imagery is a fundamental task of photogrammetry and computer vision. Modern high-resolution satellite sensors have the ability to capture optical images of nearly every geographic location on Earth from multiple viewpoints, with ground sampling distances (GSD) of 30--50$\,$cm. This imagery, combined with state-of-the-art stereo algorithms and contemporary computing infrastructure, makes it possible to reconstruct city-scale or even country-scale scenes with an impressive level of detail. High-resolution 3D models of the Earth are the geometric foundation for so-called \say{digital twins} of (parts of) the Earth. They serve as the basis for a broad range of Earth observation tasks, including topographic mapping, environmental simulations, and 3D city modeling and planning.

In the near future, several new satellite missions will be available to support the creation of accurate, low-cost, and timely 3D models of the Earth. E.g., starting in 2021, DigitalGlobe plans to deploy WorldView Legion, a constellation of six high-resolution Earth observation satellites that will capture panchromatic imagery with a GSD of 29$\,$cm. The CO3D mission~\cite{lebegue2020co3d}, developed by CNES and Airbus Defense~\&~Space, will be launched in mid-2023 with the goal to generate a worldwide, low-cost DSM with 1$\,$m resolution and relative vertical accuracy of 1$\,$m.

To process such data, several 3D reconstruction pipelines tailored to satellite images have been developed~\cite{krauss2013fully,de2014automatic,qin2016rpc,rupnik2017micmac,beyer2018ames,cournet2020ground,youssefi2020cars}. The predominant techniques in this realm are stereo matching algorithms that are adapted versions of those originally developed for terrestrial and airborne photogrammetry. For instance, all the top-performers of the \textit{IARPA Multi-View Stereo 3D Mapping Challenge 2016}\footnote{\url{https://www.iarpa.gov/challenges/3dchallenge.html}} and the multi-view semantic stereo challenge of the \textit{2019 Data Fusion Contest}~\cite{kunwar2020large} are variants of semi-global matching~\cite{hirschmuller2005accurate}, ahead of deep learning-based stereo algorithms like~\cite{treible2018learning}. Such conventional stereo reconstruction is predominantly based on image-level correspondence analysis to maximize photo-consistency (cumulatively over all pixels) while enforcing only fairly simple a-priori assumptions about the 3D scene---in essence, an explicit preference for piece-wise smooth surfaces, and in some cases, for vertical discontinuities.
With high-quality images, such as the ones acquired from terrestrial or airborne mapping platforms, stereo matching with such a basic smoothness prior yields rather accurate surfaces. On the contrary, DSMs reconstructed from satellite data tend to be comparatively noisy, blurred, and often incomplete due to the limited sensor resolution, the often sub-optimal imaging conditions, and the restricted recording geometry of satellite imaging systems. In practice, the resulting surface models must thus be cleaned up.

Early methods for automatic DSM cleaning and refinement apply low-level filtering techniques to remove inconsistent height estimates, such as geostatistical~\cite{felicisimo1994parametric}, Kalman \cite{wang1998applying}, or Gaussian~\cite{walker2006comparative} filters. However, these techniques from the world of low-level signal processing lack a higher-level understanding of the observed scene and apply the same, rudimentary filtering rule at all pixels, irrespective of their geometric and semantic context. As a consequence, they tend to neither preserve height discontinuities nor enforce sharp building edges. Later works specifically tackle the problem of refining building shapes by fitting parametric models~\cite{haala1997integrated,sirmacek2010detecting,lafarge2011building}. Typically, such methods predefine a set of primitive shapes (and composition rules) and are thus inherently limited by that \say{library} in their ability to refine complex building shapes. Furthermore, they quickly reach their limits when the initial surface estimate is corrupted by significant noise, such that the correct primitive cannot be recognized.
In summary, we posit that, to further improve DSMs and to facilitate downstream modeling applications, such as the automatic vectorization of 3D buildings~\cite{wang2021machine}, one must go beyond the simplistic smoothness assumption of conventional stereo reconstruction and inject more prior knowledge about plausible surface shapes; while at the same time maintaining enough flexibility to accommodate the variety of real surfaces beyond schematic \say{box+roof} primitives.

Deep neural networks have excelled in many image interpretation tasks, thanks to their ability to represent complex priors that cannot be described by a small set of hand-coded rules and to extract those priors from example images in a data-driven manner.
In the domain of satellite stereo, \cite{bittner2018dsm} were perhaps the first who leveraged a learning-based approach to see image pixels in context and convert raw, noisy height measurements into a more accurate DSM. Later extensions introduced a \emph{single} ortho-rectified panchromatic satellite view to guide the DSM refinement~\cite{bittner2019dsm}, an improved network architecture~\cite{bittner2019late,bittner2020long,wang2021machine}, and a multi-task loss~\cite{bittner2019multi} to jointly optimize for building shapes, roof orientations, and roof type classification. All these methods seem to imply that a rather involved network architecture (cGAN) and multiple loss terms, including adversarial losses, are needed to refine an imperfect DSM into a more accurate and detailed one.

Here, we show that a \emph{residual learning} strategy greatly simplifies the learning problem. Instead of predicting absolute heights, as in \cite{bittner2018dsm,bittner2019dsm,bittner2019late,bittner2019multi,bittner2020long}, we focus on learning only a residual correction at every pixel, which is added to the initial height estimate. I.e., we retain the imperfect input surface and learn how to modify it to refine building shapes, complete missing surface details, and remove artefacts of the preceding stereo reconstruction.
Notably, what is an undesired artefact is defined in a data-driven manner via the training data, which makes it possible to include advanced filtering options. E.g., we will show that to obtain a model of the built environment, without vegetation, one only has to supply vegetation-free training labels, and the network will implicitly learn the corresponding filter.

The proposed residual network can be interpreted as an intricate geometric prior, learned from data, that seamlessly complements the simple (piece-wise) smoothness prior of classical stereo algorithms. The learned prior comes into use where neither the measured photo-consistency nor the assumptions built into the initial stereo algorithm are able to correctly reconstruct the surface.
Furthermore, we demonstrate that the residual network performs a lot better when given the chance to jointly exploit the initial geometry \emph{and} the original image content of \emph{two} ortho-rectified satellite views. Compared to previous works, our \resdepth network has a simple, efficient \mbox{U-Net} architecture~\cite{ronneberger2015u} and minimizes a standard $\ell_1$-loss w.r.t.\ per-pixel heights. \resdepth consistently reduces the errors of state-of-the-art stereo matchers, under ideal conditions by up to $\approx$60\%. 

A current limitation of prior work on learned DSM enhancement, including our own, is that the idea has so far mostly been validated under ideal machine learning conditions, where the characteristics of the training data almost perfectly match those of the test data. Little is known about the ability to generalize to new geographic conditions, except for two qualitative examples in~\cite{bittner2020long}.

The present paper builds on our preliminary work~\cite{stucker2020resdepth}, where we have introduced the concept of residual learning for DSM refinement. In that proof-of-concept study, the \resdepth approach was validated in a setting with barely any domain shift, where the training and test regions share the same urban style, topography, and imaging conditions (viewpoint, illumination). Such a setting may already be of use in certain scenarios,%
\footnote{E.g., to regularly update the model of a fixed local region that is periodically monitored from the same orbits.}
but the learned prior will include the particular viewing conditions and urban layout seen in the training data. It may thus not transfer all that well to a different stereo geometry, let alone a distant geographic location. Here, we demonstrate that \resdepth, with suitable extensions, is able to learn a generic prior for DSM refinement, which remains valid under varying imaging conditions (lighting and atmospheric influences as well as viewing directions and stereo geometry) and across different geographic locations with varying urban planning and construction styles. Our main contributions are as follows:
\begin{enumerate}
    \item We describe \resdepth, an efficient, easy-to-use, no-frills neural architecture for DSM refinement. \resdepth removes noise and outliers from imperfect input DSMs; restores shape regularities such as vertical planes, straight and/or parallel edges, etc.; and even adds missing surface details that are visible in the images but have been lost during the initial stereo reconstruction.
    \item We show that, when appropriately trained, \resdepth is able to generalize to viewing directions, imaging conditions, urban layouts, and architectural features that differ from those seen during training. On the one hand, we demonstrate that, to a large extent, suitably varied training data already equips the model with the ability to disentangle height errors from other sources of appearance change. On the other hand, we revisit data normalization, which we found critical to obtain good geographical generalization, and suggest a local standardization scheme that introduces some invariance against the degree of terrain undulation (hilly vs.\ flat).
    \item We present an extensive set of experiments and ablation studies that empirically examine \resdepth's ability to enhance DSMs in various scenarios. Among others, we apply it to new stereo images not seen during training; to a new city that has not been seen during training and for which no reference data may be available; and we learn a model from data sampled in two different cities and show that the resulting, more general model can improve DSMs of unseen test sites in both cities. To the best of our knowledge, this is the first study that systematically and comprehensively evaluates the generalization of learned DSM refinement for various practically relevant use cases.
\end{enumerate}

\section{Related Work}
We give a brief overview of conventional stereo matching algorithms before proceeding to deep and end-to-end stereo and disparity refinement methods. Finally, we cover recent advances in using deep learning for DSM enhancement.

\subsection{Classical Stereo Matching}
Stereo methods find a dense set of correspondences that have high photo-consistency while at the same time forming a (piece-wise) smooth surface. Typically, stereo matching is formulated as a multi-stage optimization problem~\cite{scharstein2002taxonomy}, consisting of matching cost computation, matching cost aggregation, disparity calculation, and disparity refinement. Local stereo algorithms usually make implicit smoothness assumptions by aggregating the matching costs within a finite window and compute optimal disparities by determining the disparity with minimal cost per pixel. In contrast, global stereo algorithms formulate the task of disparity estimation as an energy minimization problem, where smoothness priors are explicitly imposed and efficiently approximated by using graph cuts~\cite{kolmogorov2001computing}, dynamic programming~\cite{sun2003stereo,hirschmuller2005accurate,klaus2006segment} or the PatchMatch method~\cite{bleyer2011patchmatch}. For processing high-resolution images, stereo algorithms are often employed in an iterative manner using a spatial pyramid scheme~\cite{rothermel2012sure}, whose later iterations can be seen as a refinement of a coarser initial solution. Other methods are by design iterative, such as many variational schemes~\cite{slesareva2005optic,ben2007variational} and methods based on mesh surfaces~\cite{lengagne2000_3d}.

\subsection{Deep Stereo Matching}
In the last few years, the focus has been on leveraging the power of deep learning to boost the performance of stereo matching. CNN-based stereo matching methods are first introduced in~\cite{zbontar2016stereo} to substitute hand-crafted matching cost metrics with a learned measure for patch similarity within a conventional optimization framework. Other early attempts employ CNNs for matching cost calculation~\cite{luo2016efficient,chen2015deep}, matching cost aggregation~\cite{seki2017sgm} and learned disparity refinement~\cite{gidaris2017detect,batsos2018recresnet}. Later works~\cite{mayer2016large,kendall2017end,chang2018pyramid} seamlessly integrate all steps of the traditional stereo pipeline in a learned optimization to directly regress dense disparity maps from stereo images in an end-to-end manner. Recent learning-based stereo methods significantly improve the computational efficiency and memory consumption~\cite{tulyakov2018practical,duggal2019deeppruner,xu2020aanet} and adapt the encoder-decoder architecture to handle high-resolution images~\cite{yang2019hierarchical,gu2020cascade,liu2020novel}.

\subsection{Residual Learning for Stereo Matching}
Directly regressing high-resolution disparity (or depth) maps that recover small details and thin structures is computationally expensive and challenging to train. Many state-of-the-art deep stereo methods thus adopt a residual learning strategy~\cite{pang2017cascade,jie2018left,wang2020fadnet,liang2021StereoMU} by internally splitting the computation into a first, coarse disparity estimation and a subsequent refinement network. In this way, the second network focuses solely on learning the (often complicated and highly non-linear) residual correction to improve the initial disparity estimate, which is much easier and more efficient than directly learning the fine-grained disparities in a single step. This process can also be realized hierarchically by cascading multiple refinement modules~\cite{gu2020cascade}, where each module refines the upsampled intermediate disparity estimate of the preceding module. However, these methods do not investigate the influence of the refinement step in isolation and seem to imply that an end-to-end, deep integration of the cascade is crucial. We have shown in our previous work~\cite{stucker2020resdepth} that, to infer detailed and accurate depth maps, it is actually not necessary to embed the refinement step in an end-to-end learning framework. The critical step for high-quality reconstructions appears to be the learned refinement, which can be accomplished in isolation using a simple standard architecture that needs much less training data. According to our empirical findings, depth maps obtained from classical stereo methods are sufficient as initialization.

\subsection{Learned Disparity Refinement}
Our work is closely related to recent, rather complex disparity refinement networks that operate either as the last component of end-to-end deep stereo pipelines~\cite{pang2017cascade,jie2018left} or as standalone methods~\cite{gidaris2017detect,batsos2018recresnet}. The methods mainly differ in the type of additional information used to refine an initial, coarse disparity map. The network proposed in~\cite{gidaris2017detect} exploits a confidence map and monocular image information to detect and refine regions of incorrect disparity, whereas \cite{jie2018left} uses a left-right comparison to focus the refinement on inconsistent disparities between the two views. Unlike~\cite{gidaris2017detect,jie2018left}, we refrain from explicitly selecting the pixels to be refined. Instead, we refine all pixels, following the reasoning that it is comparatively easy for a machine learning method to learn where applying an identity mapping is appropriate.

In this sense, our method is most closely related to~\cite{pang2017cascade,batsos2018recresnet}. While these methods employ a hierarchical scheme to regress and supervise residual updates across multiple scales, we leverage a large receptive field and exploit local and global long-range context to estimate a one-shot residual update at the resolution of the input. The method proposed in \cite{batsos2018recresnet} refines initial disparities using a recurrent residual network guided by a single image, claiming that a binocular setup would not be beneficial. In our preliminary work~\cite{stucker2020resdepth} and also in the present paper, we experimentally demonstrate that refinement using stereo information does outperform refinement based only on monocular image guidance. \cite{pang2017cascade} uses two input views and a geometric error map, defined as the absolute intensity differences between the left and warped right view. However, we argue that the latter is not meaningful for images captured under substantial lighting differences, such as satellite images from different days.

\subsection{Satellite-based DSM Refinement}
In the domain of satellite stereo, \cite{bittner2018dsm} were perhaps the first who proposed to learn a general prior for enhancing digital surface models. The method trains a conditional generative adversarial network (cGAN) to learn the mapping from a noisy and incomplete photogrammetric DSM towards a more realistic CAD-like DSM. Later works~\cite{bittner2019dsm,bittner2019late} complement this initial approach by leveraging the spectral information of a \emph{single} ortho-rectified panchromatic satellite image to more faithfully reconstruct building outlines and complete missing building parts. More recently, \cite{bittner2019multi,liebel2020generalized} incorporate a multi-task loss to jointly optimize for building shapes, roof orientations, and roof type classification. The latest extensions introduce skip connections in the generator~\cite{bittner2020long}, learned weights of the multi-task loss terms~\cite{bittner2020long}, and an additional attention module~\cite{wang2021machine}. In comparison, our preliminary work~\cite{stucker2020resdepth} uses stereo guidance but a much simpler \mbox{U-Net} architecture~\cite{ronneberger2015u}, trained with just a standard $\ell_1$-loss on pixel-wise height residuals. This strategy delivered very promising results, but the proof-of-concept did not consider generalization across geographic contexts and imaging conditions. Indeed, we did find that it was challenged in particular by topographic conditions not encountered during training~\cite{stucker2020resdepth_arxiv}.

Interestingly, none of the existing works considered the influence of lighting, atmospheric variations, and changes in the viewing geometry on the refinement,%
\footnote{Perhaps because multiple satellite images covering a range of conditions are required to study those factors.}
although they are fundamental for a system that can operate at large scale. As a first step in that direction, we study measures that ensure invariance across those sources of variability and provide a systematic, quantitative analysis of geographical generalization.

\section{Method}
\begin{figure*}[!t]
\centering
\includegraphics[trim={0 0 0 10}, clip, width=0.9\textwidth]{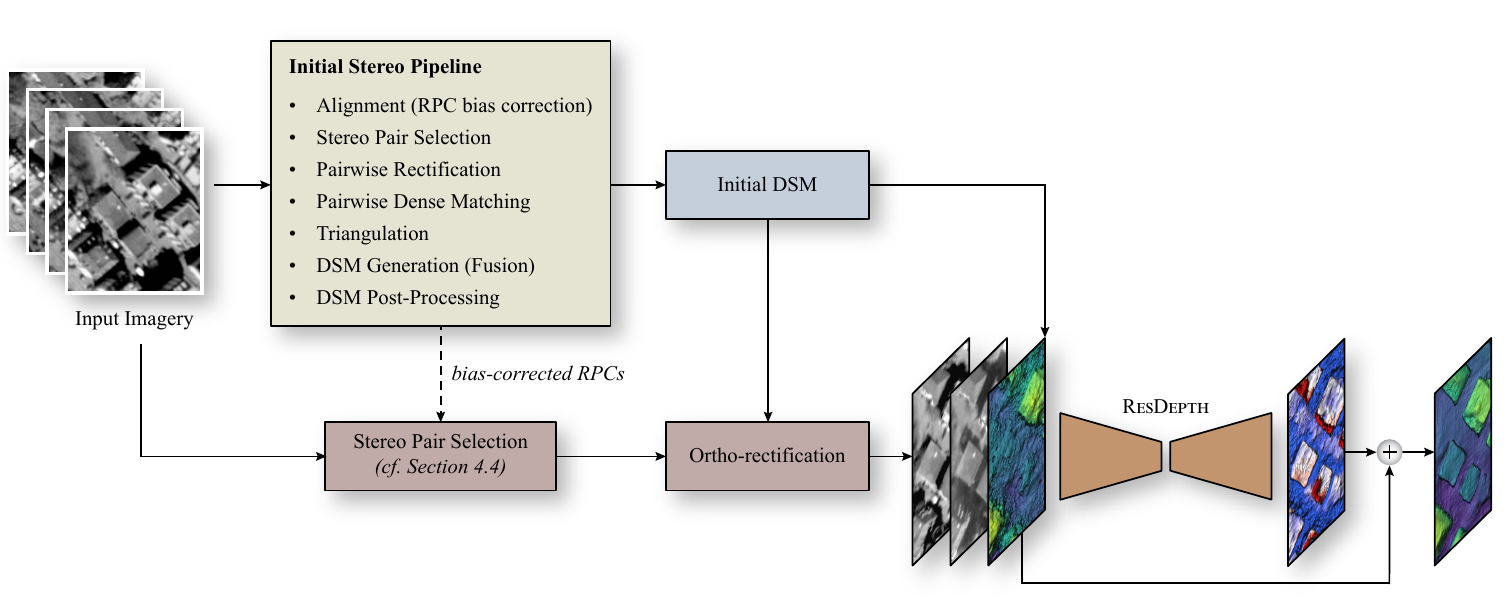}
\caption{Overview of our DSM refinement method. Given a photogrammetric DSM and two ortho-rectified panchromatic stereo images as input, a residual neural network is trained to regress a height correction for every pixel, which is added to the initial height to refine it. \resdepth is based on a simple \mbox{U-Net} architecture~\cite{ronneberger2015u} with symmetric skip connections between corresponding layers of the encoder and decoder. The detailed architecture is shown in Fig.~\ref{fig:architecture}. Thanks to its modular design, \resdepth can seamlessly complement any stereo matching pipeline to improve the reconstructed surface. The list of processing steps for the initial stereo reconstruction are representative of current best practice; individual pipelines may deviate from it.}
\label{fig:overview}
\end{figure*}

We propose to use a deep neural network to encode a geometric prior for high-quality, dense surface reconstruction. Our method starts from satellite images with overlapping fields of view and known camera poses. Moreover, we assume that an initial coarse reconstruction of the observed scene has already been generated with existing multi-view stereo matching and/or depth map fusion techniques. The core of our approach is \resdepth, a simple yet highly effective deep convolutional neural network trained to refine the initial surface estimate by regressing an additive residual correction at every pixel, as illustrated in Fig.~\ref{fig:overview}. The inputs to \resdepth are the initial reconstruction and two (typically panchromatic) satellite views that we ortho-rectify with the help of the initial surface estimate.

A key challenge when using machine learning is to reach the right balance between specificity and generality of the model: on the one hand, the learned prior should capture the relevant task-specific regularities in the training data, such that it has a positive impact on the reconstruction. On the other hand, it should not overfit to the training set and extract spurious correlations that do not generalize to the target scenes. 
In practice, it is likely that the training and test regions for 3D reconstruction from satellite imagery will lie in distinct neighborhoods of a city or even in different cities, i.e., their geographical context is not the same. Furthermore, the training and test images will exhibit varying viewing angles and lighting conditions. For practical applications, \resdepth must therefore learn to refine building shapes, complete missing surface details, and remove noise while simultaneously generalizing across viewing directions, radiometric differences, geographic locations, and variations in urban layout and architecture.
In the following, we describe the \resdepth model, building on our preliminary work~\cite{stucker2020resdepth}. We first describe the representation of the initial surface estimate and the use of ortho-rectified satellite images as additional input to \resdepth. Then, we detail the architecture of the neural network and discuss newly incorporated measures to avoid overfitting and ensure the model generalizes well across different geographic contexts and imaging conditions.

\subsection{Initial Reconstruction}
To leverage the efficiency of 2D convolutional neural networks, we use a digital surface model (DSM, i.e., a height field in raster format) as scene representation. We parametrize the DSM as a regular horizontal $x,y$-grid in the local UTM zone, with pixel values denoting ellipsoidal heights w.r.t.\ the GRS80 ellipsoid. To compute the initial DSM, we use a conventional stereo pipeline based on hierarchical semi-global matching~\cite{rothermel2012sure} (cf. Section~\ref{sec:initial_recon}) but note that \resdepth is generic and can refine any DSM, irrespective of which method was used to generate it.\footnote{After concordant training, since \resdepth does adapt to the specific biases and error patterns of the input DSM. The ability to adapt to the peculiarities of different stereo pipelines is beyond the scope of the present paper, but has been shown in~\cite{stucker2020resdepth_arxiv} (Appendix C.2) for the examples of \textit{tSGM}~\cite{rothermel2012sure} and \textit{s2p}~\cite{de2014automatic}.}

\subsection{Image Rectification}
Besides the initial DSM, \resdepth receives two pan\-chromatic satellite images as additional input. In this way, the network can exploit image textures, alignment errors between each individual image and the DSM, and discrepancies between the stereo correspondence and the 3D shapes.

We ortho-rectify the satellite images using the initial DSM to obtain (approximate) pixel-wise alignment. If the initial DSM were perfect, 2D image shapes would exactly align with 3D surface shapes, and the ortho-images would be maximally photo-consistent (except for regions occluded in one of the views). During ortho-rectification, we deliberately do not perform ray-casting to handle occlusions. Instead, we render duplicate textures if a ray intersects the surface twice, resulting in photo-metrically inconsistent and systematically repeated textures. These systematic patterns, along with disparities between the ortho-images and misalignments between each ortho-image and the initial DSM, provide a strong signal about the underlying surface shape. Therefore, \resdepth acts as a learned prior for both the 3D surface shapes themselves and for their alignment with the 2D image content.

\subsection{Deep Residual Regression Network}
Our method uses a deep convolutional neural network at its core to regress a residual height correction for every pixel of the DSM. Fig.~\ref{fig:architecture} depicts the detailed network architecture of \resdepth. We employ the well-proven \mbox{U-Net}~\cite{ronneberger2015u} architecture, consisting of symmetric encoding and decoding modules. The inputs to the encoder are 256$\times$256 pixel tiles with three channels, namely the initial photogrammetric DSM and the two ortho-images. The encoder gradually transforms that input into a discriminative multi-scale representation through a sequence of five downsampling blocks. Each block consists of a 3$\times$3 convolutional layer with stride~1, followed by batch normalization, a rectified linear unit (ReLU) as non-linear activation function, and a 2$\times$2 max-pooling layer. The decoder expands the latent representation to a single-channel output with the same spatial resolution as the network input. The structure of the upsampling blocks is the same as for the downsampling blocks, except that fractionally strided transposed convolutions replace max-pooling layers. A 3$\times$3 convolutional layer follows the last upsampling block to output the height correction for every pixel.

Through the sequence of downsampling blocks, the network progressively loses high-frequency details and localization information. Thus, skip connections from encoder to decoder levels of equal resolution are a vital component of the network to propagate high-frequency information. As mentioned, we also include a long residual connection that directly adds the initial DSM to the output of the last decoder layer, such that the network regresses residual height updates rather than absolute heights.

\subsection{Generalization}
\label{sec:generalization}
In the following, we discuss three factors that impact the network's capability to generalize across space and time, along with our associated mitigation strategies. 

\subsubsection{Acquisition Geometry and Radiometry}
\label{sec:generalization_images}
If the satellite images of the training and target scenes have been acquired under the same stereo geometry and lighting, the scene geometry is the dominant cause for disparity variations and occlusion patterns. E.g., equally tall buildings in the training and test sets will have equal disparities when captured under the same viewing angles (as in~\cite{stucker2020resdepth}). With significantly different viewing angles, the displacements during inference will deviate from those seen during training; and the same is true for unseen radiometric conditions, including seasonal changes. Hence, one would expect the performance of \resdepth to deteriorate, as the network has not learned invariance to these sources of variability. We sidestep this restriction at the level of training data: to better generalize across variations in acquisition geometry and the images' radiometry, we collect a training set of multiple pairs with varying imaging conditions. In particular, we select (monocular) training images covering the same region, split them into a training and a test set,\footnote{Such that both training and test data contain pairs with predominantly north-south as well as predominantly west-east oriented baselines (cf. Section~\ref{sec:pair_selection}).} and construct all stereo pairs in the training set that fulfill the selection criteria described in Section~\ref{sec:pair_selection}. All of these pairs are then combined with the same initial DSM to form training samples. That procedure gives \resdepth the possibility to learn from ortho-images that show the same geographic location but stem from different stereo pairs---thus containing discrepancies that are not only a function of the height error but also of the acquisition geometry and illumination. To predict the same height correction from different pairs, \resdepth must learn invariance against varying imaging conditions within the expected range.

\subsubsection{Urban Style and Architecture}
Again, we prefer to address this factor by appropriate training data: we see no viable unsupervised strategy to adapt the shape prior to a new location whose architectural characteristics are markedly different from the training area. Hence, we simply gather a diverse set of training regions to learn suitable height corrections for a variety of scenarios, ranging from small, widely spaced residential buildings to tight city centers and commercial/industrial areas. Moreover, we take care to include flat and hill regions to be able to handle different terrain and areas with a varying abundance of trees and vegetation.

\subsubsection{Data Normalization}
\label{sec:normalization}
The best practice for neural network training is to normalize the data as much as possible before feeding it to the network in order to learn a better model and also to speed up the convergence of the iterative training.
We found experimentally that the normalization is indeed critical also to obtain good generalization. Not surprisingly, when the training area is comparatively flatter than the test area or vice versa, inappropriate normalization of the DSM will introduce biases that negatively affect the prediction.
To minimize the influence of the terrain, it is important to preserve a meaningful vertical scale. We suggest the following local standardization scheme, which we found to work best for our data, which includes both a predominantly flat city (Berlin) and one with height variations on the order of 100$\,$m over distances \textless1$\,$km (Zurich). To rescale the data, we compute a single, robust scale factor across all data samples during training, so as to preserve a (relative) notion of scale. To that end, we first center each 256$\times$256 pixel DSM patch to its mean height. Then, we compute the standard deviation of the height within each patch, discard standard deviations below the \nth{5} percentile and above the \nth{95} percentile to ensure robustness, and average the remaining ones to obtain a single, robust estimate of the standard deviation. At test time, we center each DSM patch to its mean height and rescale with the standard deviation estimated during training. To normalize the panchromatic ortho-images, we globally compute the mean radiance and its standard deviation across all images used during training.

\section{Datasets}
\subsection{Study Areas}
\label{sec:area_description}
We evaluate our method on satellite images acquired over Berlin, Germany and Zurich, Switzerland. Our goal was to assess the generalization performance of \resdepth across different districts of the same city and from one city to the other. Therefore, we chose multiple areas per city and within each area geographically separate regions for training, validation, and testing.

Fig.~\ref{fig:datasets} depicts these areas for both cities. In total, we select five distinct areas, each covering 4{$\,$km\textsuperscript{2}}. Two areas are in Berlin, denoted as \berlinOne and \berlinTwo, three areas in Zurich, denoted \zurichOne, \zurichTwo, and \zurichThree. The areas in Berlin are located in the Pankow district to the north of the city center. \berlinOne includes closely spaced, detached residential buildings, terrace housing, allotments, and large fairly flat industrial buildings. The vegetation cover is moderate. \berlinTwo consists almost exclusively of tightly packed terrace houses, with only little vegetation. In both \berlinOne and \berlinTwo the terrain is entirely flat.

The areas in Zurich include both flat and rather steep terrain, with \zurichOne the most undulating region among all five areas. \zurichOne contains widely spaced, detached residential buildings, allotments, and high commercial buildings, including the tallest building of Zurich with a height of 126$\,$m. Furthermore, it includes a stretch of the river Limmat and a forested hill, two challenging surface types for dense matching. \zurichTwo is located in the heart of Zurich and comprises the historic city center, another stretch of the Limmat river, and parts of the lake. Lastly, \zurichThree consists of low residential districts, moderately high buildings, and industrial areas.

\subsection{Satellite Imagery}
For Berlin, we had at our disposal nine \mbox{WorldView-2} panchromatic images captured between 2010 and 2020. For Zurich, we use one \mbox{WorldView-3} and 14~\mbox{WorldView-2} panchromatic images acquired between 2014 and 2018. The shortest time interval between two acquisitions amounts to 44~days for Berlin and 22~days for Zurich. The average ground sampling distance is $\approx$0.5$\,$m at nadir, see Table~\ref{tab:imagery}.

\begin{table*}
    \footnotesize
    \centering
    \caption{Acquisition details of the Berlin and Zurich satellite image datasets and selection criteria used to determine the stereo pairs for dense matching (middle rows) and DSM refinement (last rows). We report the number of images and stereo pairs for each study area separately.}
    \label{tab:imagery}
    \begin{adjustbox}{max width=\textwidth}
    \begin{tabular}{@{\hspace*{1.8em}} p{4.5cm} c c @{}}
        \toprule
        & Berlin & Zurich \\
        & (\berlinOne/\berlinTwo)  & (\zurichOne/\zurichTwo/\zurichThree)    \\
        \midrule
        \multicolumn{3}{@{}l}{\textbf{Acquisition}}\\
        Sensor          & WV2         & WV2/WV3       \\
        Time period     & 2010--2020  & 2014--2018    \\
        Avg. GSD [m]    & 0.52$\,$m   & 0.50$\,$m     \\
        No. of images   & 9           & 15            \\
        \midrule
        \multicolumn{3}{@{}l}{\textbf{Initial DSM reconstruction}}\\
        Intersection angle [$^{\circ}$]     & [5, 30]  & [5, 30] \\
        Incidence angle [$^{\circ}$]        & [0, 40]  & [0, 40] \\
        Sun angle difference [$^{\circ}$]   & [0, 35]  & [0, 35] \\
        No. of images                       & 9/9      & 10/12/14 \\
        No. of stereo pairs                & 15/15    & 26/33/40 \\
        \midrule
        \multicolumn{3}{@{}l}{\textbf{DSM refinement} (\resdepth)} \\
        Intersection angle [$^{\circ}$]              & [10, 28]  & [10, 28] \\
        Incidence angle [$^{\circ}$]                 & [0, 40]   & [0, 40] \\
        Sun angle difference [$^{\circ}$]            & [0, 35]   & [0, 35] \\
        No. of images                                & 9/9       & 8/8/8 \\
        No. of stereo pairs (cf. Fig.~\ref{fig:valid_pairs}) & 15/15     & 9/9/9 \\
        \bottomrule
    \end{tabular}
    \end{adjustbox}
\end{table*}

\subsection{Initial DSM Reconstruction}
\label{sec:initial_recon}
We use a re-implementation of state-of-the-art hierarchical semi-global matching~\cite{rothermel2012sure}, tailored to satellite images, to generate the initial DSM with a grid spacing of 0.25$\,$m. As usual in satellite imaging, we solely use the panchromatic channel, which has the highest spatial resolution, for reconstruction. All available images are used in order to leverage the full redundancy and obtain the best possible baseline (initial) DSM.

The overall pipeline for reconstructing the initial DSM follows best practice for satellite-based photogrammetry. First, we apply bias correction of the supplied rational polynomial coefficients (RPC) projection models with the method of~\cite{patil2019new}. Then, we use a heuristic inspired by \cite{facciolo2017automatic} to determine suitable image pairs for dense matching. Starting from all possible pairings of images, we eliminate those whose intersection angles in object space are \textless5$^\circ$ or \textgreater30$^\circ$ (measured at the center of the region of interest), or whose incidence angles are \textgreater40$^\circ$ (mean of the two images). Furthermore, we discard image pairs whose difference in sun angle is \textgreater35$^\circ$. It has been observed~\cite{qin2019critical} that changes in sun angle are one of the most decisive factors for the quality of stereo matching, as they lead to unequal directional illumination of anisotropic and specular ground surfaces and to different shadow patterns, two aspects that are particularly challenging for stereo algorithms. Lastly, we follow \cite{krauss2019cross} and keep all remaining stereo pairs, irrespective of differences in acquisition time between the two images. Empirically, the time between two observations both in absolute and relative (modulo \textonehalf~year) terms does not seem to be a critical factor, given adequate stereo acquisition geometry.

Based on these criteria, we select between 15~(\berlinOne) and 40~(\zurichThree) image pairs for pairwise rectification and pairwise dense matching (cf. Table~\ref{tab:imagery}, middle rows). Next, we triangulate corresponding points into a 3D point cloud per image pair using the inverse RPC projection function, then fuse all point clouds into a single, coherent multi-view raster DSM by computing the cell-wise median of the $n$ highest 3D points,\footnote{To ensure that a majority of the points lie on a common, highest surface~\cite{rothermel2016}.} where $n$ is defined as the average number of 3D points per grid cell. Lastly, we follow a popular strategy implemented in several aerial and satellite-based photogrammetry packages and employ post-processing operations to denoise the initial DSM, remove spikes, and fill cells without a valid height with inverse distance weighted (IDW) interpolation.

\subsection{Stereo Pair Selection for \resdepth}
\label{sec:pair_selection}
Our network uses stereo information as guidance. This raises the question of which satellite images should be grouped to form stereo pairs as input to \resdepth. We follow similar heuristics as for the initial DSM reconstruction (see Section~\ref{sec:initial_recon}) to determine suitable stereo configurations (cf.~Table~\ref{tab:imagery}, last rows). Furthermore, we discard all stereo pairs that include at least one image with snow cover and ortho-images whose footprint does not fully overlap the study areas.

Fig.~\ref{fig:valid_pairs} depicts the remaining 15~stereo pairs for Berlin and the 9~stereo pairs for Zurich. For both cities, we split the selected stereo pairs into two mutually exclusive groups $A$ and $B$, such that each group contains pairs with predominantly north-south as well as west-east oriented baselines (see Fig.~\ref{fig:train_test_pairs}). To ensure the inference results are representative for previously unseen images (i.e., \resdepth is applied to images that were not seen during training), we use only stereo pairs in $A$ for training and only those in $B$ for testing.

\subsection{Ground Truth}
We train \resdepth in a fully supervised manner. For supervision, we use surveying-grade  DSMs rendered from the publicly available city models of Zurich\footnote{\url{https://www.stadt-zuerich.ch/ted/de/index/geoz/geodaten_u_plaene/3d_stadtmodell.html}} and Berlin.\footnote{\url{https://www.businesslocationcenter.de/en/economic-atlas/download-portal/}} The model of Zurich is provided by the municipal surveying department and was assembled semi-automatically by merging airborne laser scans, building and road boundaries (including bridges) from national mapping data, and roof models derived by interactive stereo digitization. The height accuracy is given as $\pm$0.2$\,$m for buildings and $\pm$0.4$\,$m for general terrain. The model is based on data collected before 2015. The CityGML model of Berlin was created in a similar manner, based on national mapping data and aerial images captured in 2013. 

Both city models differ from the state visible in the satellite images in a handful of places. For quantitative evaluation, we mask out areas with evident temporal differences due to construction activities. To this end, we manually draw polygons around all areas with large temporal inconsistencies between the satellite images and the ground truth DSMs and rasterize those polygons to a binary mask of pixels with unreliable reference height. Furthermore, note that vegetation is not included in either of the two city models. Consequently, \resdepth learns to filter out vegetation present in the initial stereo reconstruction.

\section{Experiments}
Deep neural networks have a vast amount of trainable parameters. This gives them the capacity to adapt to intricate details of complex training data distributions, e.g., as specific architectural styles or common building shapes and heights, but it also entails the danger of overfitting to local peculiarities that may be at odds with the goal to learn a generic prior valid for DSM refinement in unseen locations or even different, distant cities.
In the following, we train and evaluate \resdepth in a number of scenarios with increasing complexity to investigate geographical generalization \textit{(i)}~within a local region, \textit{(ii)}~across regions within the same city, and \textit{(iii)}~between Berlin and Zurich. Furthermore, we experimentally study the influence of the image information to guide the reconstruction and the method's tolerance to imaging conditions not seen during training.

\subsection{Setup}
\subsubsection{Geographical Cross-Validation}
In the first experiment, we focus on a single geographic area (\zurichOne) and use data from the same stereo pair for training and testing, so as to isolate the principle of learned DSM refinement from the choice of stereo pairs and validate the learned prior under ideal conditions. For that experiment, we chose a single best stereo pair according to three criteria, namely low intersection angle, small time difference between acquisitions (similar season), and low cloud coverage. We then perform 5-fold cross-validation over five spatially disjoint stripes of \zurichOne, see Fig.~\ref{fig:datasets}.
For a more fine-grained analysis, we also separately analyze the performance for building and terrain pixels (according to the ground truth) to better disentangle the learned corrections and identify failure cases.

\subsubsection{Influence of Image Guidance}
Next, we conduct an ablation study and use the previously selected stereo pair and the same geographic area (\zurichOne) to systematically study the impact of each input channel on the quality of the refined DSM. We construct several variants that differ w.r.t.\ the number and combination of input channels but are otherwise identical. In particular, we keep the network architecture fixed and train each variant using the same training settings and DSM samples. The network configuration based on stereo guidance represents our full model, referred to as \resdepth-stereo (or simply \resdepth, as this is our default setup). The first variant, \resdepth-0, omits both input views. It thus merely learns a geometric prior to transform a noisy and incomplete DSM to a more realistic DSM, without conditioning on any image evidence. Note, however, that this is still more powerful than a simple context-aware smoothing operation such as bilateral filtering~\cite{tomasi1998bilateral} due to the added capability to recognize and exploit long-range correlations such as straight gable lines. The next variant, \resdepth-mono, receives the initial DSM and a single view as input. Therefore, it has no access to stereo disparities but can still use the monocular image information to refine the DSM. On the one hand, this amounts to exploiting low-level information such as image gradients to localize and sharpen crease edges and height jumps, as in guided filtering~\cite{he2010guided}. On the other hand, it can also utilize high-level information implicit in the image like semantic objects or their relative layout, e.g., to smooth trees and buildings differently or to complete missing object parts. The last variant, \resdepth-stereo\textsubscript{iter}, is a cascaded refinement approach consisting of two consecutive \resdepth-stereo networks with individual weights. The second stage receives the refined DSM of the first one and updated ortho-images as input to further reduce the new, smaller height errors.

\subsubsection{Influence of Unseen Image Views}
To test the method’s sensitivity to images unseen during training, we test the performance of the previously trained \resdepth-stereo model when using a new, previously unseen stereo pair during inference. In detail, we perform the refinement once with each stereo pair in group $B$ (see the right plot of Fig.~\ref{fig:train_test_pairs}) and average the statistics over both predictions. Furthermore, we train \resdepth-stereo from scratch while leveraging all stereo pairs in group $A$ during training (see the right plot of Fig.~\ref{fig:train_test_pairs}). Again, we test the performance of the trained model using the stereo pairs in $B$ and average the statistics over both predictions.

\subsubsection{Geographical Generalization Within a City}
We train our model from scratch in one region in Berlin and once on two regions in Zurich. The training set includes all stereo pairs of group $A$ (see Fig.~\ref{fig:train_test_pairs}) to cover a range of different acquisition geometries. We then test the learned model on the held-out area of the respective (same) city, using only stereo pairs of group $B$. As the model has neither seen the region nor the images used during inference, the setting is representative to assess whether the model learns generic features that generalize across different districts of a city, or whether it overfits to the street layout or architectural style of the training locations.

\subsubsection{Geographical Generalization Across Cities}
In the last and most challenging experiment, we train our model on all areas of one city and use the learned model (without further fine-tuning) to refine the areas of the respective other city in order to determine how well \resdepth generalizes across geographically distant cities with different histories of urban planning and construction. Notably, this scenario is of practical relevance, as one may be faced with the task of enhancing the reconstruction of a city for which no reference data is available to train a location-specific model.

\subsection{Implementation Details}
We split each of the five geographic areas into five equally large and mutually exclusive stripes, see Fig.~\ref{fig:datasets}. In each area, we allocate three stripes for training, one for validation, and one for testing, except for the 5-fold cross-validation experiment in \zurichOne, where we use four stripes for training and the \nth{5} for testing. When training over multiple geographic regions, we stick to the same, fixed split into training and test stripes to ensure the quantitative results are as comparable as possible.

For each geographic area, we randomly sample 20'000 training patches of 256$\times$256 pixels (64$\times$64$\,$m in world coordinates). To increase the variability in the training data and avoid biases due to the specific topography, we augment the data by randomly rotating the training patches by $\alpha\in\{0^\circ,90^\circ,180^\circ,270^\circ\}$ as well as random flipping along the horizontal and vertical axes. When using multiple stereo pairs during training, we further randomly flip the order of the ortho-images in every training sample.

We follow the same training procedure throughout all experiments: the network is trained in a fully supervised manner by minimizing the pixel-wise absolute distance to the ground truth DSM (i.e., the $\ell_1$-loss). We use the \textsc{ADAM} optimizer~\cite{adam} with a base learning rate of 0.0002 ($\beta_1$=0.9, $\beta_2$=0.999), weight decay of $10^{-5}$, and a batch size of 20. For the experiments that train on a single stereo pair, we drop the learning rate by a factor of 10 after every 200 training epochs and stop training when the validation loss does not significantly change over 100 epochs. Depending on the variant, this takes at most 400 epochs. If multiple stereo pairs are used in training, we drop the learning rate every 50 training epochs, in which case convergence is reached after 50--150 epochs, depending on the training region.

We have implemented \resdepth in PyTorch and run it on a NVIDIA GeForce~GTX~1080~GPU (8~GB RAM). Training takes $\approx$3.5~min per epoch (20'000 DSM patches and one stereo pair). The cost of inference is one forward pass per DSM patch. With our current implementation, which has not been optimized for speed, the refinement takes \textless5~sec/{$\,$km\textsuperscript{2}}. The main computational burden when applying \resdepth is actually the ortho-rectification of the images with the initial DSM, prior to running the refinement network. Source code is available at \url{https://github.com/stuckerc/ResDepth}.

\subsection{Evaluation Metrics}
To quantify the quality of the refined DSM, we report the mean absolute error (MAE), the root mean square error (RMSE), and the median absolute error (MedAE), computed over per-pixel deviations from the reference heights:
\begin{equation}
    \centering
    \mathrm{MAE} = \frac{1}{N} \sum_{i=1}^{N} \left ( \hat{y}_i - y_i \right ),
\end{equation}
\begin{equation}
    \centering
    \mathrm{RMSE} = \sqrt{ \frac{1}{N} \sum_{i=1}^{N} \left ( \hat{y}_i - y_i   \right )^2 },
\end{equation}
\begin{equation}
    \centering
    \mathrm{MedAE} = \mathrm{median} \left( \mid \hat{y}_i - y_i \mid \right),
\end{equation}
where $\hat{y}_i$ denotes the predicted height at pixel $i$, $y_i$ the corresponding reference height, and $N$ the total number of pixels. Moreover, we measure the bias of the prediction with the median error:
\begin{equation}
    \centering
    \mathrm{median} \left( \hat{y}_i - y_i \right).
\end{equation}
Under this definition, a positive bias indicates that the prediction is higher than the ground truth data and conversely for a negative bias.

\section{Results}
\begin{figure}[!t]
    \centering
    \includegraphics[trim={0 5 0 5}, clip,width=\columnwidth]{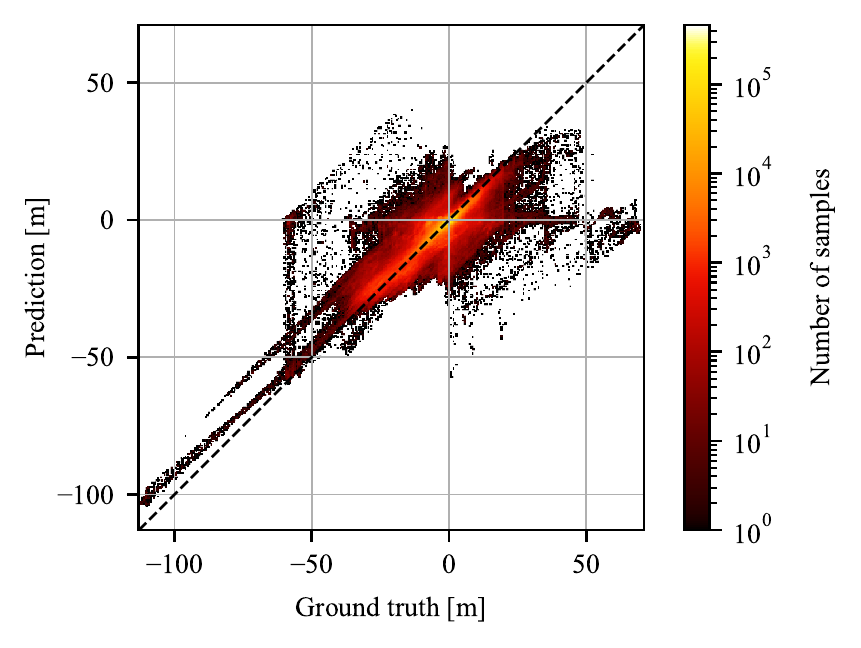}
    \caption{Ground truth residuals vs.\ DSM height corrections predicted by \resdepth (output of the last decoder layer) for test stripe~2 of \zurichOne.}
    \label{fig:crossval_regression_output}
\end{figure}

\subsection{Effect of \resdepth on DSM Quality}
We start by analyzing the impact of the \resdepth prior in isolation on area \zurichOne. In this way, its behavior under ideal conditions is separated from the ability to generalize across different viewpoints and geographic contexts,\footnote{Training, validation, and test stripes are of course mutually exclusive but lie near each other.} which will be examined in subsequent experiments. 

As typical for satellite-based stereo, the initial stereo DSM (Fig.~\ref{fig:ablation}, \nth{1}~row) is comparatively noisy due to the limited sensor resolution, the image quality, and the large sensor-to-object distance. It has a MAE of 3.89$\,$m and a MedAE of about 1.6$\,$m. The RMSE is $\approx$2$\times$ higher than the MAE due to a small number of very large matching errors (outliers); see Table~\ref{tab:ablation} for details.
After applying \resdepth, the MAE is reduced to 1.53$\,$m, the MedAE and RMSE both exhibit similar drops. I.e., the learned correction reduces the errors of the DSM to less than half their initial values. The difference is also reflected in a clearly improved visual quality, as illustrated by the examples in Fig.~\ref{fig:ablation}.

We go on to study the corrections predicted by the \resdepth network in more detail. The median error of that initial DSM is \mbox{-1.10$\,$m}, with the $\pm$25\% quantiles lying at \mbox{-3.65$\,$m} respectively 0.10$\,$m; indicating an underestimation bias, as one would expect from conventional stereo matching. Fig.~\ref{fig:crossval_regression_output} compares these errors with the corrections predicted by \resdepth for one of the five cross-validation folds. Overall, the predicted corrections agree well with the actual residual errors of the initial DSM but tend to be slightly too small (i.e., the underestimation is greatly mitigated but not completely compensated).

\begin{figure*}[!t]
\centering
\includegraphics[trim={0 10 0 5}, clip, width=\textwidth]{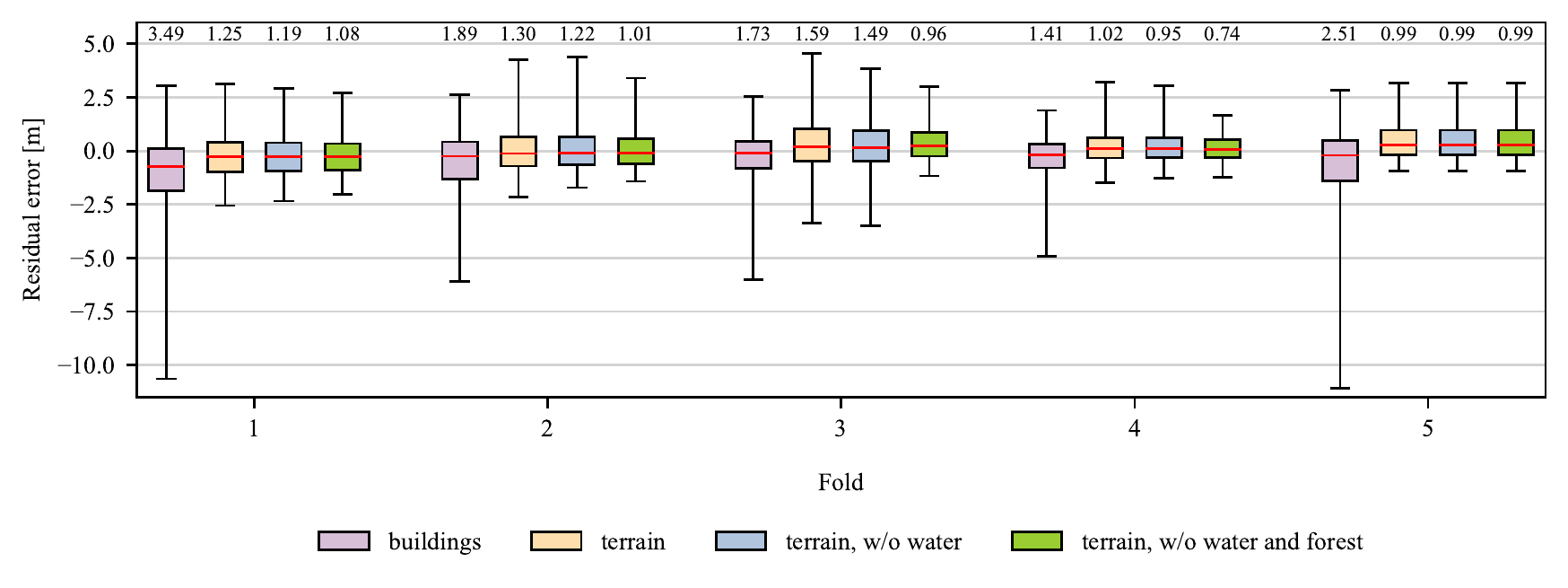}
\caption{Analysis of residual errors on \zurichOne (5-fold cross-validation). Positive error means that the predicted height is larger than the reference value. The boxplots show the median, the quartiles, and the \nth{5} and \nth{95} percentile. The label above each box specifies the corresponding MAE in meters.}
\label{fig:crossval_errors}
\end{figure*}

Fig.~\ref{fig:crossval_errors} shows the remaining errors w.r.t.\ ground truth after applying \resdepth, separated for the five folds. We use building masks derived from the city model of Zurich to analyze the errors for buildings and terrain separately.\footnote{Note that beyond houses, the building masks include further man-made non-terrain objects like bridges.} For computing the object-specific metrics, we dilate the building masks by two pixels (0.5$\,$m) to avoid aliasing at vertical walls.

The residual errors of terrain pixels are consistently smaller than of building pixels and are fairly constant across the folds. The third fold, with the biggest area of forested terrain (and water), exhibits the largest terrain errors, with a MAE of 1.6$\,$m. After excluding water bodies and densely forested areas from the evaluation, the MAE reduces to $\approx$1$\,$m, which is in accordance with the other folds. Overall, we observe almost no bias in the estimated terrain heights. As expected, the largest errors occur in densely forested areas, where the model has to guess the terrain height under a closed canopy; and at water pixels, which are not suited for (multi-temporal) stereo reconstruction.
For building heights, we observe a significant, systematic underestimation across all folds, with MAE between 1.4$\,$m and 3.5$\,$m, depending on the building heights. More specifically, building heights up to 40$\,$m are accurately predicted with only a minor bias (median error of \mbox{-0.4$\,$m}), see Fig.~\ref{fig:residual_errors_building}. The median error drastically increases for buildings taller than 40$\,$m, with extreme values of almost \mbox{-90$\,$m} for buildings taller than 100$\,$m.
These errors are due to the data-driven nature of machine learning: tall buildings are very rare in Zurich (fewer than 1\% of the building heights exceeding 40$\,$m) and therefore not adequately represented in the training data. Consequently, the learned model exhibits a bias towards at most moderately high buildings. That bias is particularly evident for very high buildings, which \resdepth tends to squeeze to the average building height seen during training, see Fig.~\ref{fig:prime_tower}.

Except for very tall buildings and dense forests, the results of our proposed refinement are consistent across the folds, suggesting that the model indeed learns a generic prior for building shapes and urban layout, which is invariant to the variations between different stripes. Furthermore, \resdepth is robust against label noise and reconstructs realistic scenes even in the presence of previously unseen building shapes and arrangement of buildings (see \mbox{\ref{sec:label_noise}}). In such cases, where no higher-level shape information could be learned from the training data, the refinement falls back to straightening the building outlines and making walls vertical.

\subsection{Influence of Image Guidance}
\begin{figure*}[!p]
    \def\mywidth{0.3\textwidth}  % ratio:   16:9
    \setlength{\tabcolsep}{0.2em}
    \centering
    \begin{tabular}{l m{\mywidth} m{\mywidth}  m{\mywidth}}
        \rotatebox[origin=c]{90}{\footnotesize Initial DSM} &  
        \includegraphics[width=\mywidth,trim={0 0 0 0},clip]{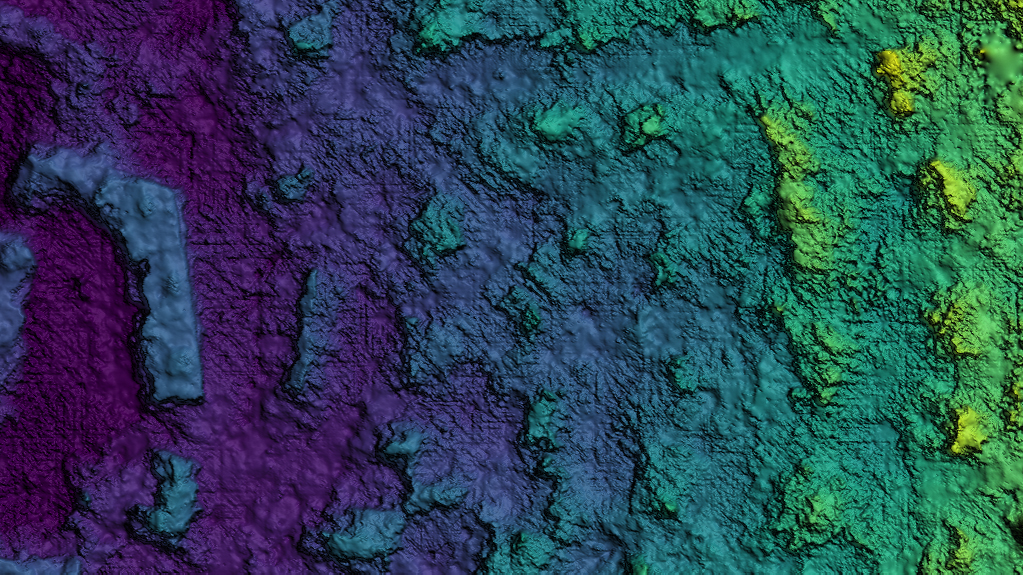} &
        \includegraphics[width=\mywidth,trim={0 0 0 0},clip]{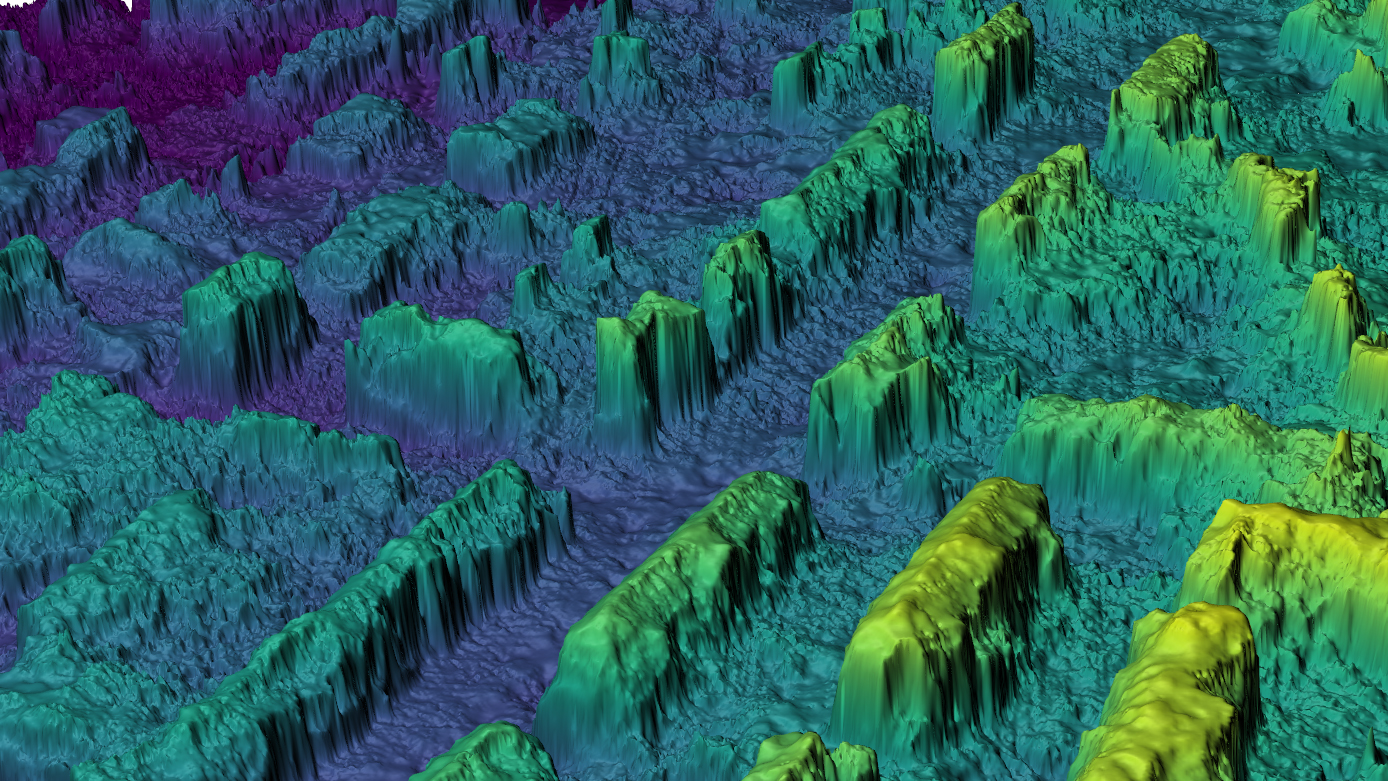} &
        \includegraphics[width=\mywidth,,trim={0 0 0 0},clip]{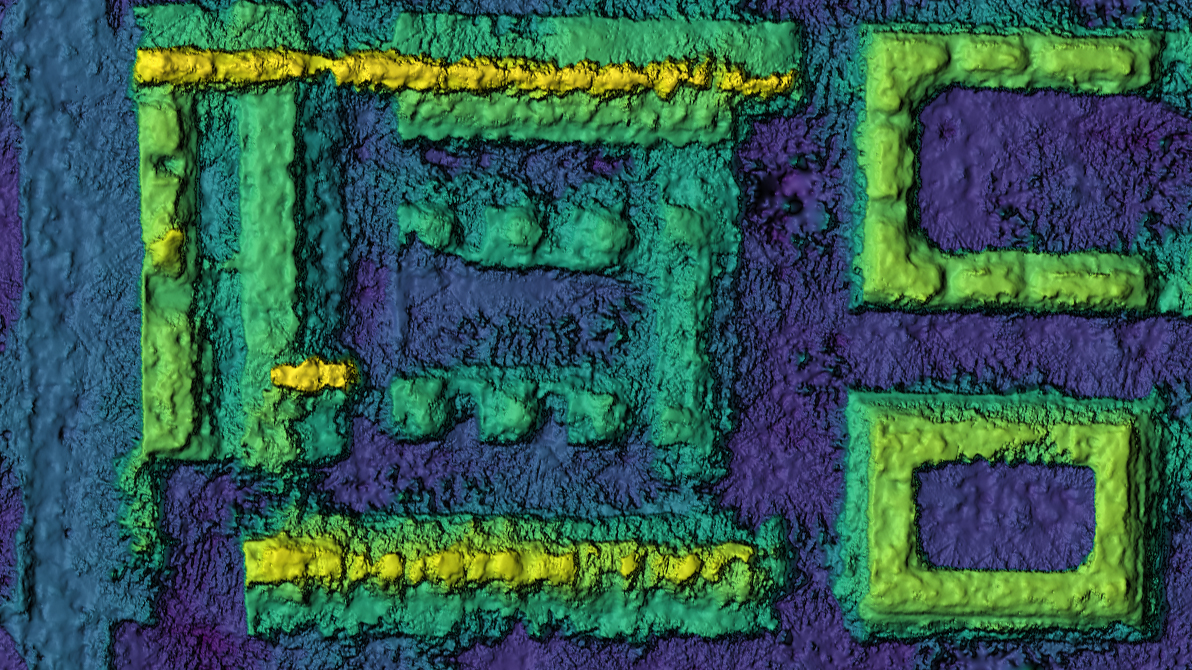}
        \\
        \rotatebox[origin=c]{90}{\footnotesize \resdepth-0} &
        \includegraphics[width=\mywidth,trim={0 0 0 0},clip]{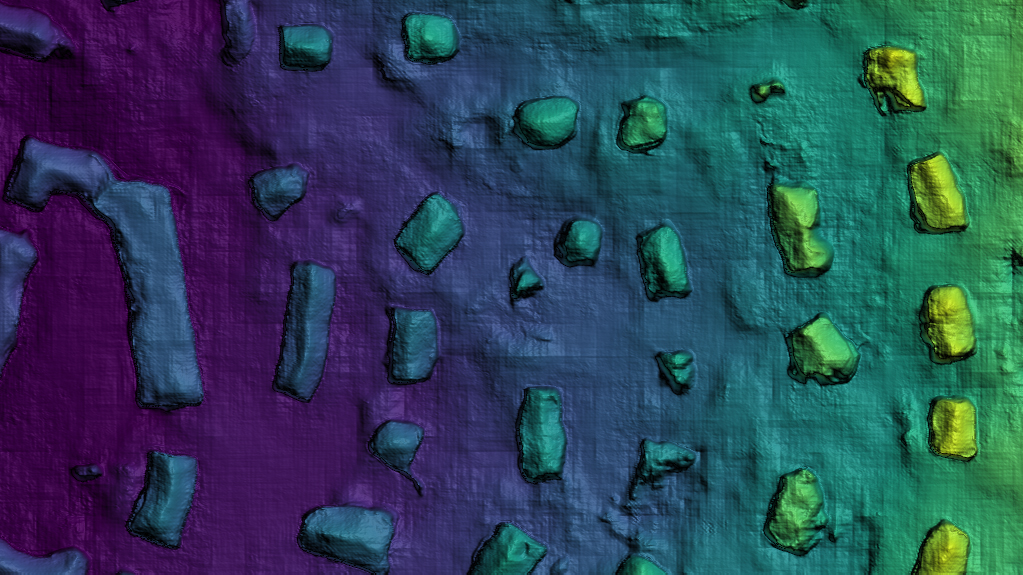} & 
        \includegraphics[width=\mywidth,trim={0 0 0 0},clip]{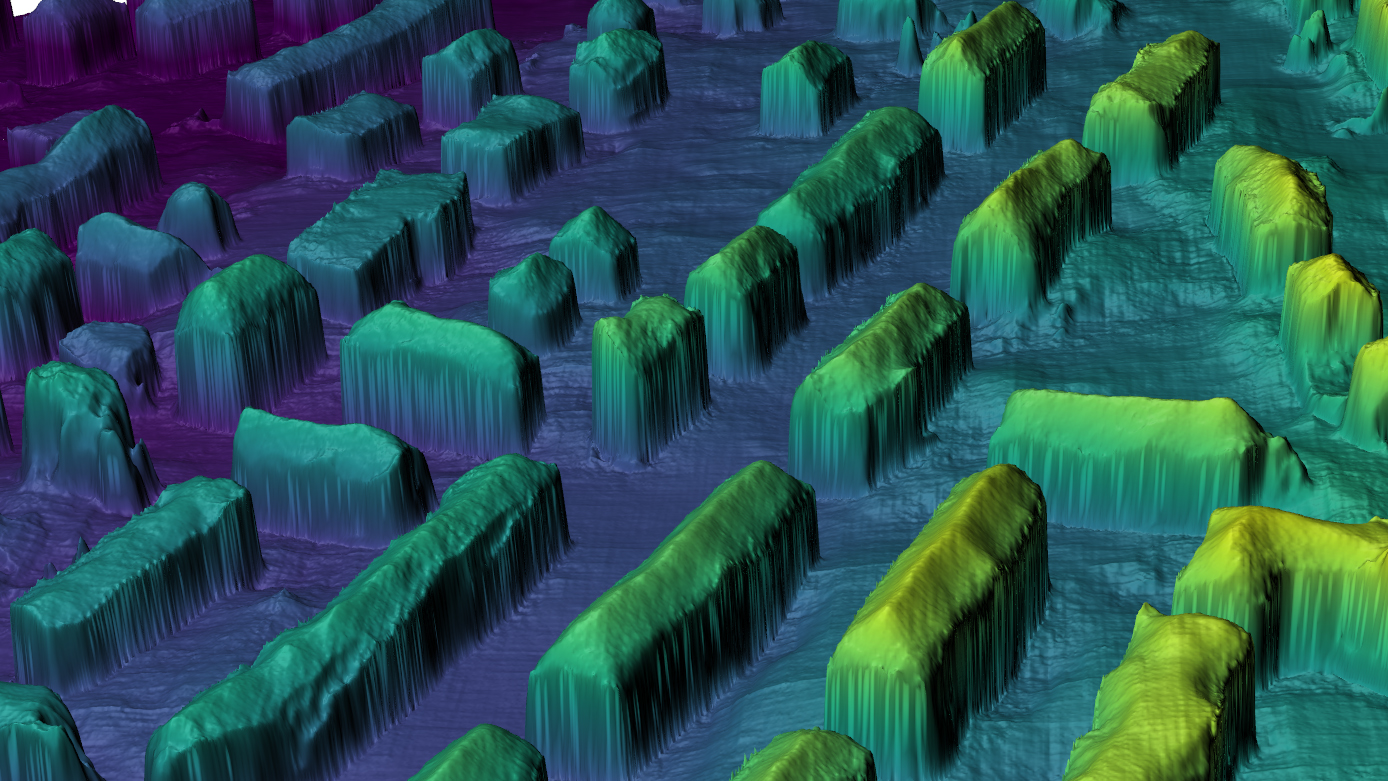} & 
        \includegraphics[width=\mywidth,trim={0 0 0 0},clip]{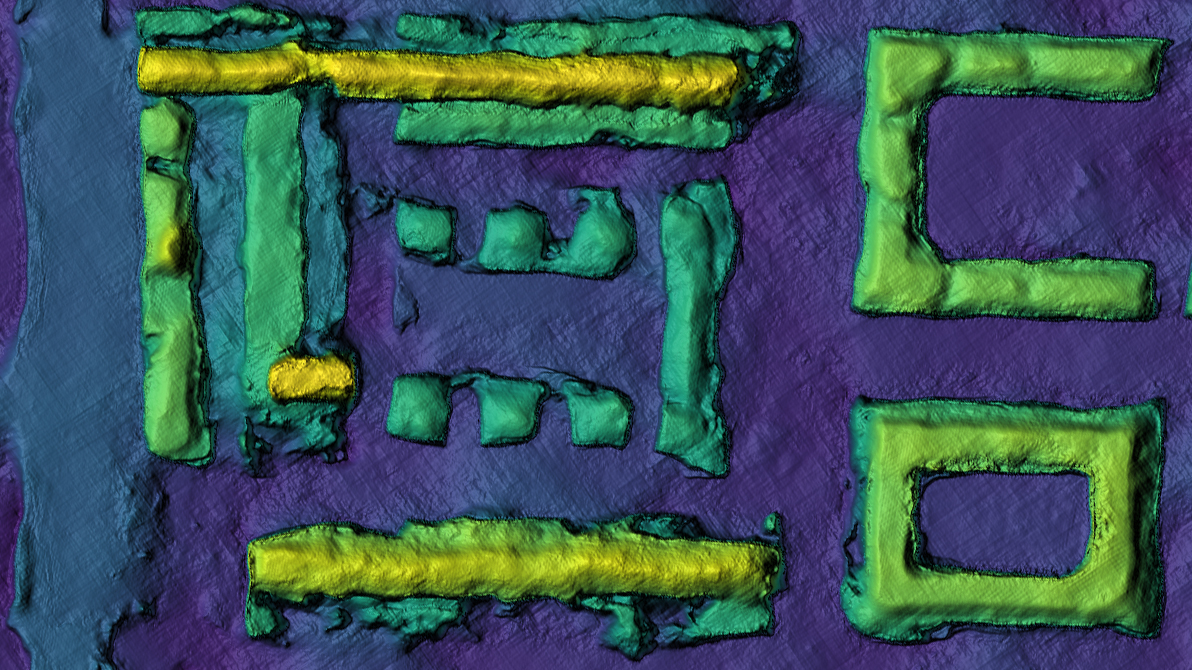}
        \\
        \rotatebox[origin=c]{90}{\footnotesize \resdepth-mono} &
        \includegraphics[width=\mywidth,trim={0 0 0 0},clip]{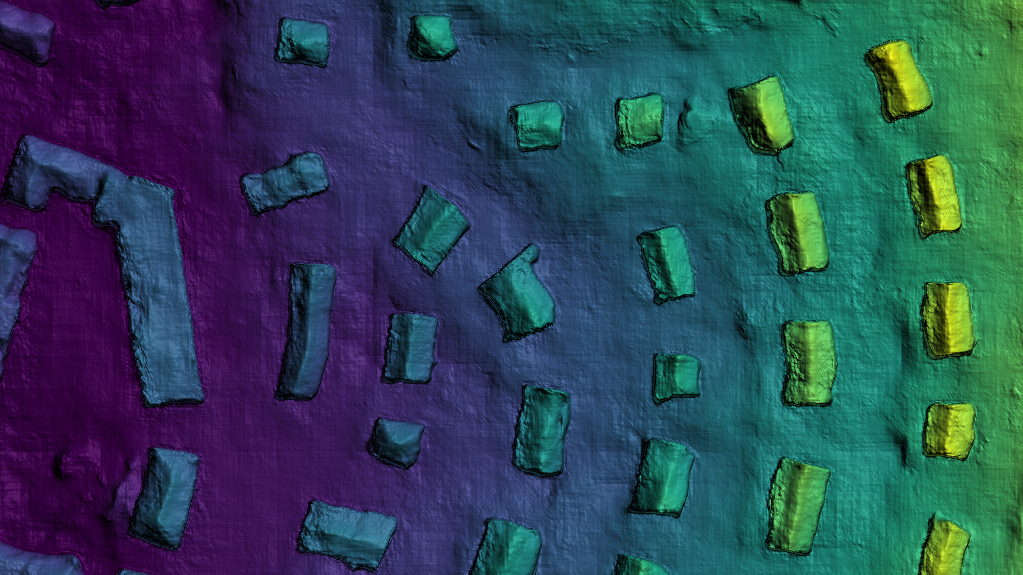} & 
        \includegraphics[width=\mywidth,trim={0 0 0 0},clip]{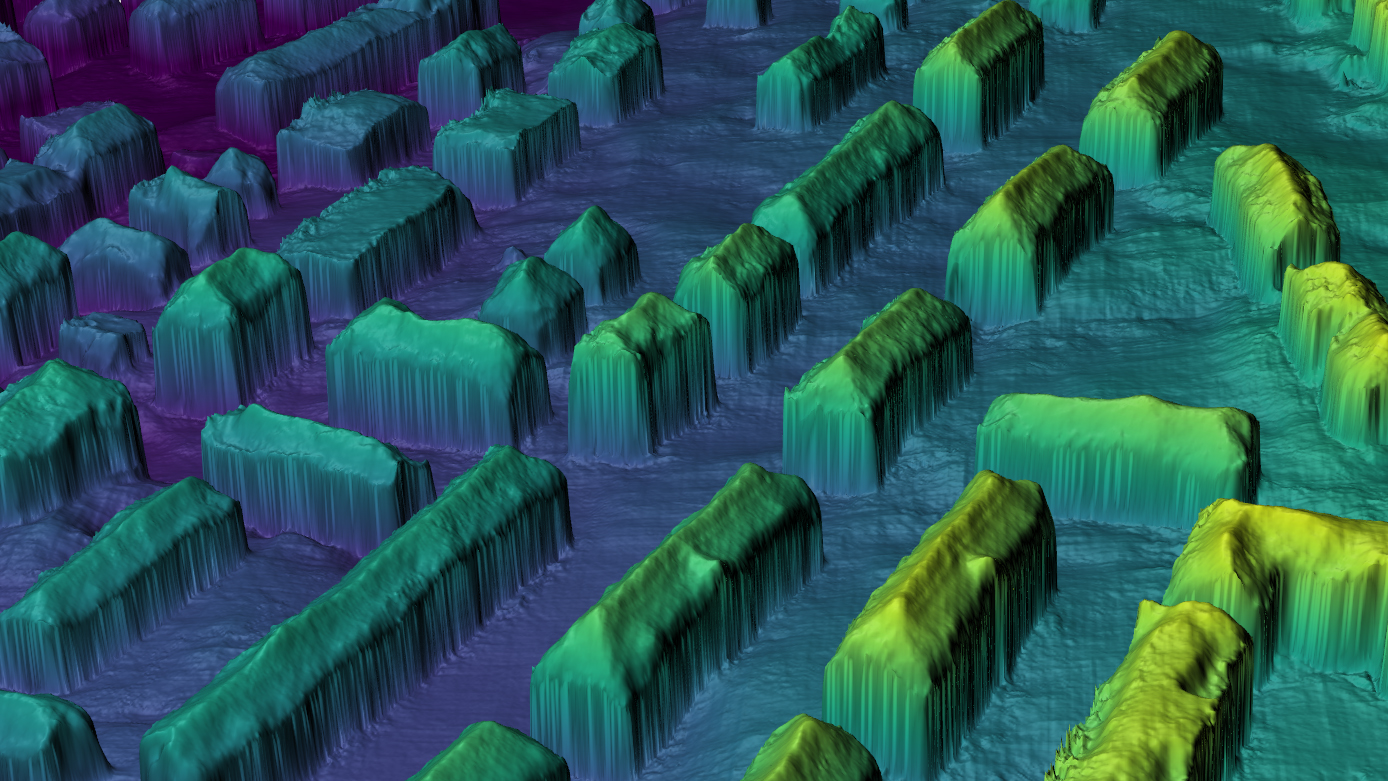} & 
        \includegraphics[width=\mywidth,trim={0 0 0 0},clip]{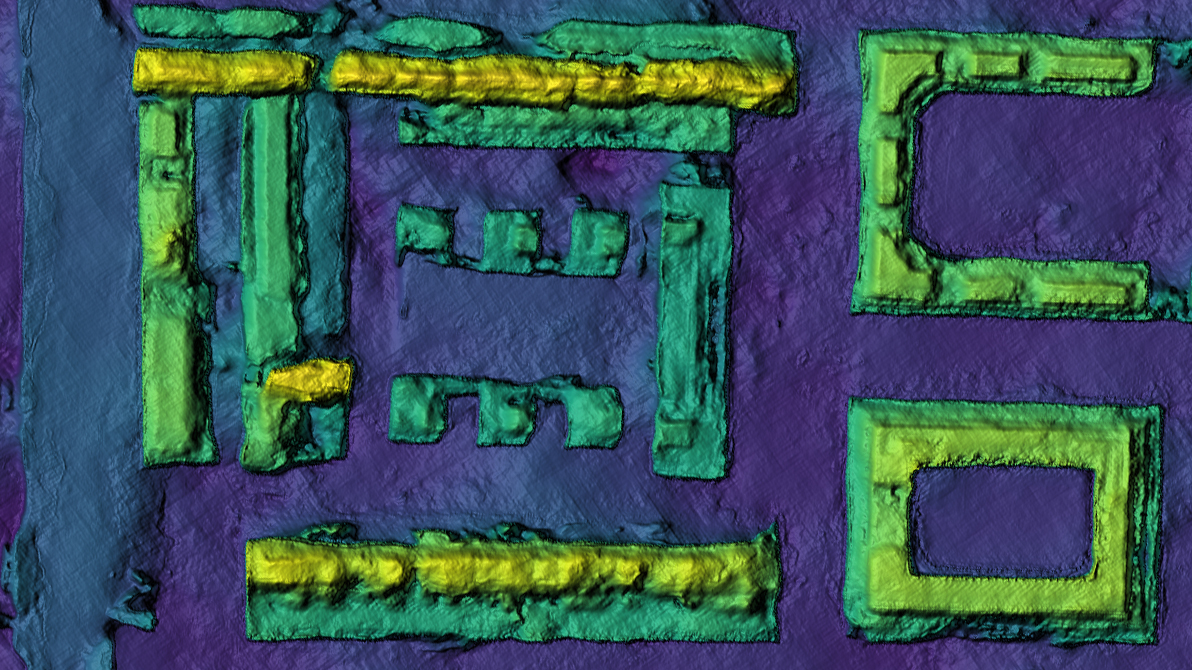}
        \\
        \rotatebox[origin=c]{90}{\footnotesize \resdepth-stereo} &
        \includegraphics[width=\mywidth,trim={0 0 0 0},clip]{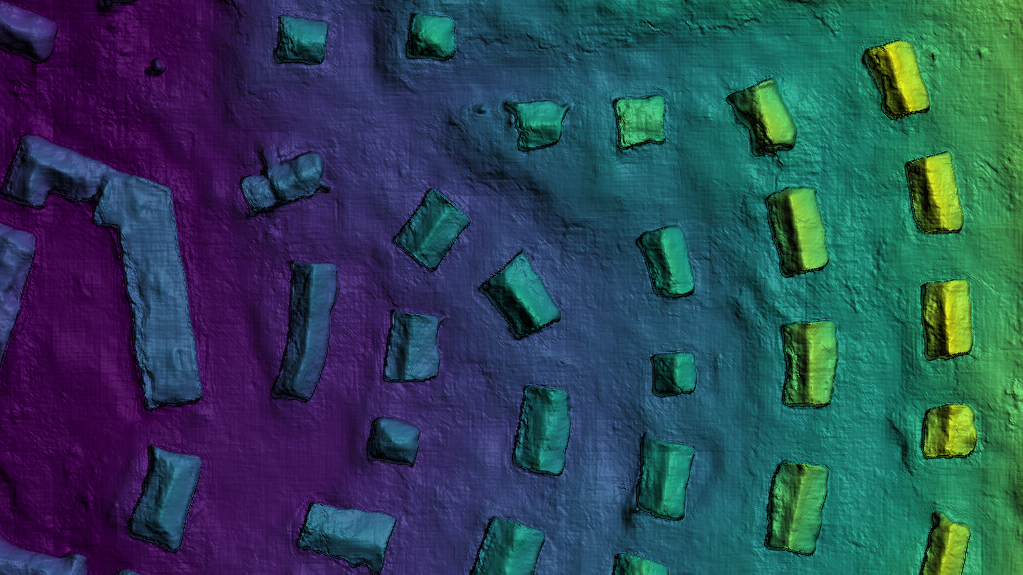} & 
        \includegraphics[width=\mywidth,trim={0 0 0 0},clip]{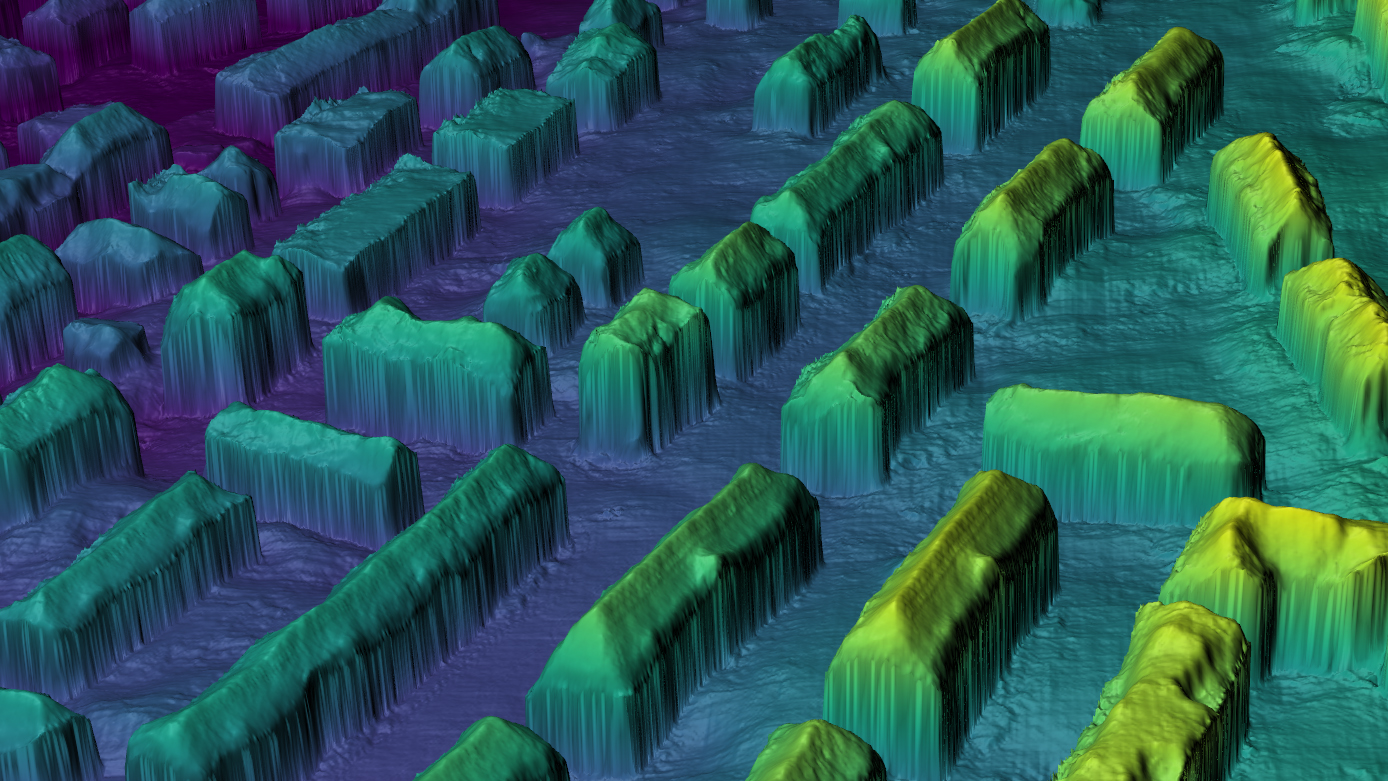} & 
        \includegraphics[width=\mywidth,trim={0 0 0 0},clip]{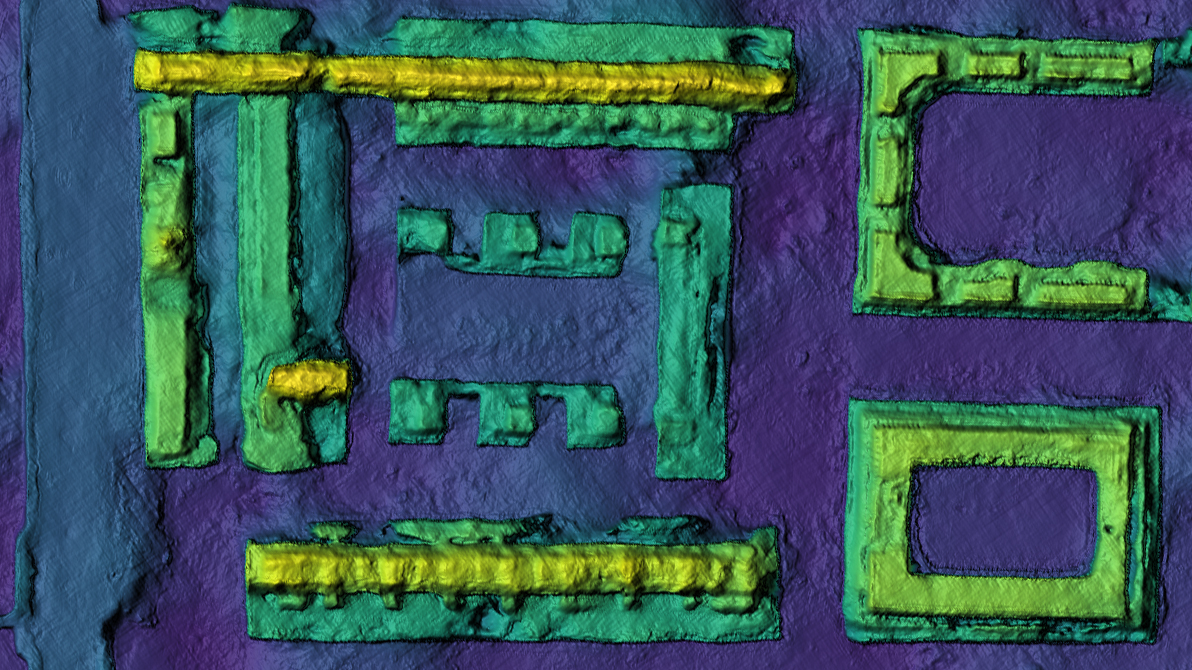}
        \\
        \rotatebox[origin=c]{90}{\footnotesize \resdepth-stereo\textsubscript{iter}} &
        \includegraphics[width=\mywidth,trim={0 0 0 0},clip]{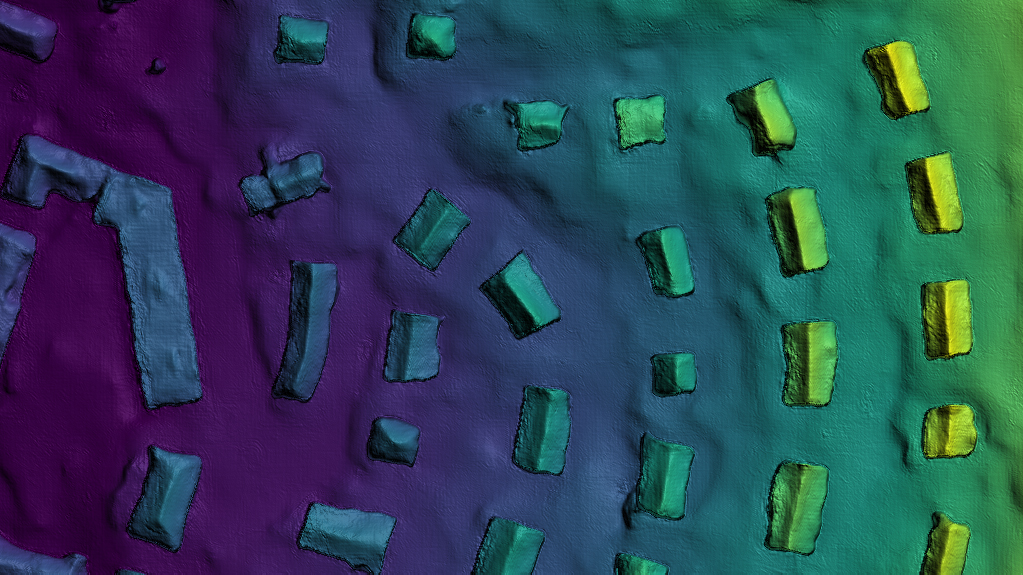} & 
        \includegraphics[width=\mywidth,trim={0 0 0 0},clip]{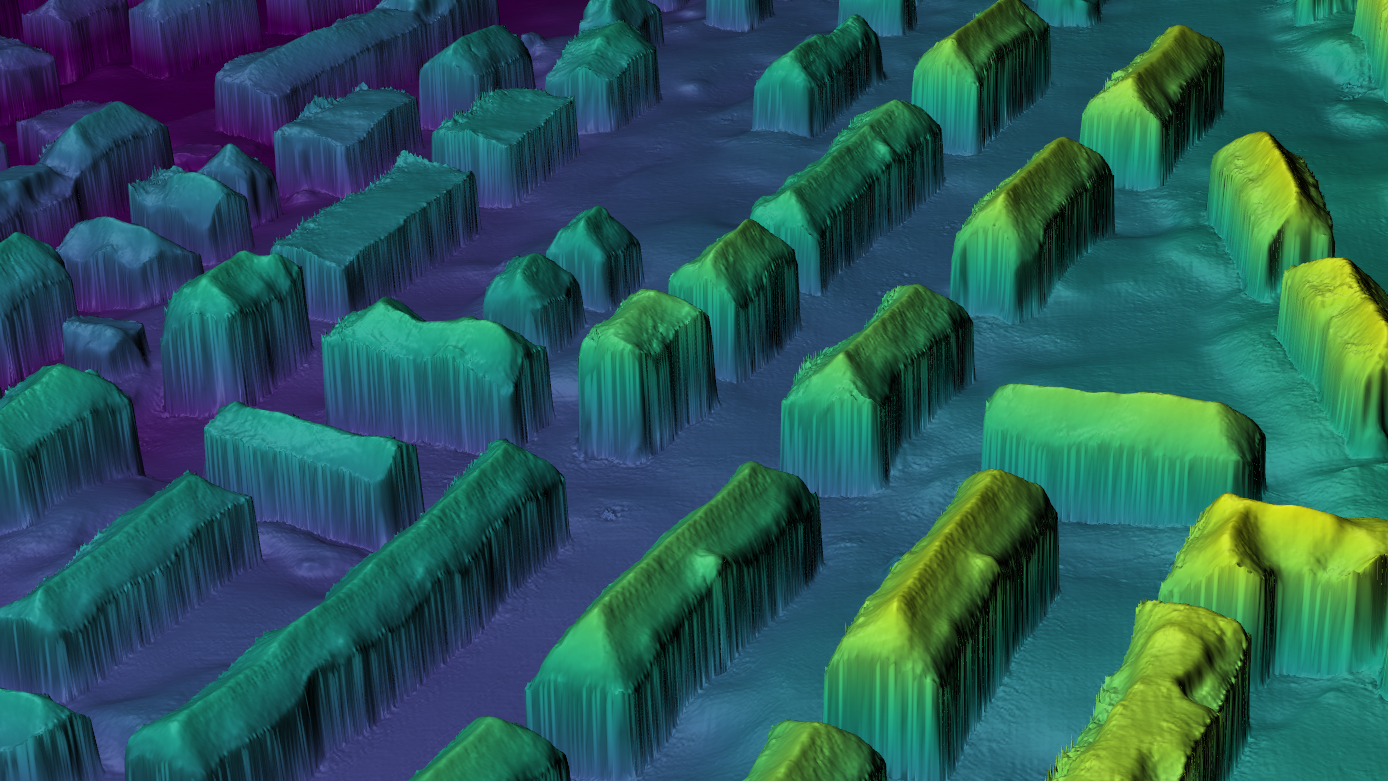} & 
        \includegraphics[width=\mywidth,trim={0 0 0 0},clip]{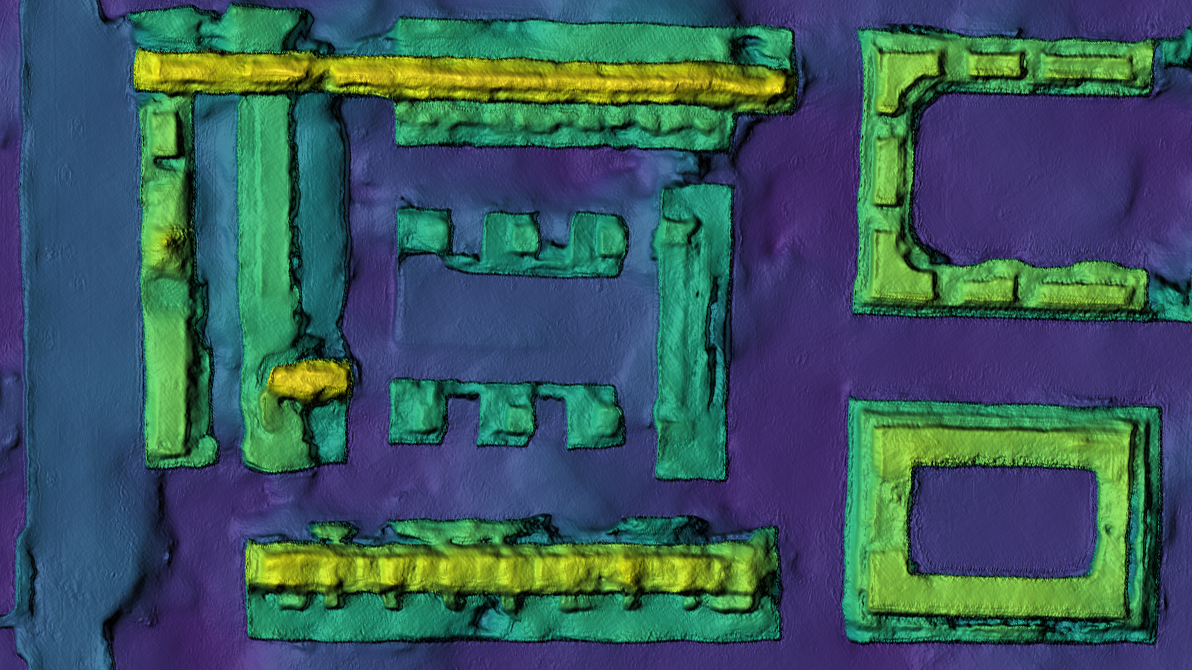}
        \\
        \rotatebox[origin=c]{90}{\footnotesize Ground truth DSM} &
        \includegraphics[width=\mywidth,trim={0 0 0 0},clip]{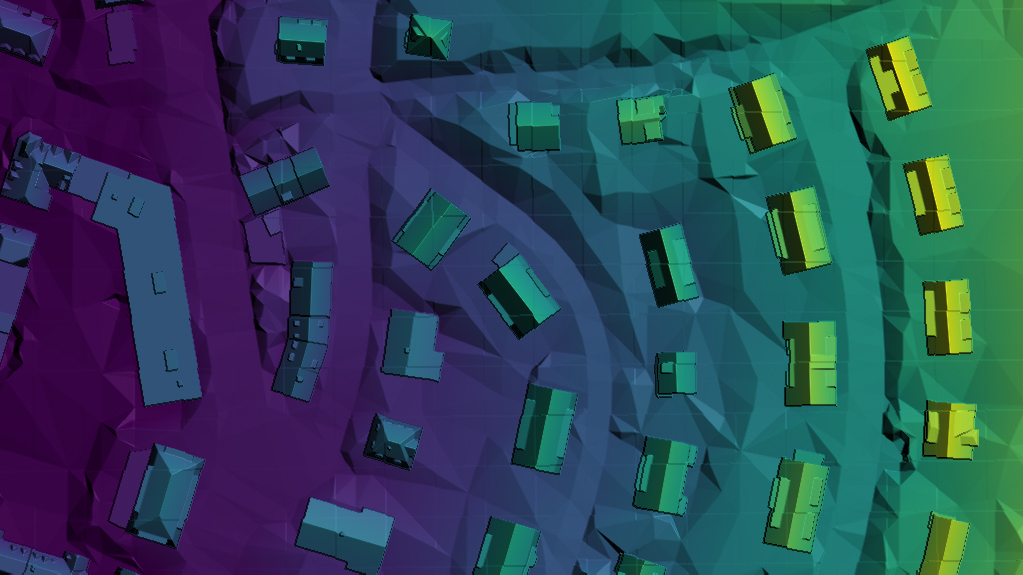} & 
        \includegraphics[width=\mywidth,trim={0 0 0 0},clip]{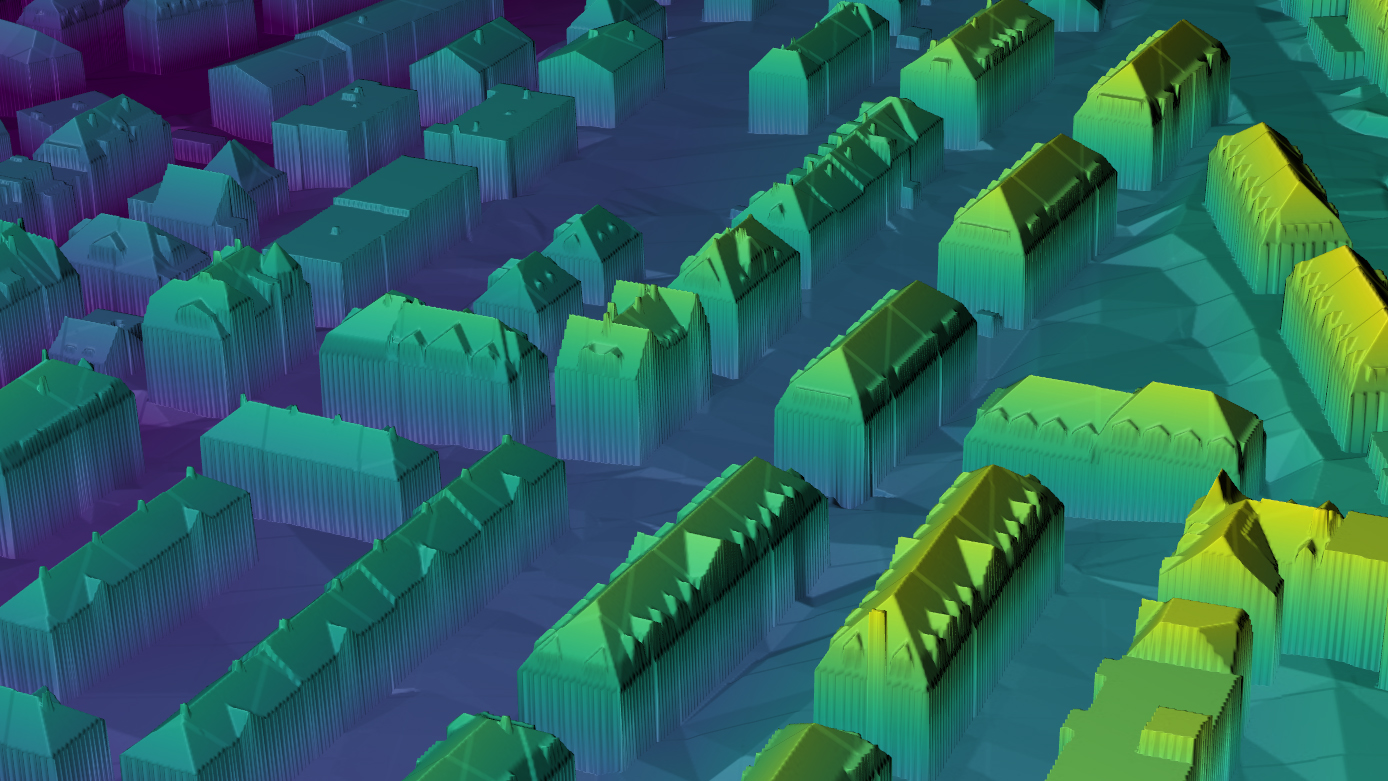} & 
        \includegraphics[width=\mywidth,trim={0 0 0 0},clip]{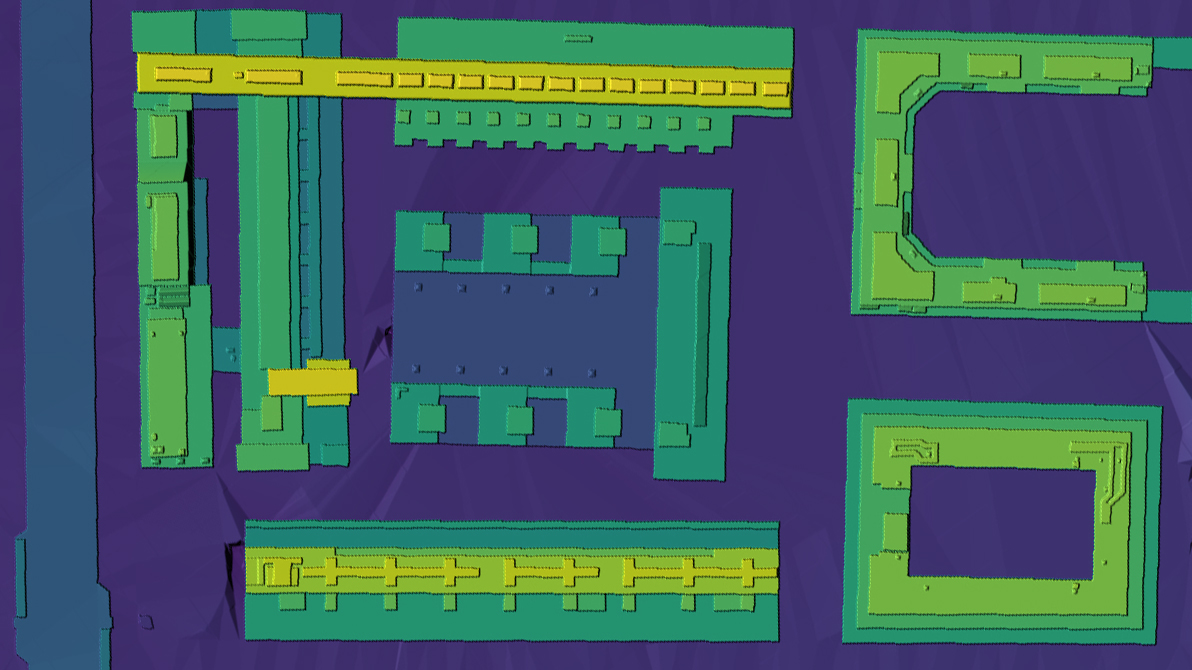}
        \\
        \rotatebox[origin=c]{90}{\footnotesize Residual errors} &
        \includegraphics[width=\mywidth,trim={0 0 0 0},clip]{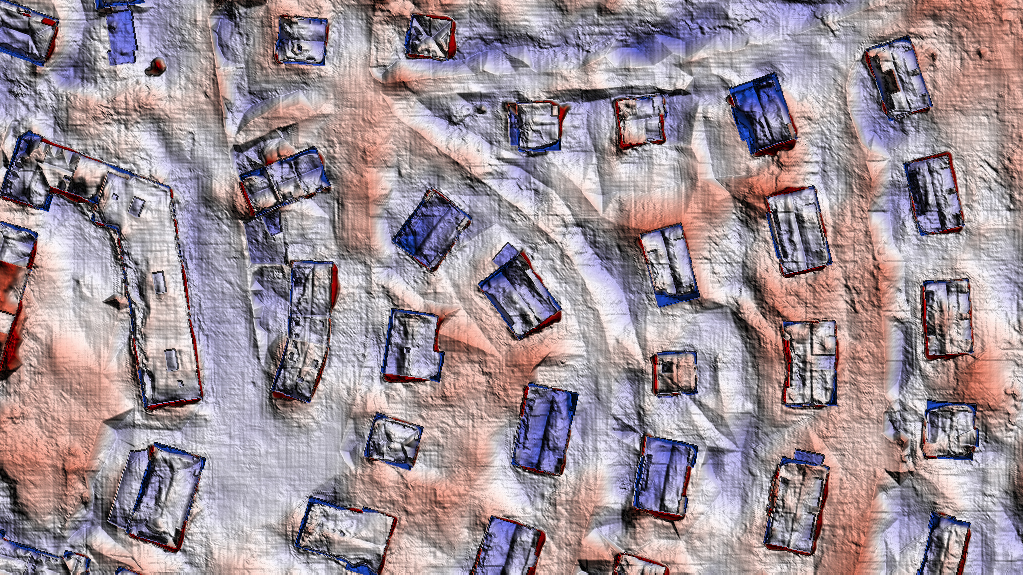} & 
        \includegraphics[width=\mywidth,trim={0 0 0 0},clip]{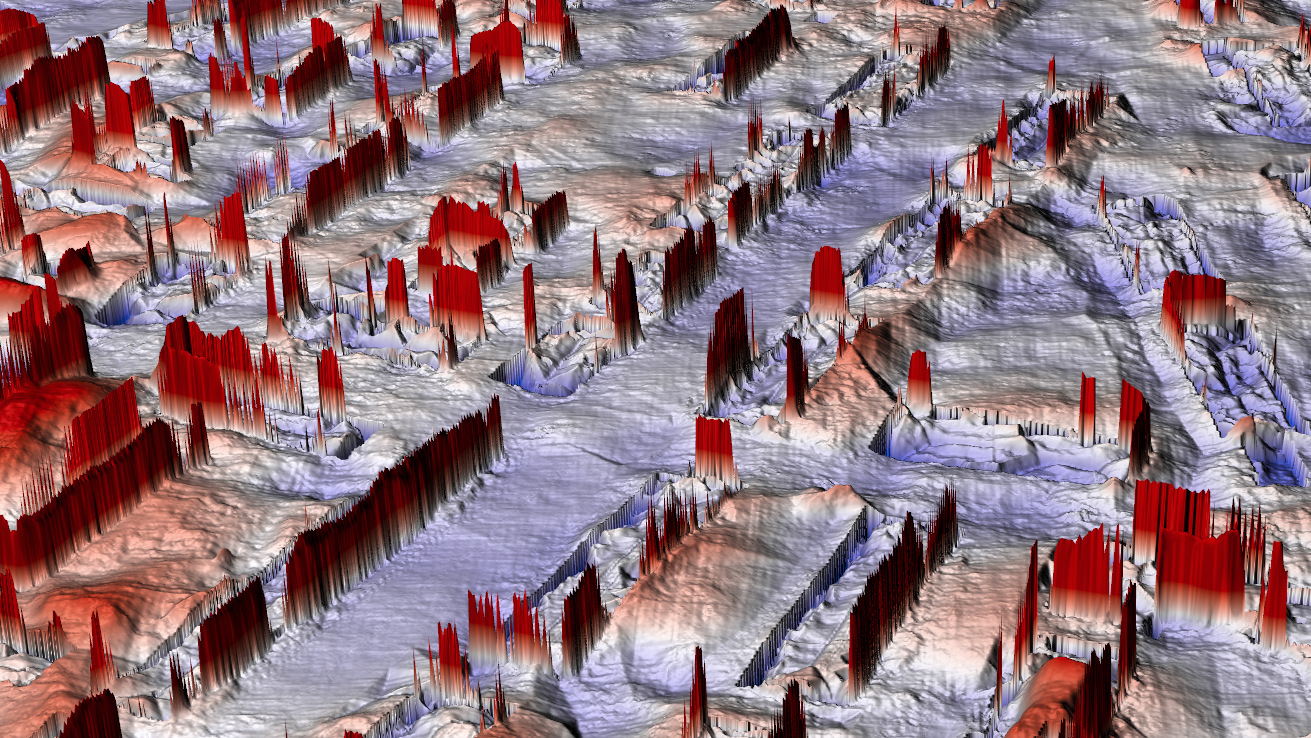} &  
        \includegraphics[width=\mywidth,trim={0 0 0 0},clip]{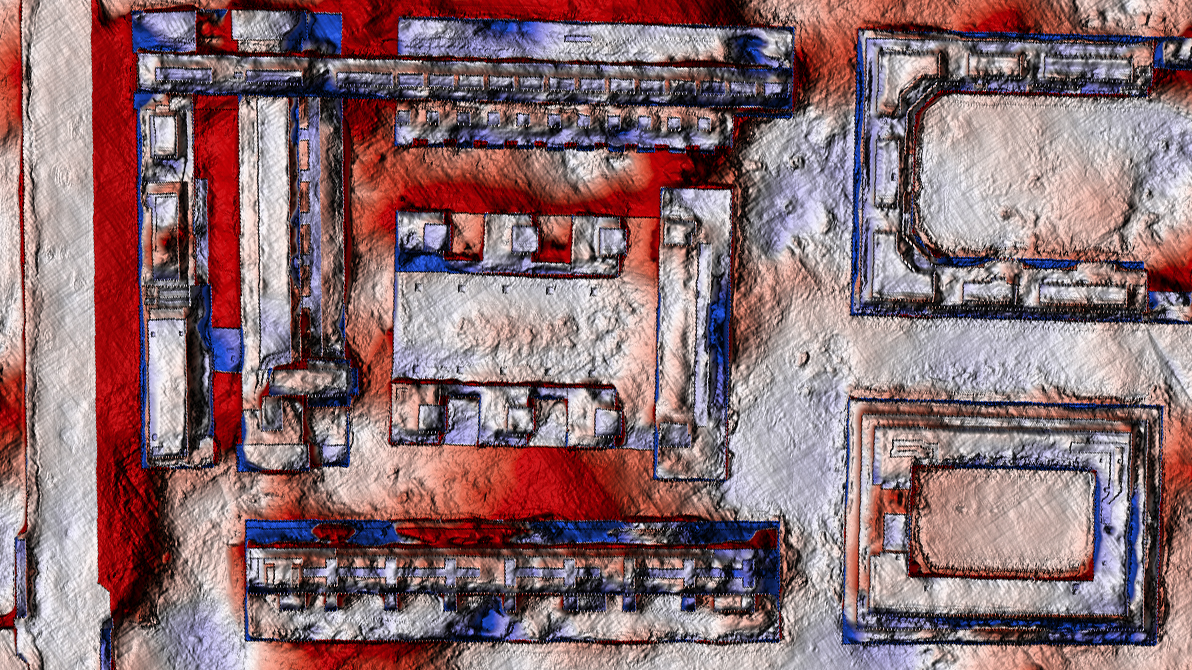}
        \\
    \end{tabular}
    \caption{Visual comparison of different \resdepth variants over selected areas of \zurichOne. Heights are color-coded from blue to green to yellow. The last row shows pixel-wise residual errors of the \resdepth-stereo model, color-coded from -4$\,$m (blue) to 4$\,$m (red).}
    \label{fig:ablation}
\end{figure*}

\newcommand*{\rowstyle}[1]{% sets the style of the next row
  \gdef\@rowstyle{#1}%
  \@rowstyle\ignorespaces%
}

\newcolumntype{=}{% resets the row style
  >{\gdef\@rowstyle{}}%
}

\newcolumntype{+}{% adds the current row style to the next column
  >{\@rowstyle}%
}

\begin{table*}[!t]
    \setlength\dashlinedash{2pt}
    \setlength\dashlinegap{1.5pt}
    \setlength\arrayrulewidth{0.3pt}
    \centering
	\caption{Quantitative results for \zurichOne with different \resdepth variants. The satellite stereo sets $A$ and $B$ refer to Fig.~{\upshape\ref{fig:train_test_pairs}} (right) in the appendix. To train and test the different \resdepth variants, we use a single image (pair) from group $A$.}
	\label{tab:ablation}
	\begin{adjustbox}{max width=0.99\textwidth}
        \begin{tabular}{@{}=l +c +c +c +c +c +c +c + c +c +c +c@{}}
    	    \toprule
    		Reconstruction & \multicolumn{2}{c}{Satellite view(s)} &
    		\multicolumn{3}{c}{Overall} & \multicolumn{3}{c}{Buildings} & \multicolumn{3}{c}{Terrain}\\
    		\cmidrule(lr){2-3}\cmidrule(lr){4-6}\cmidrule(lr){7-9}\cmidrule(l{\tabcolsep}){10-12} 
    		& Train & Test & 
    		MAE & RMSE & MedAE & MAE & RMSE & MedAE & MAE & RMSE & MedAE \\
    		& & & 
    		[m] & [m] & [m] & [m] & [m] & [m] & [m] & [m] & [m] \\
    		\midrule
    		Initial DSM & & & 
    		3.89 & 7.03 & 1.59 & 3.02 & 5.02 & 1.47 & 4.29 & 7.78 & 1.65 \\
    		Median filtered & & & 
    		3.88 & 7.02 & 1.57 & 3.01 & 5.01 & 1.46 & 4.28 & 7.77 & 1.63 \\
    		\midrule
    		\resdepth-0  & - & - & 
    		2.21 & 4.20 & 0.98 & 2.84 & 5.34 & 1.10 & 1.93 & 3.56 & 0.94 \\
    		\resdepth-mono                       & single img & single img & 
    		1.65 & 3.22 & 0.77 & 2.07 & 4.18 & 0.91 & 1.45 & 2.67 & 0.72 \\
    		\resdepth-stereo                     & single pair & single pair & 1.53 & 2.97 & 0.74 & 1.91 & 3.93 & 0.82 & 1.35 & 2.41 & 0.71 \\
    		\resdepth-stereo\textsubscript{iter} & single pair & single pair & 1.49 & 2.95 & 0.71 & 1.90 & 3.98 & 0.79 & 1.30 & 2.34 & 0.69  \\
    		\midrule
    		\resdepth-mono  (no skip) & single img & single img & 1.63 & 3.14 & 0.80 & 2.11 & 4.14 & 1.00 & 1.41 & 2.55 & 0.73 \\
    		\resdepth-stereo (no skip)  & single pair & single pair & 1.56 & 3.00 & 0.78 & 2.03 & 4.03 & 0.95 & 1.35 & 2.38 & 0.72 \\
    		\midrule
            \resdepth-stereo & single pair & $B$ & 2.82 & 4.99 & 1.29 & 3.02 & 5.29 & 1.40 & 2.72 & 4.85 & 1.25 \\
            \resdepth-stereo & $A$ & $B$ & 2.11 & 3.99 & 1.02 & 2.83 & 5.40 & 1.14 & 1.78 & 3.14 & 0.98 \\
    		\bottomrule
    	\end{tabular}
    \end{adjustbox}
\end{table*}

We go on to ablate the influence of the image input, still for area \zurichOne and using a single satellite stereo pair.
We compare different DSM refinement strategies that use either the stereo pair, or only a single satellite image, or no image evidence at all. As a baseline for the latter case, we also tested median filtering (with kernel size 5$\times$5 pixels) as an established method for DSM cleaning without additional evidence but found that it barely improves the initial reconstruction. Quantitative results are reported in Table~\ref{tab:ablation}, visual examples are shown in Fig.~\ref{fig:ablation}. 

Already \resdepth-0, without any image evidence, greatly improves the reconstruction quality. It reduces the MAE by more than 40\% to 2.2$\,$m and the MedAE to 1.0$\,$m, with the largest gain in accuracy for terrain pixels. We note that, in our setting, vegetation is not included in the ground truth DSM. Hence, even without looking into the images, the network can learn to recognize and filter out trees, which is in part responsible for the lower error metrics. \resdepth-0 mainly acts as a context-aware, learned smoothing filter. It reduces the level of noise markedly, sometimes at the expense of blurred roof features (Fig.~\ref{fig:ablation}, \nth{2}~row, \nth{3}~column). Notably, it has learned to differentiate between noise, vegetation, and buildings and even reconstructs buildings that are hardly visible in the initial reconstruction (Fig.~\ref{fig:ablation}, \nth{2}~row, \nth{1} and \nth{2}~column). However, the refined building shapes are not always realistic.

\resdepth-mono can additionally exploit correlations between intensity patterns of a single satellite view and geometric patterns in the initial DSM, and in this way manages to improve the reconstruction quality significantly. Building shapes and previously missed buildings are more faithfully reproduced (Fig.~\ref{fig:ablation}, \nth{3}~row, \nth{1}~column). Gable lines and building outlines are sharper (Fig.~\ref{fig:ablation}, \nth{3}~row, \nth{3}~column) and straighter (Fig.~\ref{fig:ablation}, \nth{3}~row, \nth{2}~column), but rectangular building footprints are sometimes inaccurate. The overall MAE is 0.6$\,$m lower compared to \resdepth-0.

Adding a second satellite view further improves the refinement. \resdepth-stereo further sharpens and straightens building outlines and better reconstructs thin structures. 
An iterative variant \resdepth-stereo\textsubscript{iter}, with a second round of corrections (with separately learned weights), yields another improvement, albeit with greatly diminishing returns. That model mostly filters the remaining noise, particularly well visible for terrain and at transitions between buildings and terrain in the 3D view of Fig.~\ref{fig:ablation}.

The two stereo-enabled variants yield visibly crisper and more detailed 3D geometry and can even recover details that were missed by the initial stereo reconstruction, such as small dormers (Fig.~\ref{fig:ablation}, \nth{4}~row, \nth{2}~column).  The final MAE of our best model is 1.9$\,$m for buildings and 1.3$\,$m for terrain. The overall MedAE is 0.7$\,$m. The largest errors occur in occluded areas and at building outlines (Fig.~\ref{fig:ablation}, last row), the latter possibly due in part to remnant co-registration errors.

\subsection{Residual vs.\ Absolute Height}
Next, we evaluate our strategy of learning residual height updates for every DSM pixel rather than regressing absolute heights. We retrain \resdepth-mono and \resdepth-stereo (again for area \zurichOne and using the same satellite image pair as before) but omit the long skip connection that adds the initial DSM to the output of the last decoder layer. We observe that without that skip connection training is noticeably less stable. Despite the rather shallow architecture, which would in principle make it easy to memorize the input, the reconstruction of buildings markedly deteriorates. For \resdepth-stereo, their MAE is 6\% higher, whereas their MedAE is 16\% higher. I.e., the difference is not due to few outliers but rather caused by a moderate performance drop across the majority of pixels. See Table~\ref{tab:ablation}.

\subsection{Generalization Across Viewpoints}
\label{sec:generalization_new_views}
So far, we have validated \resdepth under ideal conditions, where the training and test regions are cropped from the same satellite stereo pair. In practice, one will not want to retrain the model for a specific image pair but rather apply an existing model trained under viewing and lighting conditions that do not exactly match those encountered during inference.
When testing the model used in the previous experiments in that setting, the performance decreases significantly (Table~\ref{tab:ablation}, \nth{2} last row). The overall MAE increases from 1.5$\,$m to 2.8$\,$m compared to the result under ideal conditions (Table~\ref{tab:ablation}, \nth{5}~row), and there is barely any improvement for the building pixels.
In keeping with other deep machine learning models, \resdepth does overfit to the specific setting if given a chance, such that the corrections learned from a single stereo pair are tailored to its specific image characteristics and acquisition geometry. Consequently, the prior is most accurate under that particular stereo configuration and (relative) radiometry and does not transfer directly to new viewing directions and illumination conditions.

Fortunately, the issue can also be addressed in a straightforward manner by increasing the variability of the relevant parameters in the training set.
To that end, we combine the same initial DSM with multiple different stereo pairs to obtain a training set that includes a range of different, realistic stereo configurations and illumination conditions.
Stereo pairs for training and testing are generated from mutually exclusive image sets $A$ and $B$ (see \mbox{\ref{sec:valid_pairs}}) to test the generalization across viewpoints. 

When mixing multiple stereo pairs of the same geographic location during training, \resdepth can learn to disentangle the factors contributing to the observed discrepancies between ortho-images of a pair: the acquisition geometry and illumination induce fluctuations in image appearance that vary between different stereo pairs, whereas patterns common to all pairs must stem from actual height errors in the initial DSM.
The resulting DSM (Table~\ref{tab:ablation}, last row) is significantly better than the initial one and also than the output of the naive generalization experiment described above. 
There is still a moderate performance penalty compared to the ideal scenario, mostly for building pixels: since we only have access to a small number of images, respectively to four viable training stereo pairs, \resdepth cannot completely disentangle imaging conditions from height errors.
The results will likely improve further and approach the ideal setting if a larger variety of stereo pairs is used than we had in our study.

\begin{figure*}[!p]
    \def\mywidth{0.71\textwidth}
    \newcolumntype{C}[1]{>{\centering\arraybackslash}m{#1}}
    \centering
    \begin{tabular}{l C{\mywidth}}
        \rotatebox[origin=c]{90}{\parbox{3cm}{\centering \footnotesize Example panchromatic image}} &  
        \includegraphics[width=\mywidth,trim={0 0 0 0},clip]{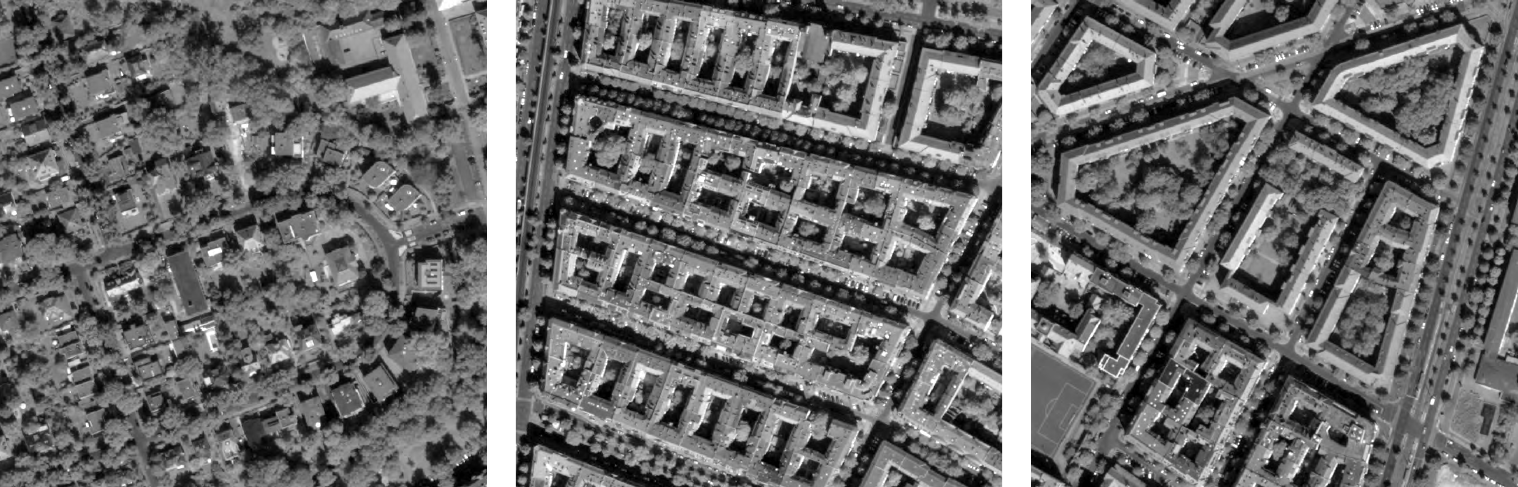}
        \\
        \rotatebox[origin=c]{90}{\parbox{1cm}{\centering \footnotesize Initial DSM}} &  
        \includegraphics[width=\mywidth,trim={0 0 0 0},clip]{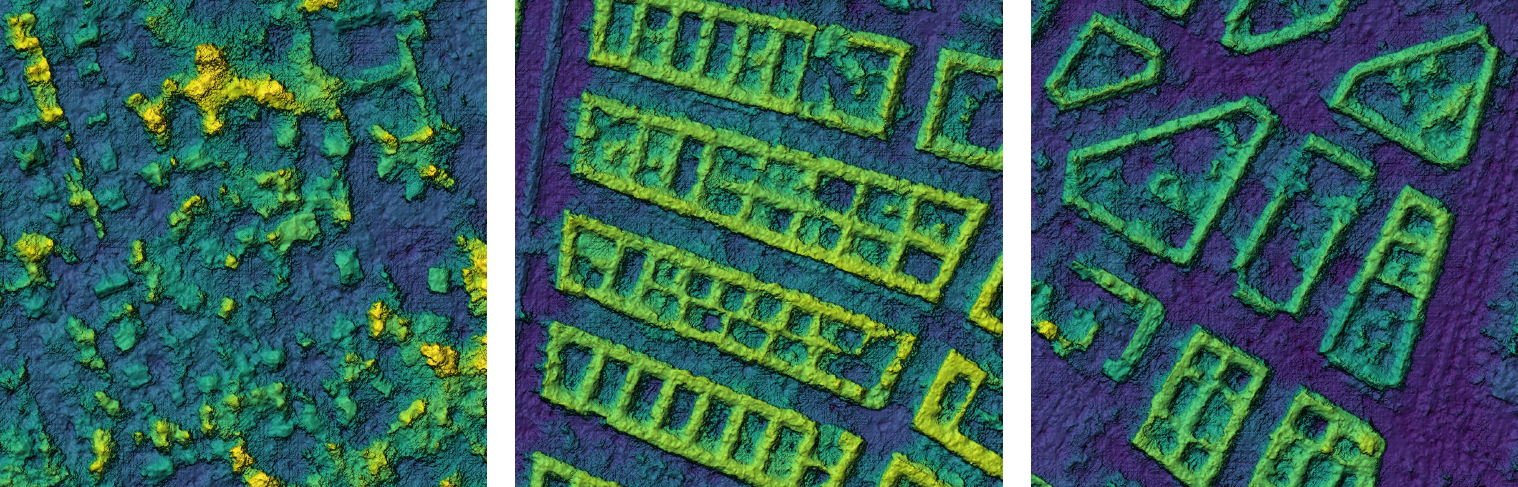}
        \\
        \rotatebox[origin=c]{90}{\parbox{3cm}{\centering \footnotesize Within-city generalization}} &  
        \includegraphics[width=\mywidth,trim={0 0 0 0},clip]{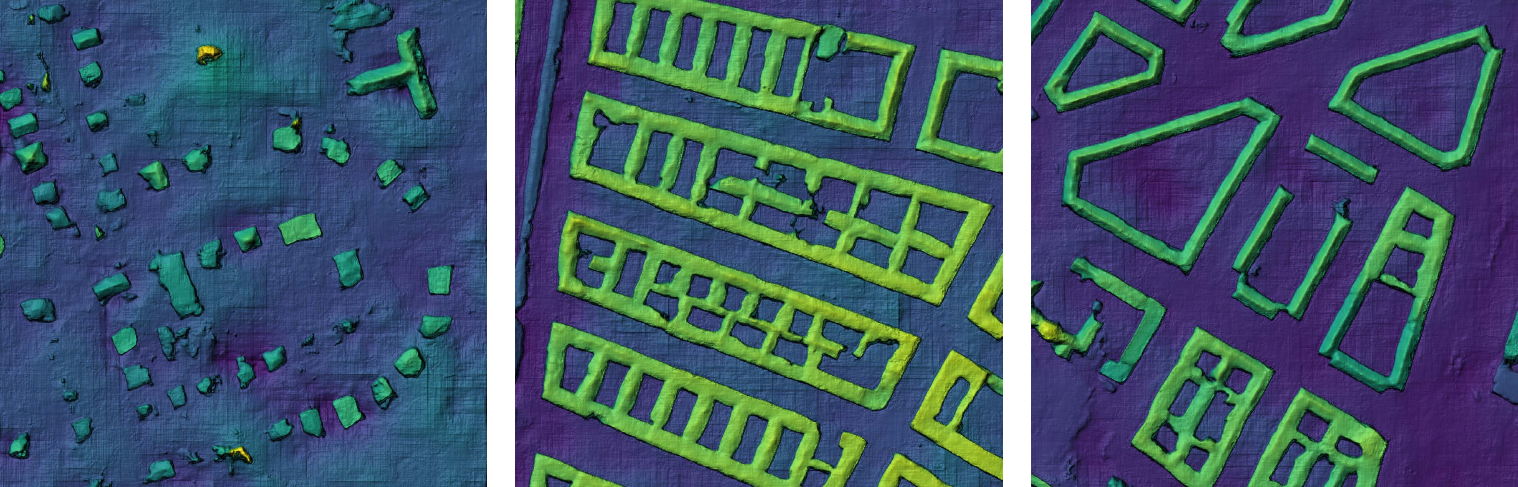}
        \\
        \rotatebox[origin=c]{90}{\parbox{3cm}{\centering \footnotesize Across-city generalization}} &  
        \includegraphics[width=\mywidth,trim={0 0 0 0},clip]{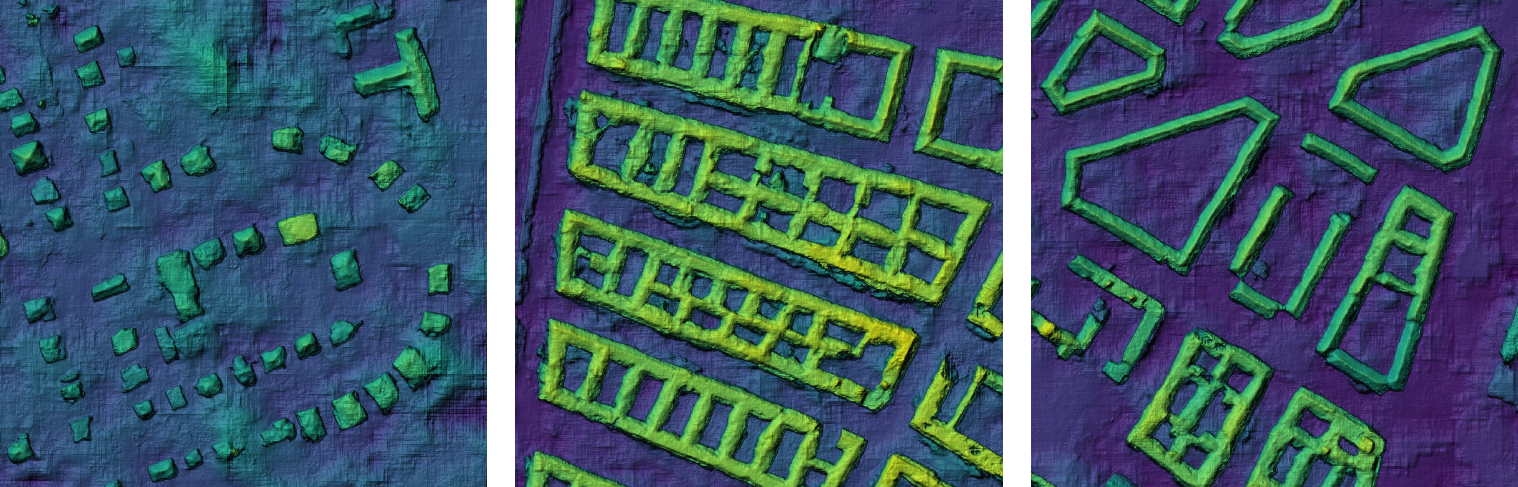}
        \\
        \rotatebox[origin=c]{90}{\parbox{2cm}{\centering \footnotesize Ground truth DSM}} &  
        \includegraphics[width=\mywidth,trim={0 0 0 0},clip]{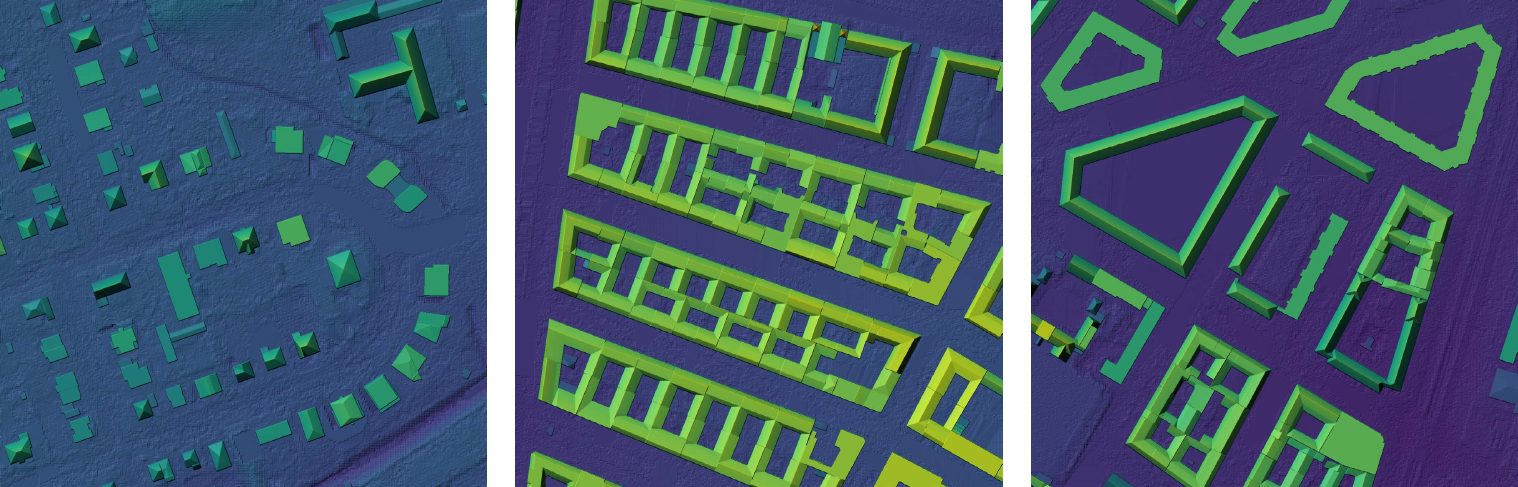}
        \\
    \end{tabular}
    \caption{Geographical generalization between different areas in Berlin. Heights are color-coded from blue to green to yellow. We show results for both a \resdepth model trained on a different part of Berlin (within-city) and of a model trained on Zurich (across-city).}
    \label{fig:generalization_berlin}
\end{figure*}
\begin{figure*}[!ht]
    \def\mywidth{0.9\textwidth}
    \newcolumntype{C}[1]{>{\centering\arraybackslash}m{#1}}
    \centering
    \begin{tabular}{l C{\mywidth}}
        \rotatebox[origin=c]{90}{\parbox{3cm}{\centering \footnotesize Example panchromatic image}} &
        \includegraphics[width=\mywidth,trim={0 0 0 0},clip]{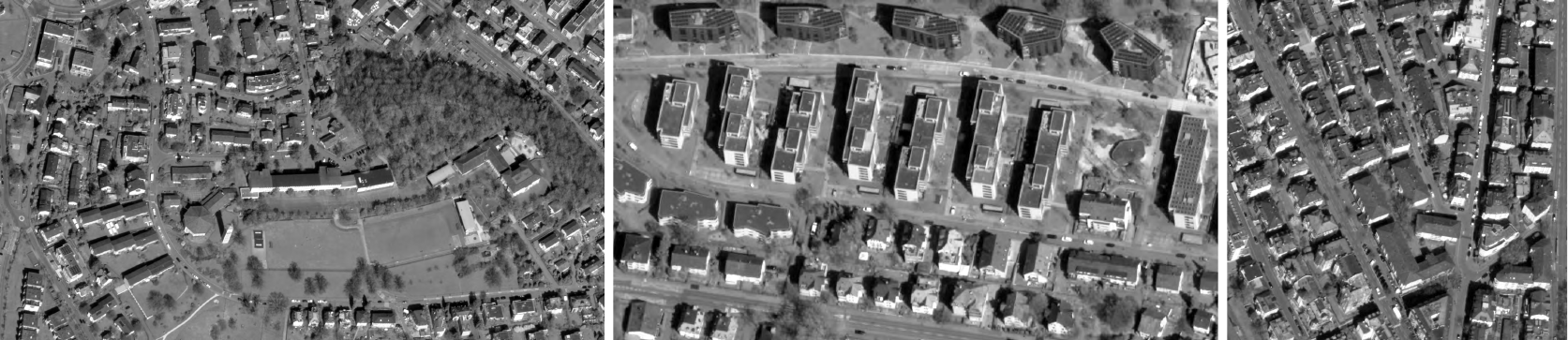}
        \\
        \rotatebox[origin=c]{90}{\parbox{1cm}{\centering \footnotesize Initial DSM}} &  
        \includegraphics[width=\mywidth,trim={0 0 0 0},clip]{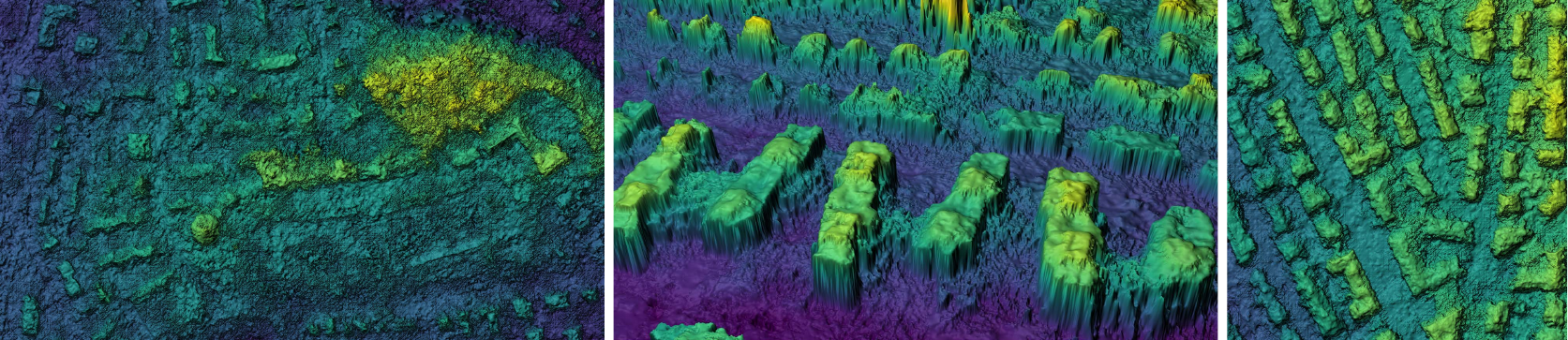}
        \\
        \rotatebox[origin=c]{90}{\parbox{3cm}{\centering \footnotesize Within-city generalization}} &  
        \includegraphics[width=\mywidth,trim={0 0 0 0},clip]{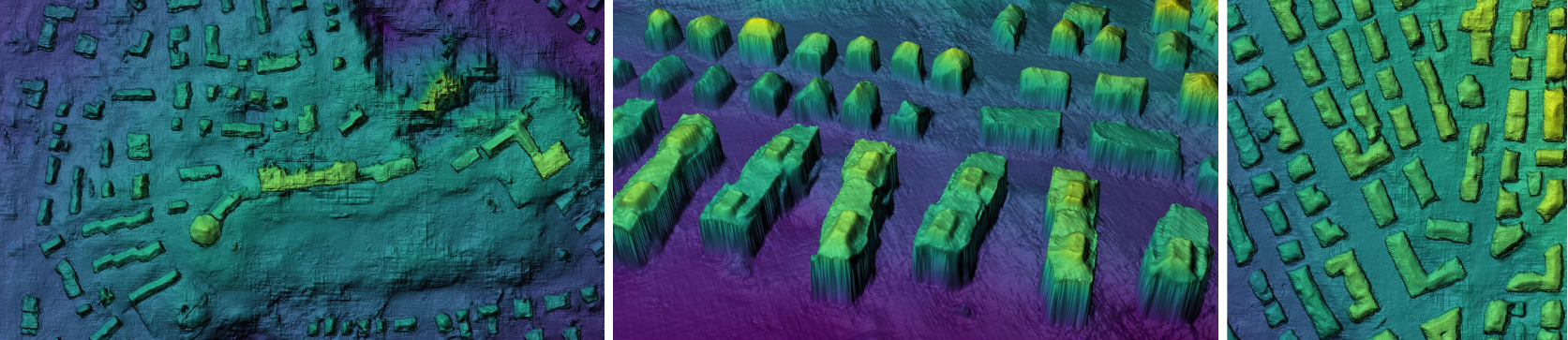}
        \\
        \rotatebox[origin=c]{90}{\parbox{3cm}{\centering \footnotesize Across-city generalization}} &  
        \includegraphics[width=\mywidth,trim={0 0 0 0},clip]{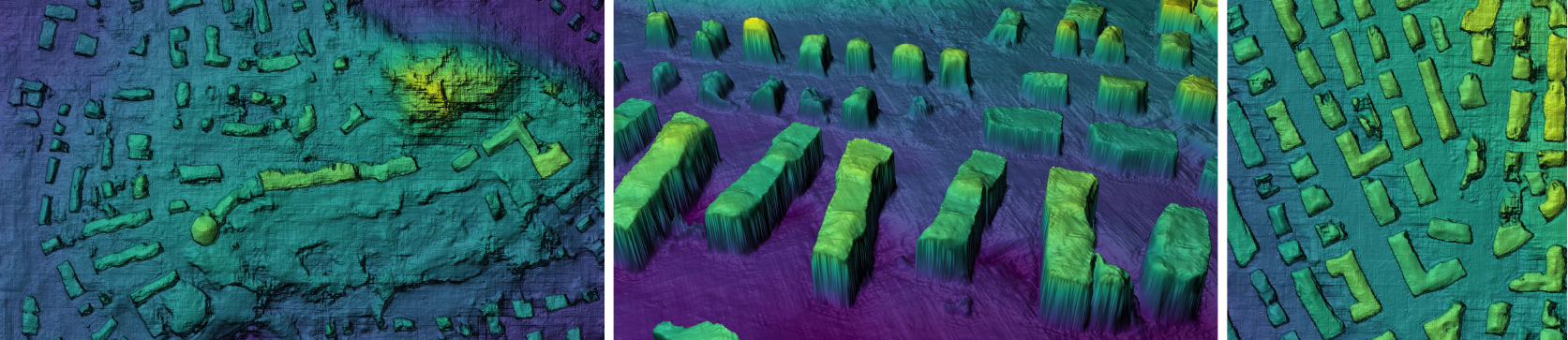}
        \\
        \rotatebox[origin=c]{90}{\parbox{2cm}{\centering \footnotesize Ground truth DSM}} &  
        \includegraphics[width=\mywidth,trim={0 0 0 0},clip]{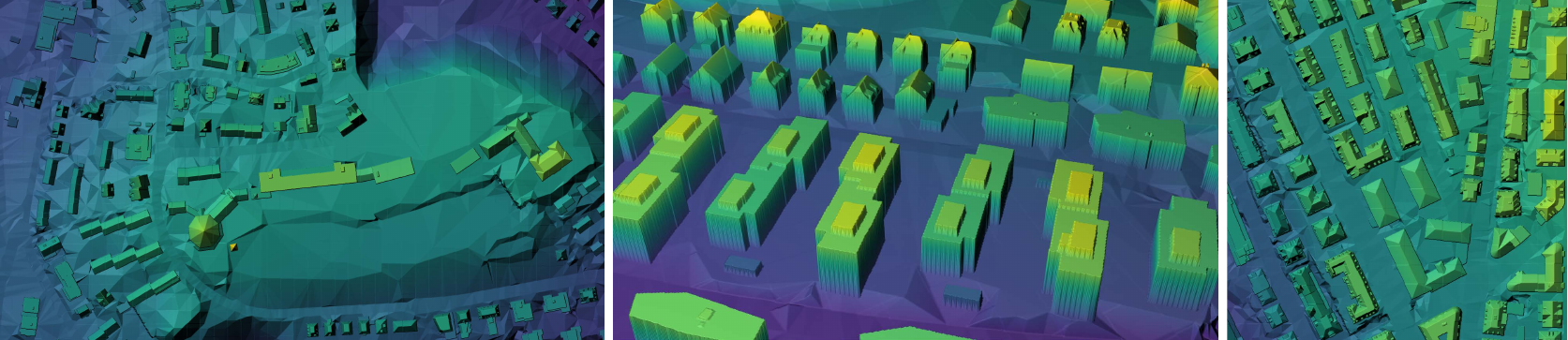}
        \\
    \end{tabular}
    \caption{Geographical generalization between different areas in Zurich. Heights are color-coded from blue to green to yellow. We show results for both a \resdepth model trained on a different part of Zurich (within-city) and of a model trained on Berlin (across-city).}
    \label{fig:generalization_zurich}
\end{figure*}

\subsection{Geographical Generalization Within a City}
\label{sec:generalization_within}
\begin{table*}[!t]
    \footnotesize
    \centering
    \caption{Geographical generalization (i) across regions within the same city and (ii) between Berlin and Zurich. We report the MAE averaged over all pixels and over only building pixels; as well as the bias, measured as the median error over all pixels.}
    \label{tab:generalization}
    \begin{adjustbox}{max width=\textwidth}
        \begin{tabular}{@{}=l +l +l +c +c +c @{}}
            \toprule
            \multirow{2}{2cm}{Reconstruction\\ type} & \multirow{2}{2cm}{Refinement\\ method} & Train area $\rightarrow$ Test area & \multicolumn{2}{c}{MAE [m]} & Bias [m] \\
    		\cmidrule(lr){4-5}
        	& & & Overall & Buildings &  \\
        	\midrule
        	\multirow{5}{*}{Initial} & & \berlinOne   & 5.20 & 2.60 & 2.43 \\
        	                         & & \berlinTwo   & 4.27 & 2.75 & 1.36 \\
        	                         & & \zurichOne   & 3.89 & 3.02 & 0.78 \\
        	                         & & \zurichTwo   & 3.92 & 3.34 & 1.21 \\
        	                         & & \zurichThree & 3.65 & 4.63 & -1.26 \\
            \midrule
        	\multirow{5}{2cm}{Within-city\\ generalization}
        	& \multirow{5}{*}{\resdepth~(ours)} & \berlinTwo $\rightarrow$ \berlinOne                   & 1.80 & 3.20 & -0.33 \\
        	& & \berlinOne $\rightarrow$ \berlinTwo                   & 2.06 & 3.20 & -0.14 \\
        	& & \{\zurichTwo, \zurichThree\} $\rightarrow$ \zurichOne & 2.71 & 2.80 & 0.75 \\
        	& & \{\zurichOne, \zurichThree\} $\rightarrow$ \zurichTwo & 2.77 & 3.48 & -0.25 \\
        	& & \{\zurichOne, \zurichTwo\} $\rightarrow$ \zurichThree & 3.07 & 4.64 & -2.05 \\
        	\midrule
            \multirow{10}{2cm}{Across-city\\ generalization}
            & \multirow{5}{*}{\resdepth~(ours)} 
            & \zurich $\rightarrow$ \berlinOne   & 1.90 & 2.45 & 0.51 \\
    		& & \zurich $\rightarrow$ \berlinTwo   & 2.30 & 2.78 & -0.07 \\
            & & \berlin $\rightarrow$ \zurichOne   & 2.80 & 3.16 & -0.41 \\
    		& & \berlin $\rightarrow$ \zurichTwo   & 3.58 & 5.20 & -0.88 \\
    		& & \berlin $\rightarrow$ \zurichThree & 3.41 & 5.45 & -2.00 \\
    		\cmidrule(lr){2-6}
    		& \multirow{5}{2cm}{\resdepth w/o\\ topographic \\ normalization} & \zurich $\rightarrow$ \berlinOne   & 2.45 & 2.54 & 0.92\\
    		& & \zurich $\rightarrow$ \berlinTwo   & 2.40 & 2.70 & 0.15 \\
            &  & \berlin $\rightarrow$ \zurichOne   & 7.27 & 5.12 & 1.41 \\
    		& & \berlin $\rightarrow$ \zurichTwo   & 5.60 & 6.41 & -0.39 \\
    		& & \berlin $\rightarrow$ \zurichThree & 4.70 & 6.42 & -2.44 \\
        	\bottomrule
        \end{tabular}
    \end{adjustbox}
\end{table*}

Next, we test how well \resdepth trained in a given training region generalizes to other parts of the same city.
Cross-validation across the different regions of both Zurich and Berlin are shown in Table~\ref{tab:generalization} (rows~\mbox{6--10}). \resdepth consistently improves the reconstruction quality in each case. Compared to the initial reconstructions (first five rows in Table~\ref{tab:generalization}), it reduces the overall MAE by \mbox{15--30\%}, to an average of 2.9$\,$m for Zurich and by more than 50\% to $\approx$1.9$\,$m for Berlin. Furthermore, in almost all cases, it also mitigates the height bias.

While, overall, \resdepth visually and quantitatively improves the DSMs also when generalizing over larger distances, there is little quantitative gain at building pixels; in the case of Berlin, the numbers even deteriorate. The examples in Fig.~\ref{fig:generalization_berlin} show that the network indeed learns meaningful features of urban design and street layout that are valid across different city districts. However, these mostly improve the terrain pixels, where the results exhibit a lower noise level, vegetation has been removed, and built-up areas are separated into plausible individual units. We speculate that the reconstruction of the intricate shape details within the buildings suffers more strongly from the imperfect generalization across viewing and illumination conditions (cf. Section~\ref{sec:generalization_new_views}) due to the mutually exclusive sets of training and test images (see above).

\subsection{Geographical Generalization Across Cities}
Going one step further, we test geographical generalization between Berlin and Zurich. We train two \resdepth models from scratch, where in one case, the training data is sampled from both regions in Berlin (\berlin), and in the other one, from all three regions of Zurich (\zurich). These models are then tested separately for each region in the respective other city. Results are shown in Table~\ref{tab:generalization} (rows~\mbox{11--15}), and visual examples are displayed in Fig.~\ref{fig:generalization_berlin} for Berlin and in Fig.~\ref{fig:generalization_zurich} for Zurich.\footnote{See Fig.~\ref{fig:close-ups} in the appendix for additional detailed views.} 

In all tested scenarios, \resdepth generalizes reasonably well across different geographic contexts.\footnote{And imaging conditions, too. The training and test stereo pairs are assembled from mutually exclusive subsets of images.} Even though the model has never seen any DSM sample from the test city or any of the test images, it reduces the MAE of the initial reconstruction by up to $\approx$30\% for Zurich and up to $\approx$60\% for Berlin. Consistent with our previous findings, it achieves the biggest improvement for terrain pixels, whereas the MAE for building pixels drops only marginally or even deteriorates, especially when generalizing from Berlin to Zurich. Visually, however, the geometric fidelity of the refined buildings is considerably better than suggested by the quantitative evaluation. Overall, building shapes become crisper and more complete. Furthermore, low-rise residential buildings missed by the original matcher are faithfully reproduced, even in the presence of strong vegetation cover (see \nth{1}~column in Fig.~\ref{fig:generalization_berlin} and Fig.~\ref{fig:generalization_zurich}). In a few cases, buildings are removed too aggressively (see Fig.~\ref{fig:generalization_zurich}, \nth{3}~column).

As expected, the model's performance is inferior to a model trained on districts of the target city, due to geographic differences between Berlin and Zurich. We observe a quantitative drop of at most 0.8$\,$m in terms of overall MAE. For building pixels, we observe the largest loss in accuracy for \zurichTwo, indicating that the urban style of Berlin does not generalize well to the intricate and closely spaced buildings of the historic city center of Zurich. Strikingly, the MAE for building pixels in Berlin is lower when the model is trained on Zurich rather than on Berlin. Visually, \resdepth yields reliable reconstructions with a surprising level of detail, even when trained on a city different from the target scene (compare the \nth{3} and \nth{4} rows in Fig.~\ref{fig:generalization_berlin} and Fig.~\ref{fig:generalization_zurich}). However, we observe a minor reconstruction bias that likely stems from the characteristics and construction practices of the training city. When trained on Berlin data, \resdepth tends to reconstruct buildings with flat roofs and flat terrain. In contrast, when trained on Zurich data, \resdepth better preserves low-rise residential buildings (see Fig.~\ref{fig:generalization_berlin}, \nth{1}~column) and sharpens roof details (see Fig.~\ref{fig:generalization_zurich}, \nth{2}~column) but tends to produce noisier terrain.

\begin{figure*}[!th]
    \centering
    \subfloat[\resdepth w/o topographic normalization]{\includegraphics[width=0.23\textwidth]{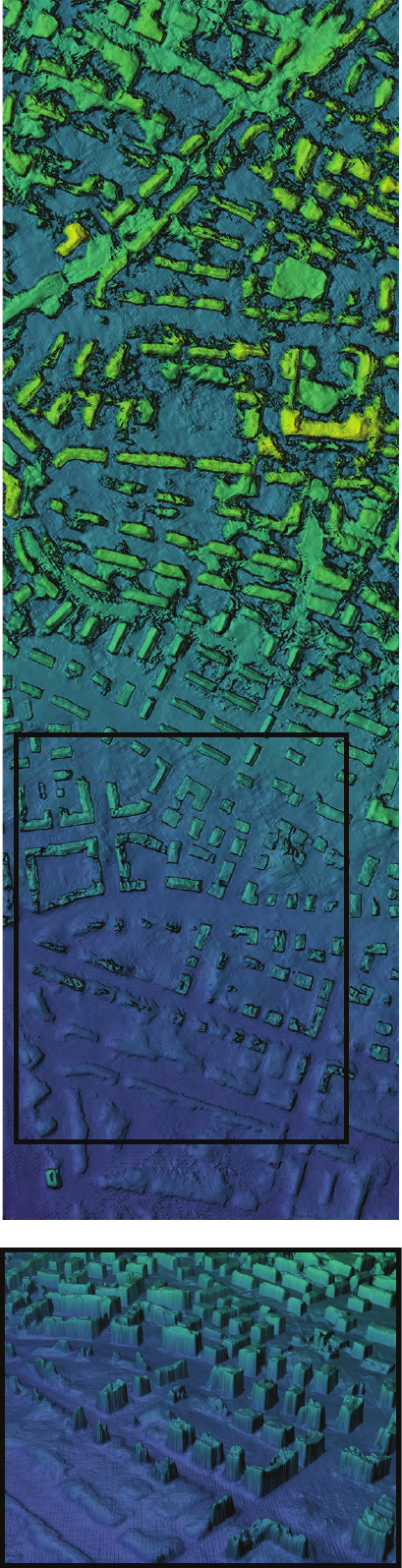}}
    \hfil
    \subfloat[\resdepth (ours)]{\includegraphics[width=0.23\textwidth]{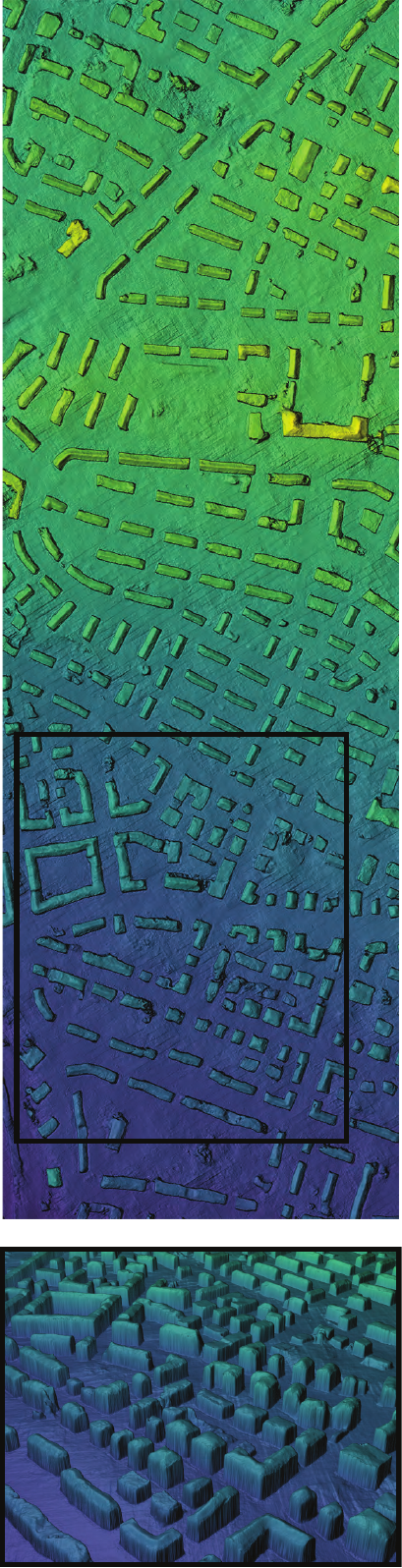}}
    \hfil
    \subfloat[Ground truth DSM]{\includegraphics[width=0.23\textwidth]{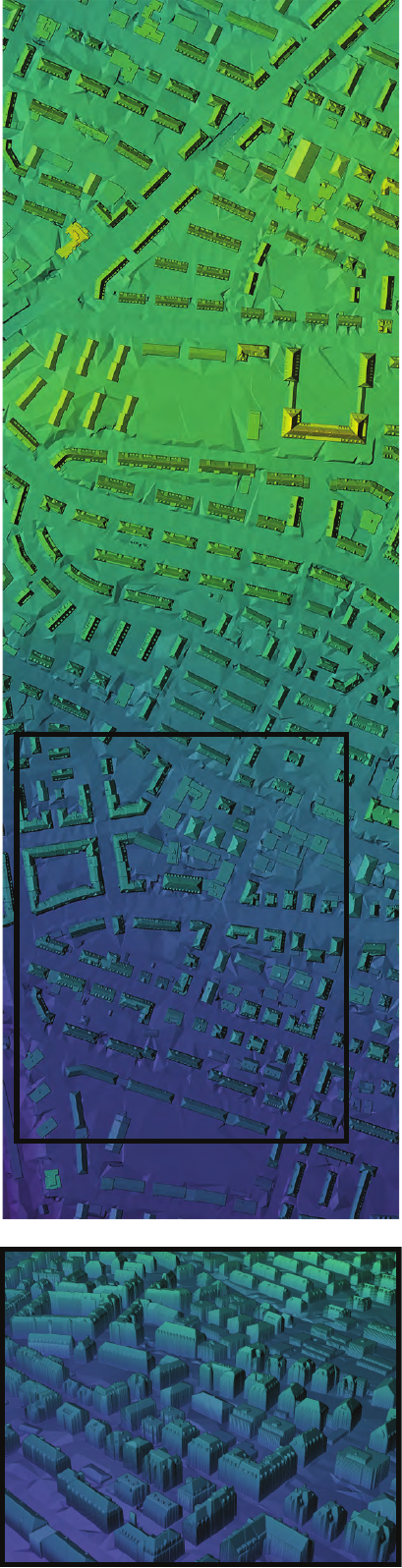}}
    \hfil
    \subfloat[Google Earth view]{\includegraphics[width=0.23\textwidth]{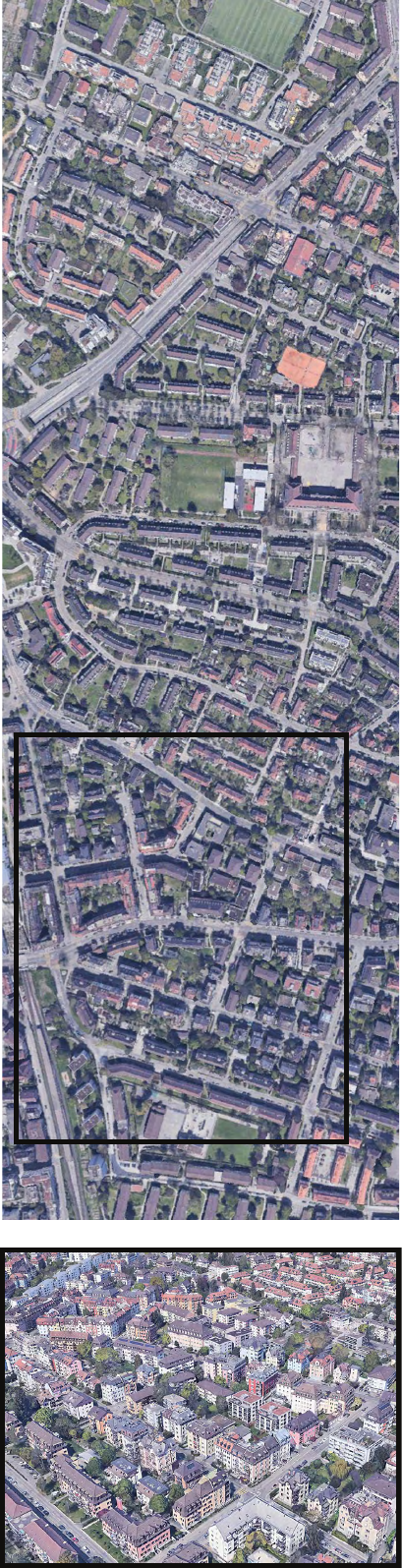}}
    \caption{Influence of DSM normalization on generalization. We train \resdepth on a (flat) region of Berlin and apply the model to \zurichOne in Zurich, which features height variations on the order of 100$\,$m over distances \textless1$\,$km. Heights are color-coded from blue to green to yellow. Without topographic normalization, the refinement fails to preserve the terrain height (upper part of images), and buildings at lower altitude are almost completely lost (lower part and zoomed oblique view).}
    \label{fig:effect_normalization}
\end{figure*}

\subsection{Influence of DSM Normalization}
Next, we examine the effect of our local DSM normalization scheme, which we have introduced as a measure to support generalization across different geographic contexts (cf.~Section~\ref{sec:normalization}). We repeat the geographical generalization experiment between Berlin and Zurich. We use the same network architecture, training settings, and data as before but replace the normalization scheme with the ubiquitous standard approach, namely whitening with the global mean height and standard deviation over all training pixels (as also used in \cite{stucker2020resdepth}). Indeed, the quantitative results presented in Table~\ref{tab:generalization} (last five rows) and the visual example in Fig.~\ref{fig:effect_normalization} demonstrate that with such a generic data normalization, the method is not able to handle topographic conditions not encountered during training. Compared to our proposed local normalization scheme, the overall MAE increases by up to $\approx$30\% when generalizing from Zurich to Berlin and by $\approx$160\% when generalizing from flat (Berlin) to hilly (Zurich) terrain.

\subsection{Multi-City Model}
\newcommand{\red}[1] {\color{Red}#1}
\newcommand{\green}[1] {\color{Green}#1}

\begin{table}[!t]
    \centering
    \caption{Joint multi-city model learned with training data from Berlin and Zurich. We report the MAE for the test portion of each scene, averaged over all pixels and over building pixels only.}
    \label{tab:multi-city}
    \begin{adjustbox}{max width=\columnwidth}
        \footnotesize
        \begin{tabular}{@{}l l c c @{}}
            \toprule
            Train Area & Test area & \multicolumn{2}{c}{MAE [m]} \\
            \cmidrule(l{\tabcolsep}){3-4} 
            & & Overall & Buildings\\
        	\midrule
        	\multirow{5}{*}{Initial}
        	& \berlinOne   & 5.20 & 2.60 \\
        	& \berlinTwo   & 4.27 & 2.75\\
        	& \zurichOne   & 3.89 & 3.02\\
            & \zurichTwo   & 3.92 & 3.34\\
            & \zurichThree & 3.65 & 4.63\\
            \midrule
            \multirow{5}{*}{\{\berlin, \zurich\}}
            & \berlinOne   & 1.41 & 2.22\\
    		& \berlinTwo   & 1.60 & 2.33\\
            & \zurichOne   & 2.08 & 2.34\\
    		& \zurichTwo   & 2.40 & 3.03\\
    		& \zurichThree & 3.00 & 4.59\\
        	\bottomrule
        \end{tabular}
    \end{adjustbox}
\end{table}
Finally, we simulate the situation where diverse and geographically well-distributed training samples are available, e.g., when training on existing high-quality data with the aim of improving future model updates in the same country. To that end, we train a joint model for all regions of interest by sampling from the training portions of all five regions in Berlin and Zurich. We randomly sample 10'000~DSM patches per geographic region for a total of 50'000~training patches. That all-round model is then evaluated on the test stripes of each region, see Table~\ref{tab:multi-city}. 
The single model over all five different regions indeed appears to learn a less city-specific, more universal prior for DSM refinement. It exhibits satisfactory performance, even a bit better than the models trained from different regions of the same city.
The overall MAE is in the order of 1.5$\,$m for test data from Berlin and varies between 2$\,$m and 3$\,$m for test data from Zurich. 
We note that the comparison should be taken with a grain of salt, since, for every individual region, the training does contain samples that are geographically closer to the test stripe than in the within-city generalization experiments. 

Still, the good performance is an indication that \resdepth has enough model capacity to cover the variations in architecture and urban planning between the two cities. While they certainly share a number of historical and cultural factors, Berlin and Zurich nevertheless lie in different countries and $\approx$700$\,$km apart, differ by an order of magnitude in size, and have very different topography.
The small size of our model, compared to contemporary deep neural networks, suggests that one can capture a much larger diversity of geographic conditions and urban styles in a single (larger) network. Given a sufficient amount of suitable training data, it seems realistic to learn a single model for country-wide or perhaps even continent-wide use.

\section{Conclusion}
We have presented \resdepth, a simple yet highly expressive prior of 3D scene geometry, to complement dense stereo matching and reconstruct high-quality DSMs from satellite imagery. The prior is encoded implicitly in a deep neural network and learned from data. \resdepth follows a residual learning strategy, i.e., it is trained to refine an imperfect input DSM by regressing a per-pixel correction to the height, using both the DSM and (ortho-rectified) stereo images as input. In this way, it can leverage not only the local distribution of heights but also image textures, alignment errors between the images and the DSM, and discrepancies between the stereo correspondence and the 3D shapes.
\resdepth indeed learns to restore shape regularities such as straight and crisp surface creases and height discontinuities and aligns them with contrast edges in the images. This goes beyond smoothing the DSM and often successfully adds or completes surface details that were missed by the initial stereo reconstruction but are visible in the images. Additionally, \resdepth also removes artefacts of the preceding stereo reconstruction and can be trained to filter out specific content that may be unwanted in the target DSM, such as vegetation.

We have experimentally demonstrated that the prior that \resdepth learns is fairly generic and remains valid under varying imaging conditions and for distant geographic locations with different styles of urban planning and construction. As expected, the highest accuracy is still achieved if the training and test regions are located near each other and are observed under similar lighting and viewpoints. In that ideal setting, the method reaches a MAE in the order of 1.5$\,$m and an absolute median error of 0.7$\,$m---quite remarkable values given the image resolution of $\approx$0.5$\,$m and the uncertainty of the satellite poses on the order of 0.5$\,$m on the ground. 

With suitable training data, \resdepth can also learn to generalize to unseen viewing directions, lighting conditions, and urban styles. It consistently improves the input DSM in the more general scenario where different images are used for training and testing and where the training and test regions lie further apart. Naturally, the gain will typically be smaller than in the ideal setting, depending on how large the domain gap is between training and test data.

Furthermore, we have also shown that a more general \resdepth model, which has been optimized jointly for two different cities, is able to refine DSMs of unseen test sites in both locations. If a sufficient amount of training data can be assembled from a larger, representative sample of cities, it seems realistic to learn a single, \say{general} network that is applicable across entire countries or even continents.

A natural extension of the \resdepth idea is to move to multi-view stereo with \textgreater2 input images/rays per pixel. Technically, this is straightforward, as the initial DSM and all input views (after ortho-rectification) already share the same planimetric coordinate system and can be stacked into a multi-channel image and fed into a \resdepth network with a suitably adapted first layer. We have performed preliminary tests with such a setup but found that it did not outperform the two-view case. Whether and how the multi-view redundancy can be exploited to learn an even more robust prior is an interesting question for future research.

\appendix

\section{Network Architecture}
\begin{figure*}[!ht]
\centering
\includegraphics[trim={0 5 0 0}, clip, width=\textwidth]{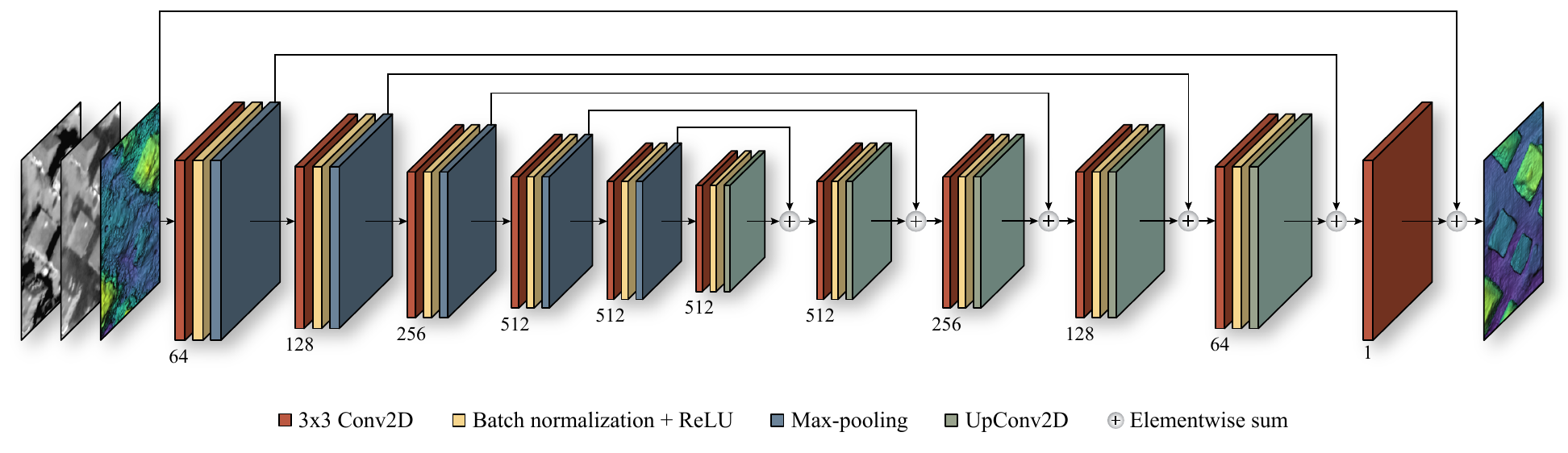}
\caption{Network architecture of \resdepth. Each convolutional layer (except for the last one) is followed by batch normalization and ReLU activation. The number below each convolutional layer specifies the number of filter channels. Note the long residual connection that directly adds the initial DSM to the output of the last decoder layer, so that the network learns to generate residual updates instead of absolute heights.}
\label{fig:architecture}
\end{figure*}

The network design of \resdepth is based on the established \mbox{U-Net}~\cite{ronneberger2015u} architecture, see Fig.~\ref{fig:architecture}. The encoder and decoder are symmetric and fully convolutional. We employ max-pooling to downsample the data in the encoder and fractionally strided, transposed convolutions for upsampling in the decoder. We use 64 filters at full resolution and gradually double the number of filters in every encoder block up to a maximum filter depth of 512. As usual, skip connections propagate high-frequency information from encoder to decoder levels of equal resolution. Recall that we further add a long residual connection that directly adds the initial DSM to the output of the last decoder layer, such that the network regresses residual height corrections rather than absolute heights. We found tiles of size 256$\times$256 pixels to work best for the spatial resolution of our data and DSM.\footnote{Note that in an early version~\cite{stucker2020resdepth} we had used only 128$\times$128 pixels.}

\section{Study Areas}
\begin{figure*}[!t]
    \centering
    \subfloat[Berlin]{\includegraphics[width=0.48\textwidth]{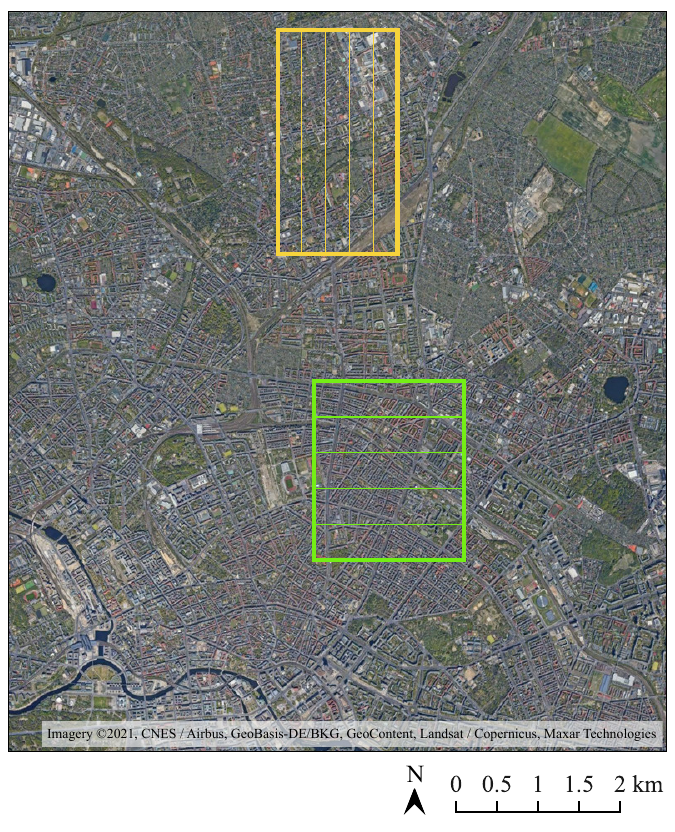}}
    \hfill
    \subfloat[Zurich]{\includegraphics[width=0.48\textwidth]{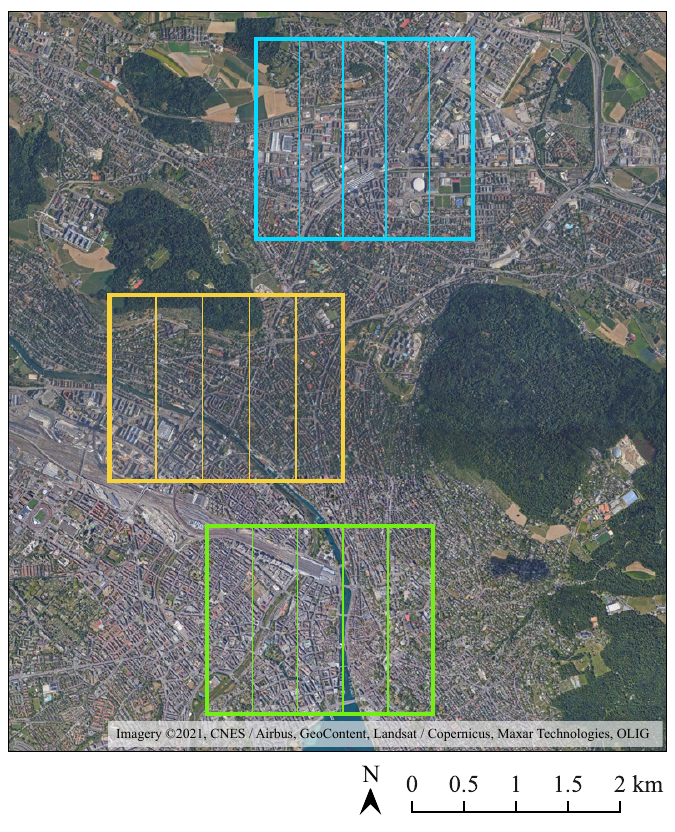}}
    \smallskip 
    \caption{Location of the study areas. (a) We define two areas \berlinOne (yellow) and \berlinTwo (green) in Berlin. (b) For Zurich, we use the areas \zurichOne (yellow), \zurichTwo (green), and \zurichThree (blue). We split each geographic area into five equally large and mutually exclusive stripes and reserve three stripes for training, one for validation, and one for testing (except for ablation studies over a single region, which employ 5-fold cross-validation).}
    \label{fig:datasets}
\end{figure*}

Fig.~\ref{fig:datasets} depicts the chosen areas per city, each covering 4{$\,$km\textsuperscript{2}}. Within each area, we define geographically separate stripes for training, validation, and testing. Please refer to Section~\ref{sec:area_description} for a detailed description of the areas.

\section{Satellite Image Distribution}
\label{sec:valid_pairs}
\begin{figure}[!ht]
    \centering
    \subfloat[Stereo pairs that fulfill the selection criteria.]{\includegraphics[width=0.5\columnwidth]{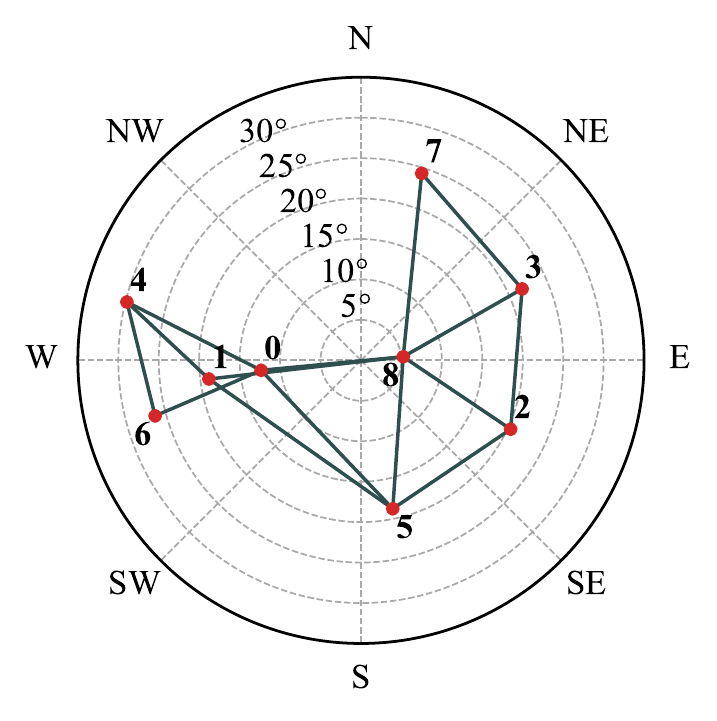}\includegraphics[width=0.5\columnwidth]{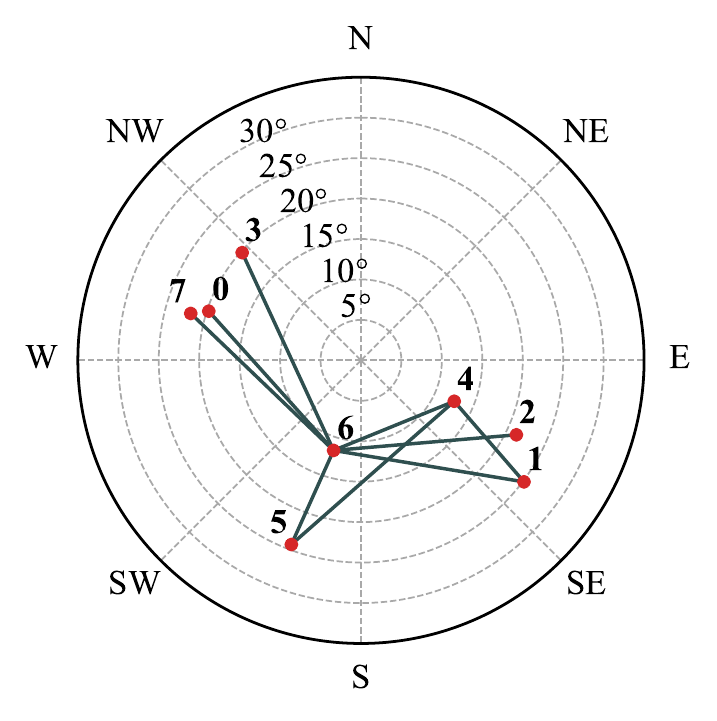}\label{fig:valid_pairs}}
    \\
    \subfloat[Stereo pairs used during training (green,~$A$) and testing (blue,~$B$). ]{\includegraphics[width=0.5\columnwidth]{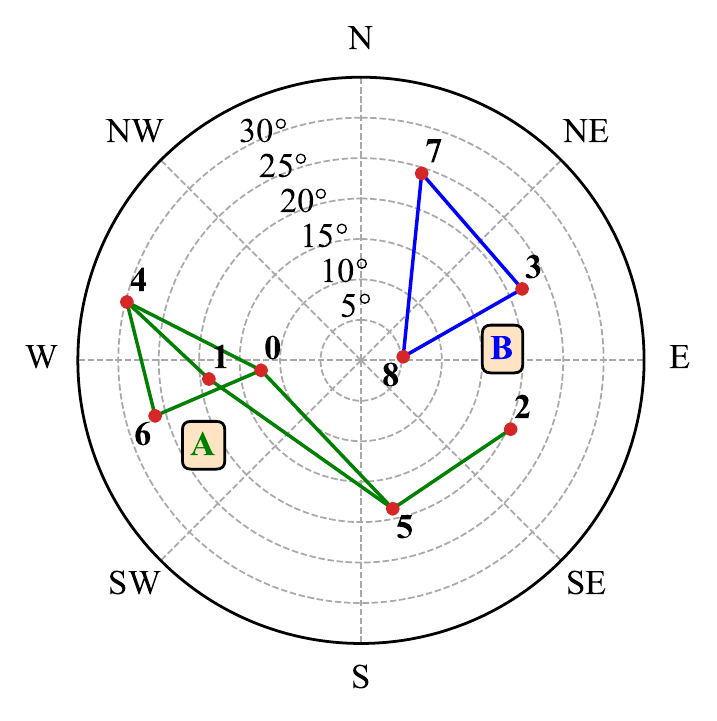}\includegraphics[width=0.5\columnwidth]{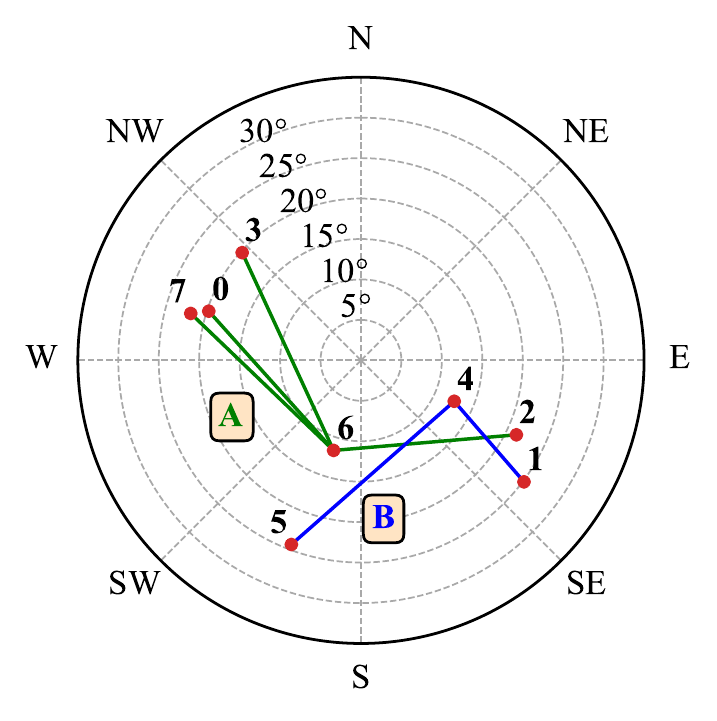}\label{fig:train_test_pairs}}
    \smallskip
    \caption{Satellite position (red marker) for each image acquisition over Berlin (left) and Zurich (right). The circular orientation refers to the azimuth angle and the radial distance from the center to the off-nadir angle. The images are sorted and labeled in ascending order (from older to newer acquisitions). Stereo pairs are connected by a line.}
    \label{fig:stereo_pairs}
\end{figure}

Fig.~\ref{fig:valid_pairs} shows the stereo pairs that fulfill the selection criteria described in Section~\ref{sec:pair_selection}. We partition the stereo pairs into two
mutually exclusive groups $A$ and $B$ (see Fig.~\ref{fig:train_test_pairs}), such that each group contains pairs with predominantly north-south as well as west-east oriented baselines. We use the pairs in $A$ for training and the ones in $B$ for testing, such that no images are shared between the training and test sets.

\section{Label Noise}
\label{sec:label_noise}

\begin{figure}[!ht]
    \centering
    \subfloat[]{
        \begin{minipage}[t]{\columnwidth}
            \includegraphics[width=0.48\textwidth]{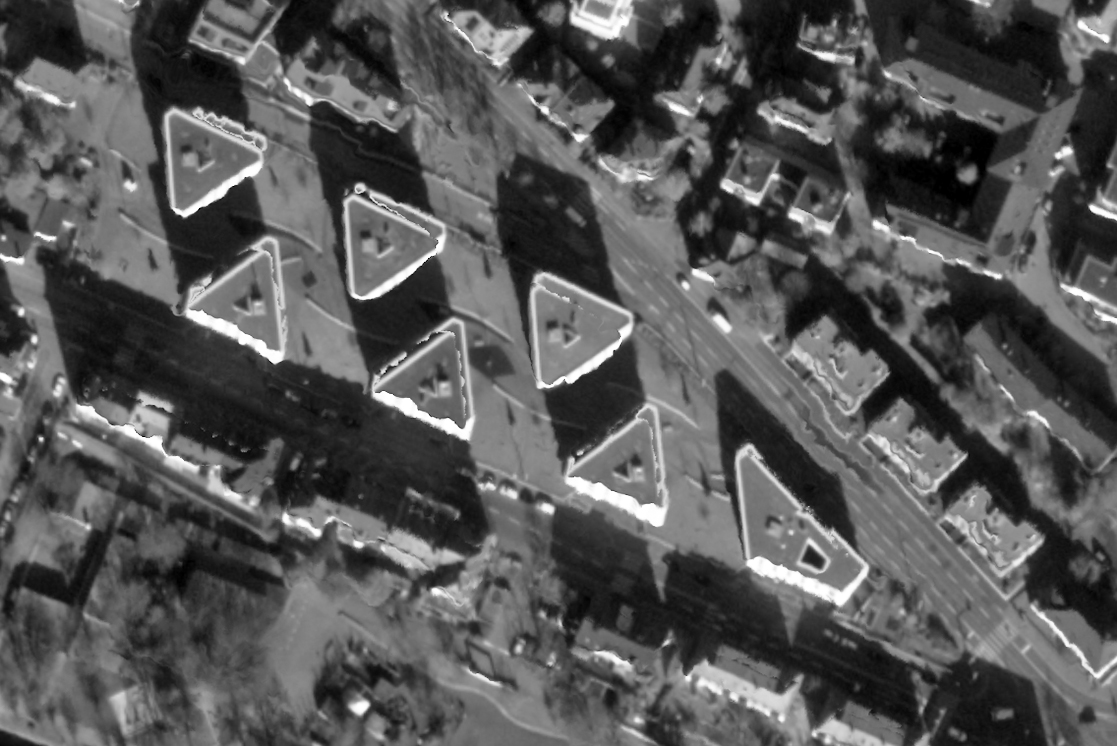}
            \includegraphics[width=0.48\textwidth]{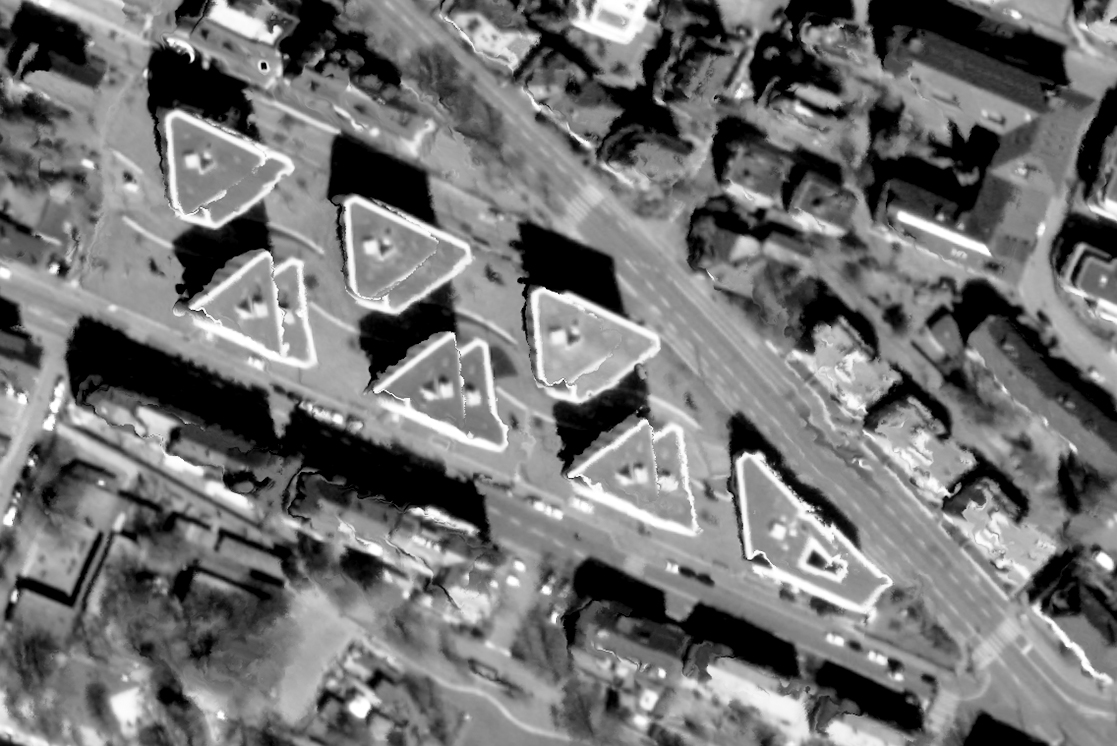}
        \end{minipage}
    }
    \\
    \subfloat[]{\includegraphics[width=0.48\columnwidth]{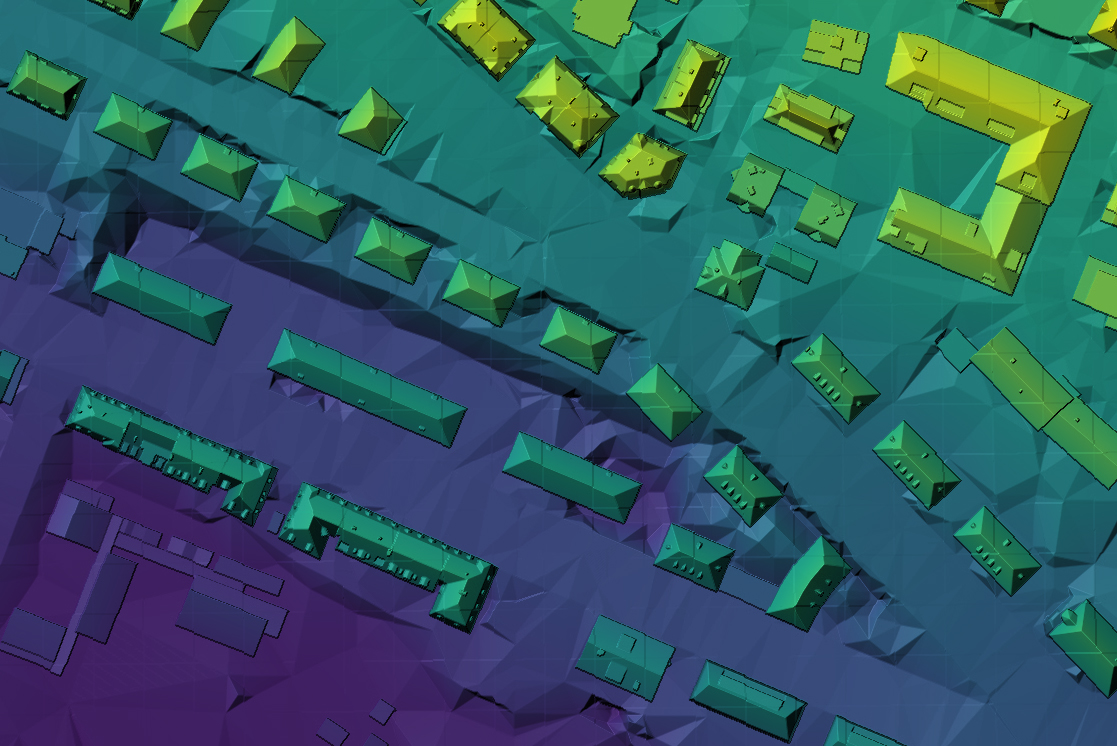}\label{fig:wipkingen_gt}}
    \hfil
    \subfloat[]{\includegraphics[width=0.48\columnwidth]{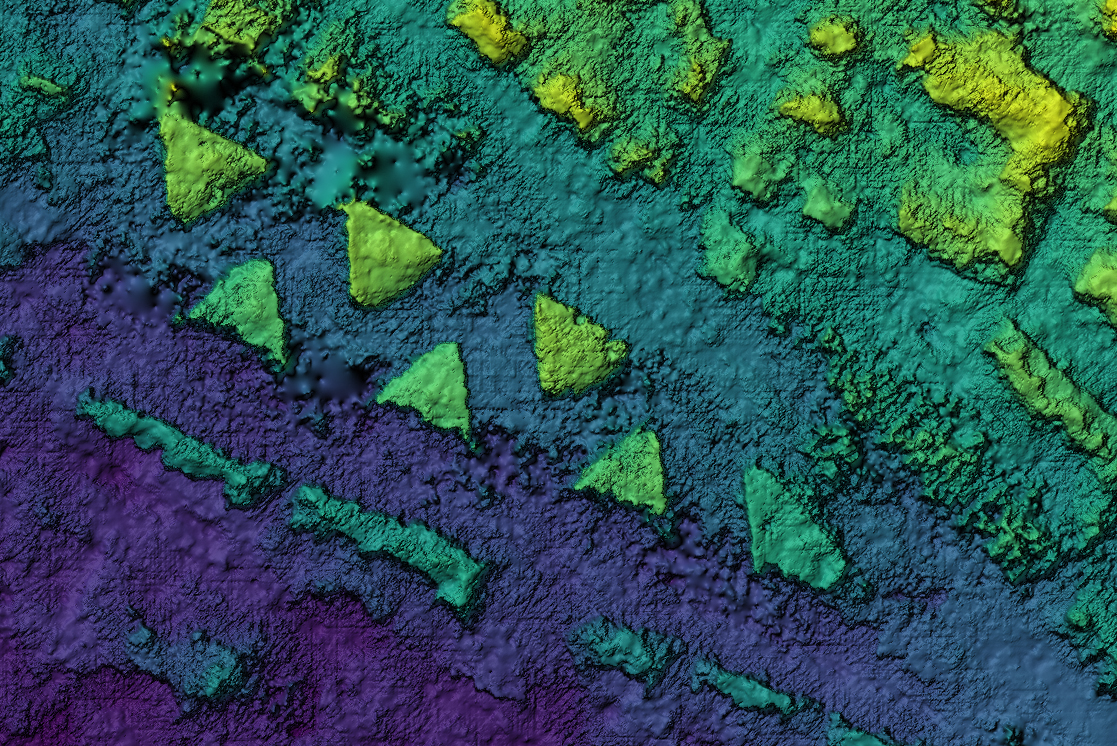}\label{fig:wipkingen_initial}}
    \hfil
    \\
     \subfloat[]{\includegraphics[width=0.48\columnwidth]{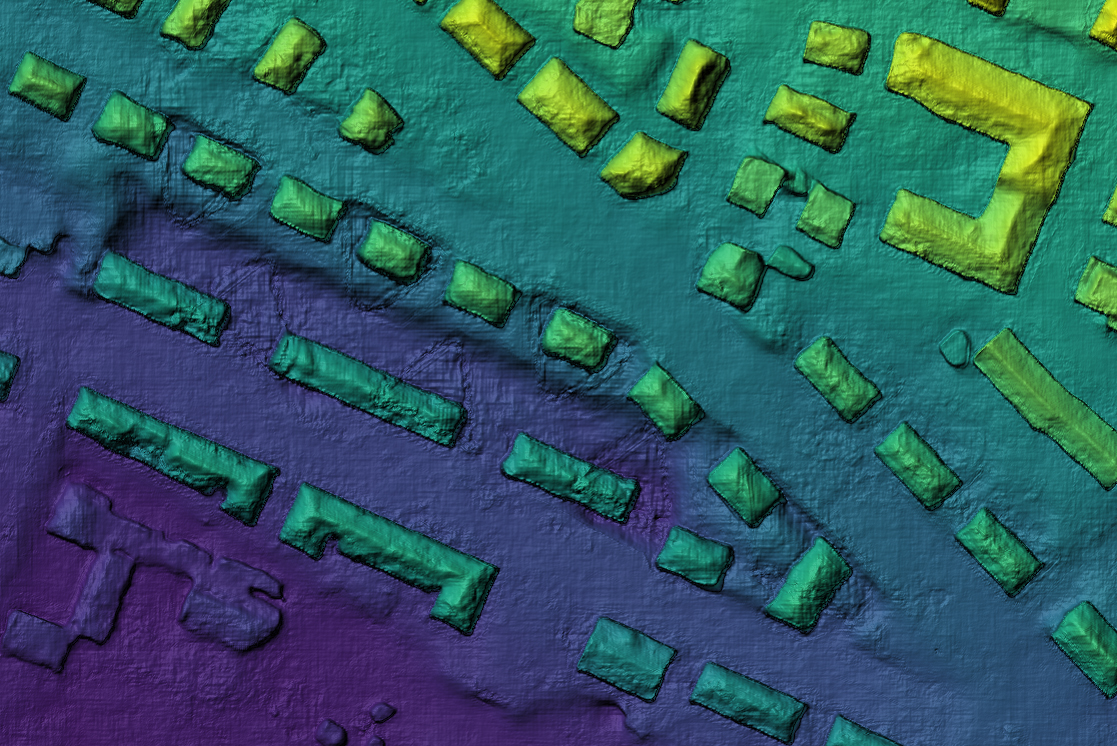}\label{fig:wipkingen_seen}}
     \hfil
    \subfloat[]{\includegraphics[width=0.48\columnwidth]{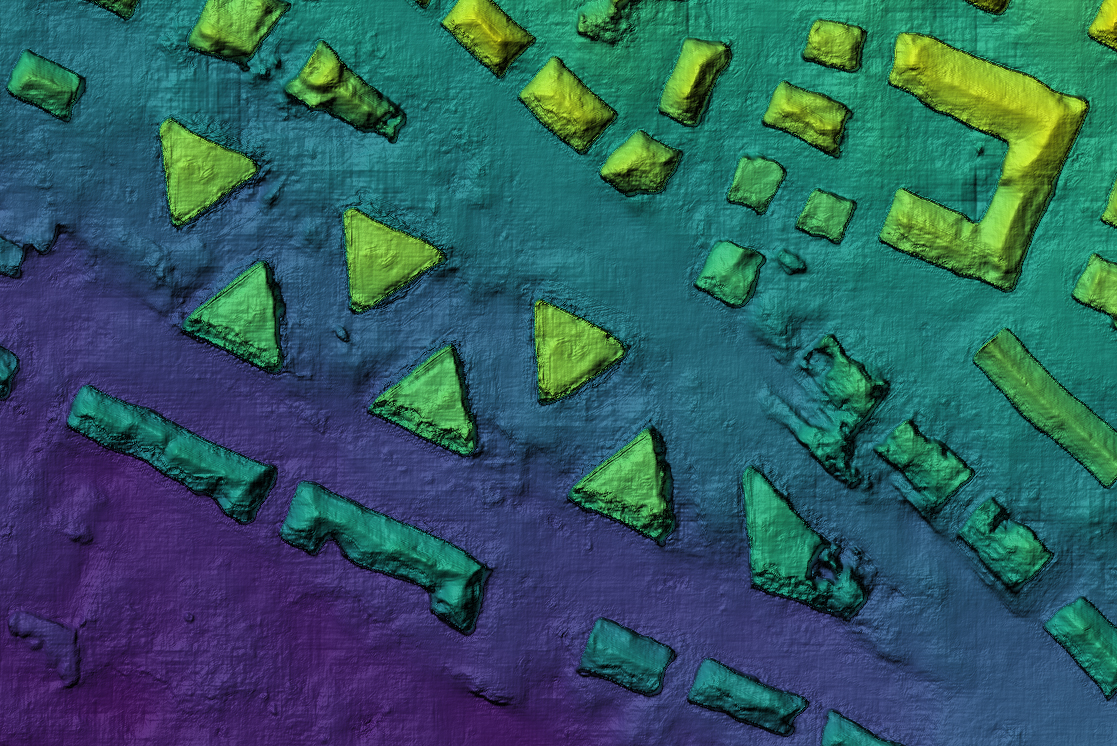}\label{fig:wipkingen_unseen}}
    \caption{Influence of label noise. (a)~ortho-rectified stereo pair, note replicated textures due to occlusions; 
    (b)~ground truth DSM with obsolete buildings demolished since the DSM was created; (c)~initial DSM derived from satellite images;
    (d)~refinement result if used as training data; (e)~refinement result if used as test data: the actually visible building shapes are correctly restored.}
    \label{fig:label_noise}
\end{figure}

As with any learning-based system, \resdepth has to find the right balance between capturing task-specific patterns and regularities in the training data and refraining from overfitting to spurious correlations that do not generalize to unseen target scenes. As urban environments continuously change over time, the state captured in the images is not always consistent with the city models used as reference data. Therefore, \resdepth must be able to deal with label noise.

We study the sensitivity of \resdepth to label noise on a small area in \zurichOne. In that area, there are known differences between the state visible in the satellite images and the reference DSM due to recent construction activities. Moreover, the area contains the only instances of equilateral triangular buildings, see Fig.~\ref{fig:label_noise}.

We test two different versions of the network. In the first run, the area is included in the training data. Once trained, the model is applied to the same area to refine the initial DSM.
\resdepth has seen the exact same context during training and learned to regard the unfamiliar triangular shapes as incorrect; hence, the reconstruction reproduces the obsolete DSM.

In the second run, we exclude the area from the training set and then perform the same refinement, this time using the area as unseen test data. Not having seen the incorrect ground truth, \resdepth has no motivation to remove the triangular buildings and sticks to the image evidence; thus, it straightens their outlines and aligns the walls with the image gradients, as desired.

Overall, the experiment indicates that the label noise did not negatively affect the model output in unseen test regions, or in other words, the learned prior is flexible enough to handle unfamiliar building shapes. On the other hand, the fact that the model can memorize the necessary \say{correction} for the training data is evidence that it learns long-range context beyond local smoothing and straightening.

\section{Building Height Bias}
\label{sec:height_bias}

\begin{figure}[!t]
    \centering
    \includegraphics[trim={0 5 0 0}, clip, width=0.95\columnwidth]{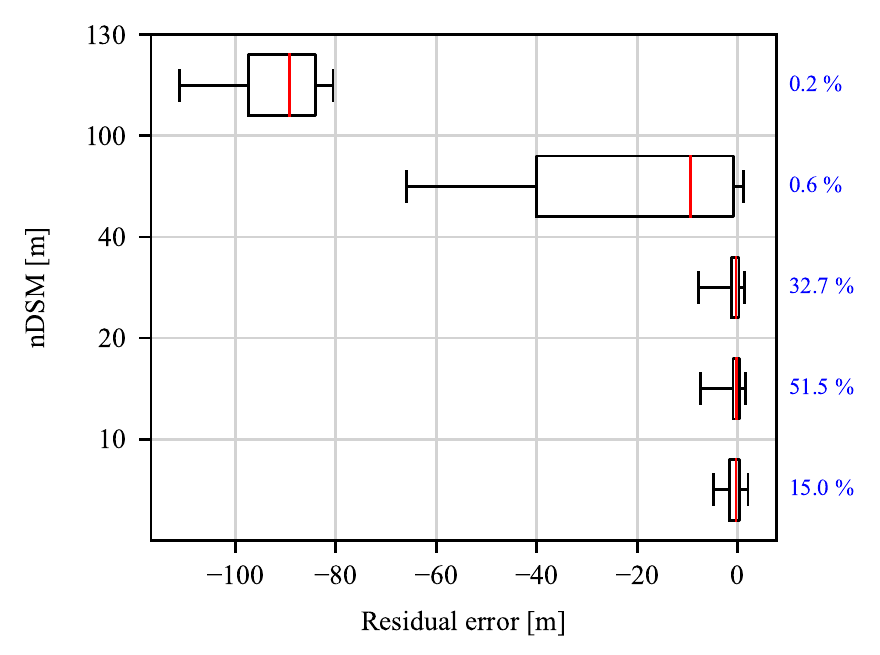}
    \caption{Analysis of residual errors of non-terrain pixels in \zurichOne, grouped by reference height above terrain (average over 5-fold cross-validation). A positive error means that the predicted height is larger than the reference value. The boxplots show the median, the quartiles, and the \nth{5} and \nth{95} percentile. The blue label to the right indicates the percentage of off-terrain training pixels within the respective height range.}
    \label{fig:residual_errors_building}
\end{figure}

\begin{figure*}[!t]
    \centering
    \subfloat[]{\includegraphics[width=0.32\textwidth]{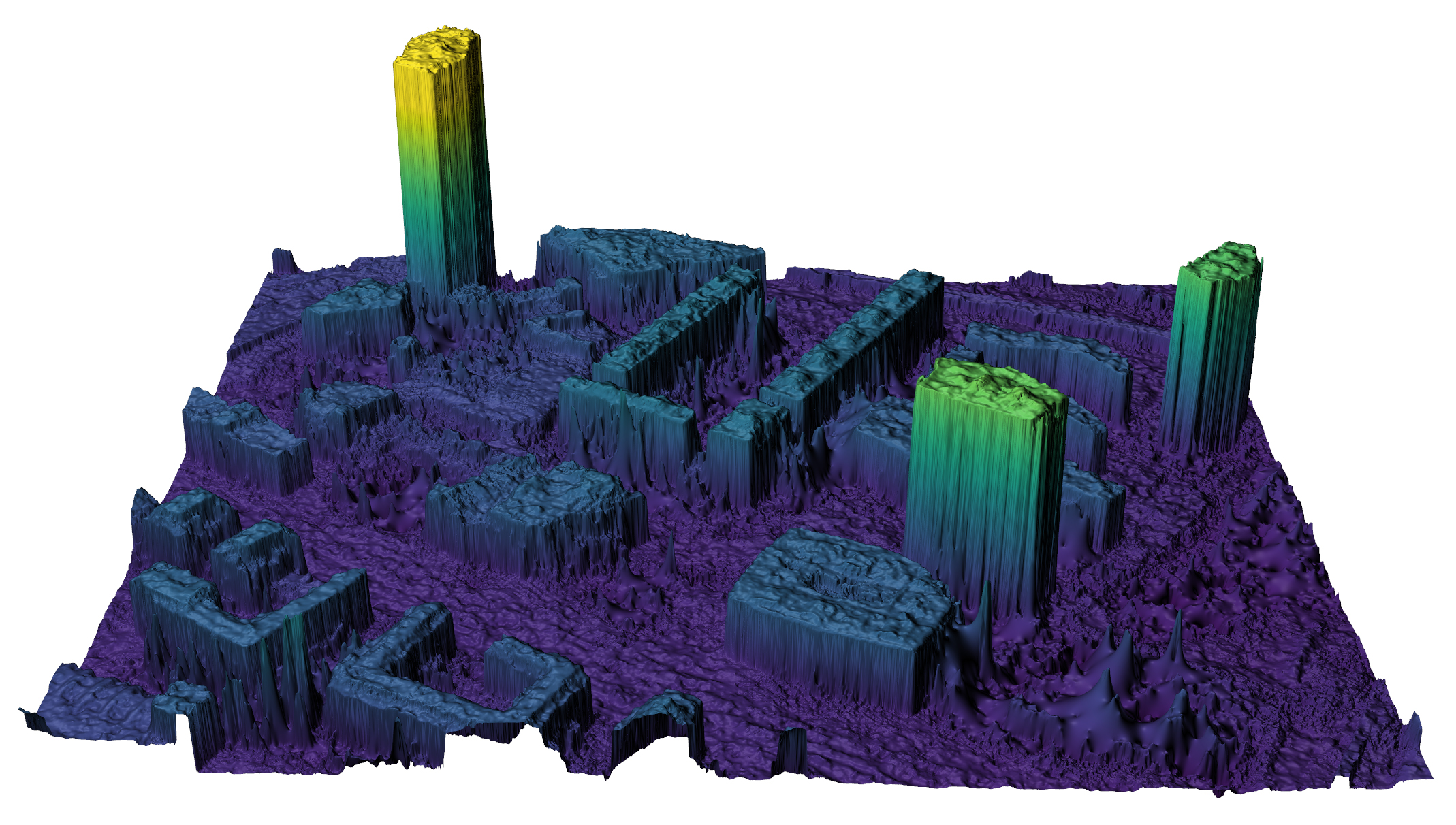}}
    \hfil
    \subfloat[]{\includegraphics[width=0.32\textwidth]{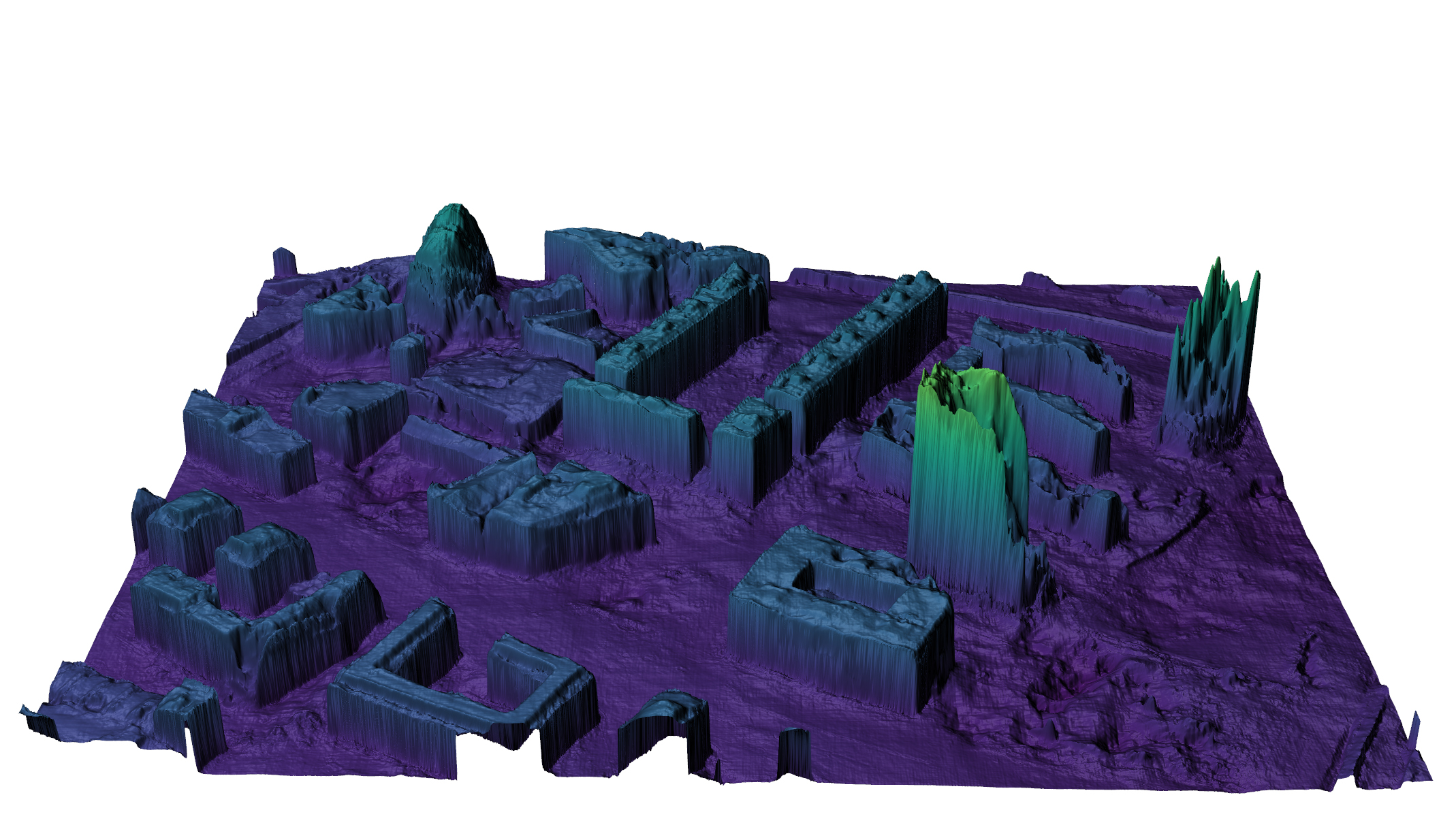}}
    \hfil
    \subfloat[]{\includegraphics[width=0.32\textwidth]{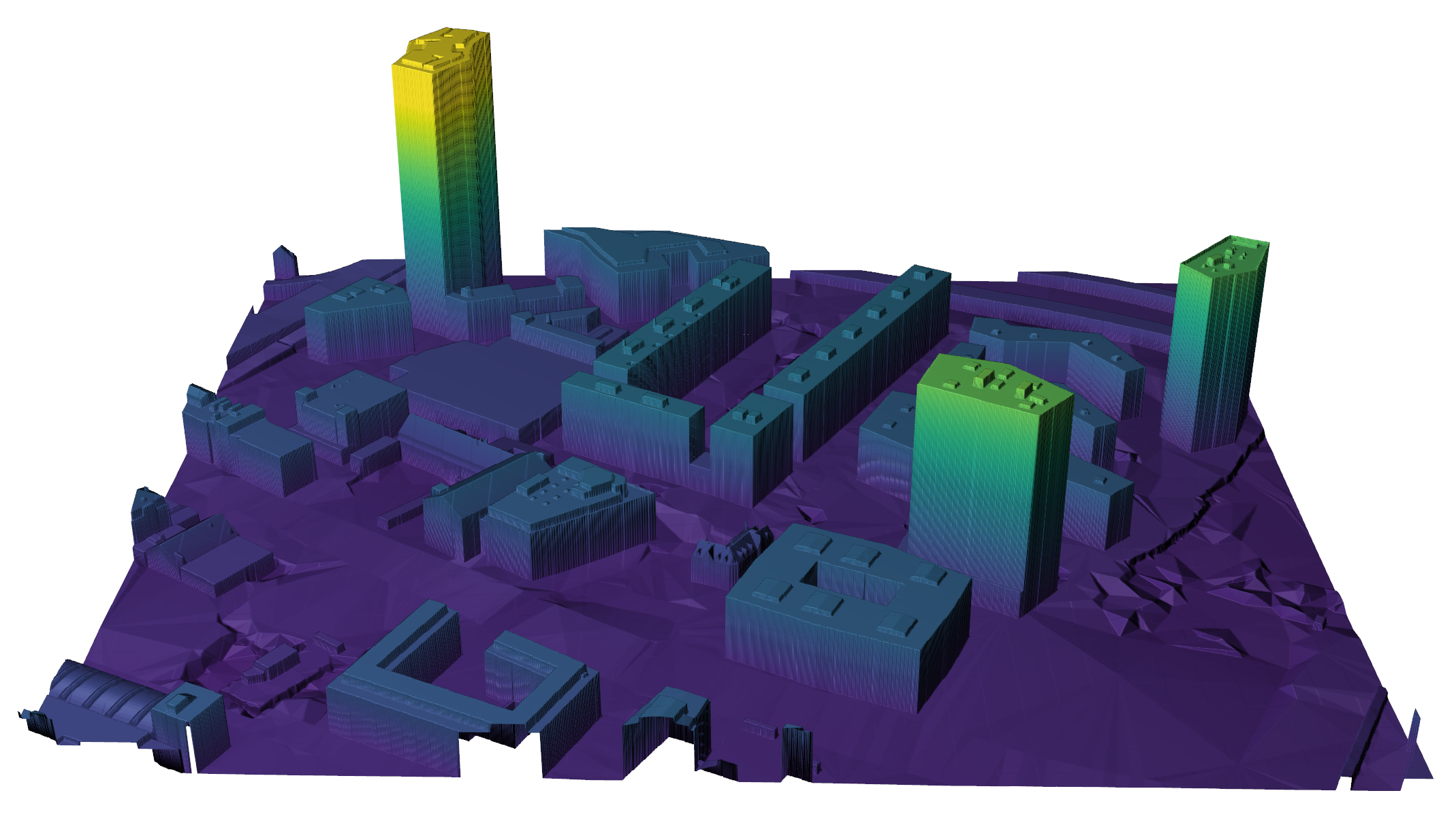}}
    \smallskip 
    \caption{Bias against unusually tall buildings. (a)~initial DSM, (b)~refined DSM, (c)~ground truth DSM. \resdepth tends to squeeze very tall buildings that exceed the height range seen during training.\bigskip}
    \label{fig:prime_tower}
\end{figure*}

Since the ground truth provides a terrain height for every pixel, we can readily compute the height above ground (nDSM) for every non-terrain pixel. We utilize this information to examine the dependency between building heights and prediction errors in \zurichOne, see Fig.~\ref{fig:residual_errors_building}. Buildings up to 40$\,$m are accurately refined, with a median error of at most \mbox{-0.4$\,$m}. For buildings taller than 40$\,$m, the median error increases drastically, reaching nearly \mbox{-90$\,$m} for buildings taller than 100$\,$m. These errors are an immediate effect of the extremely imbalanced distribution of building heights: \textless1\% of the building pixels in the region exceed 40$\,$m. Consequently, the learned prior regards such high buildings as extremely unlikely and fails to refine them correctly. Instead, it tends to squash them towards the average building height. An extreme example is shown in Fig.~\ref{fig:prime_tower}, including the tallest building of Zurich (which is tens of meters higher than any building the model has seen during training).

\section{Additional Visualizations}
\begin{figure*}[!ht]
    \def\mywidth{0.23\textwidth}
    \setlength{\tabcolsep}{0.2em}
    \centering
    \footnotesize
    \begin{tabular}{c c c c}
        Initial DSM & Within-city generalization & Across-city generalization & Google Earth view \smallskip\\
        \includegraphics[width=\mywidth,trim={0 0 0 0},clip]{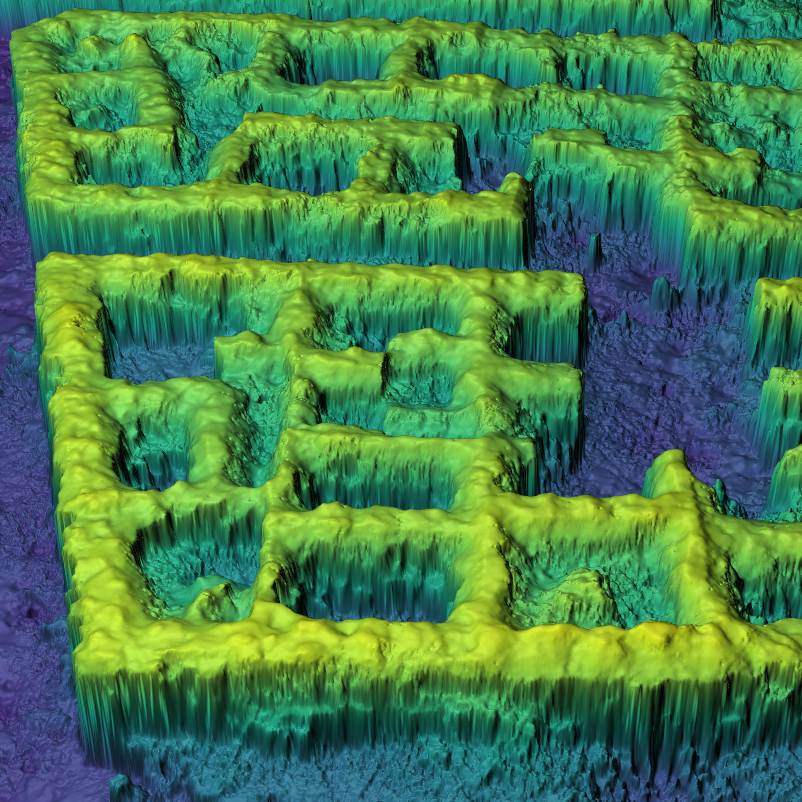} &
        \includegraphics[width=\mywidth,trim={0 0 0 0},clip]{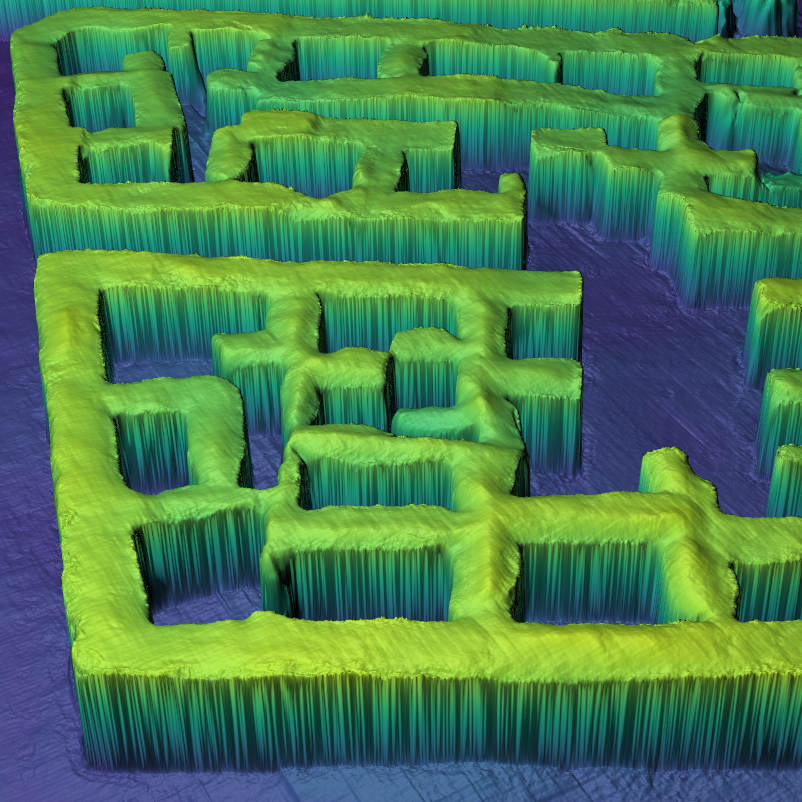} &
        \includegraphics[width=\mywidth,trim={0 0 0 0},clip]{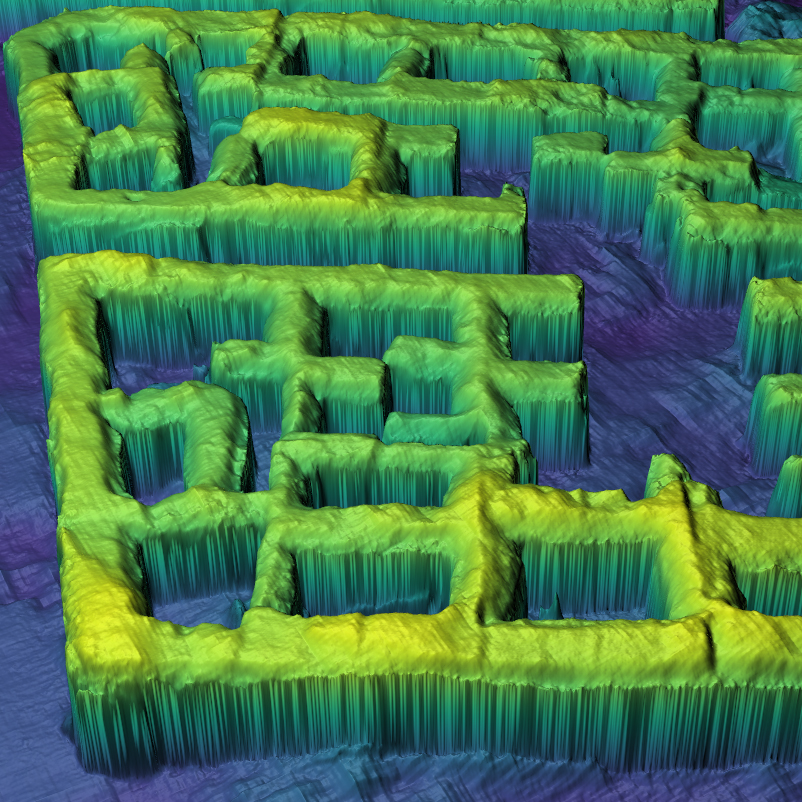} &
        \includegraphics[width=\mywidth,trim={0 0 0 0},clip]{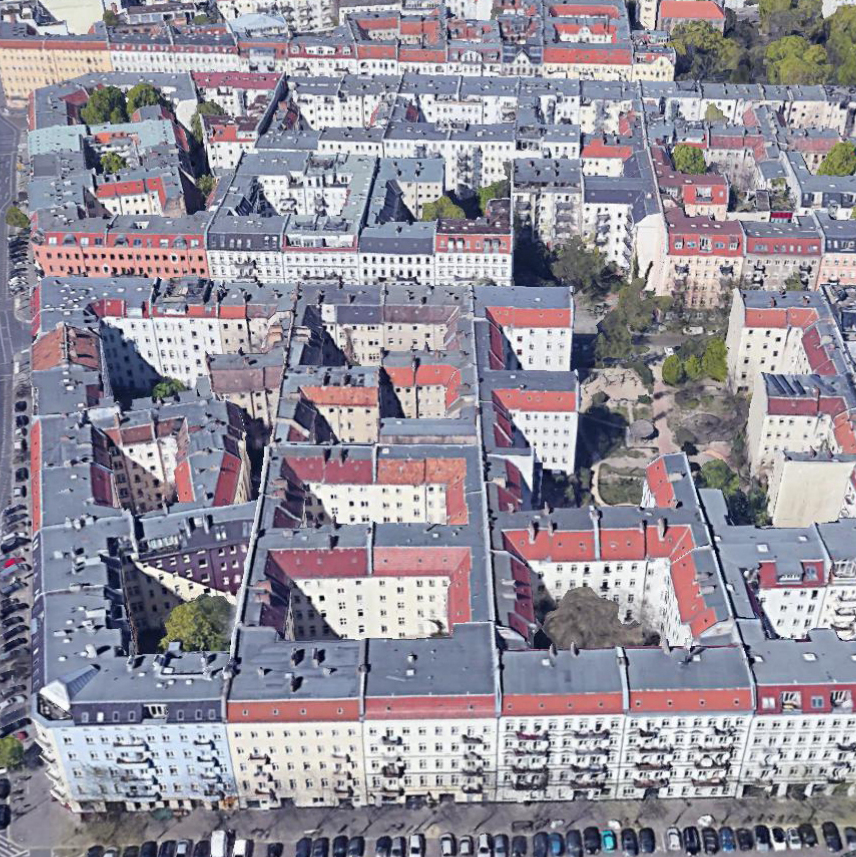}
        \smallskip
        \\
        \includegraphics[width=\mywidth,trim={0 0 0 0},clip]{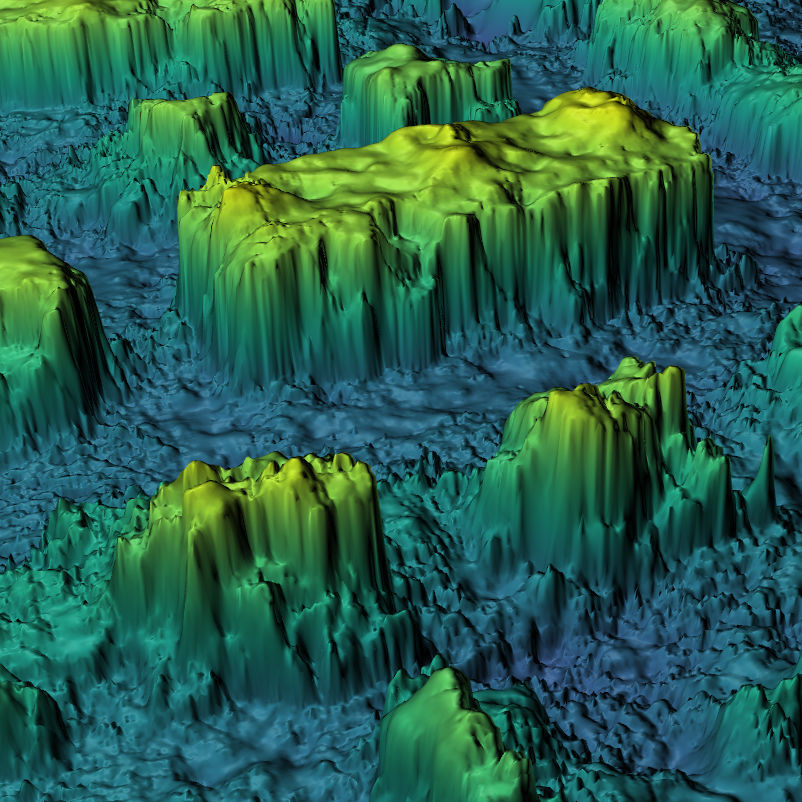} &
        \includegraphics[width=\mywidth,trim={0 0 0 0},clip]{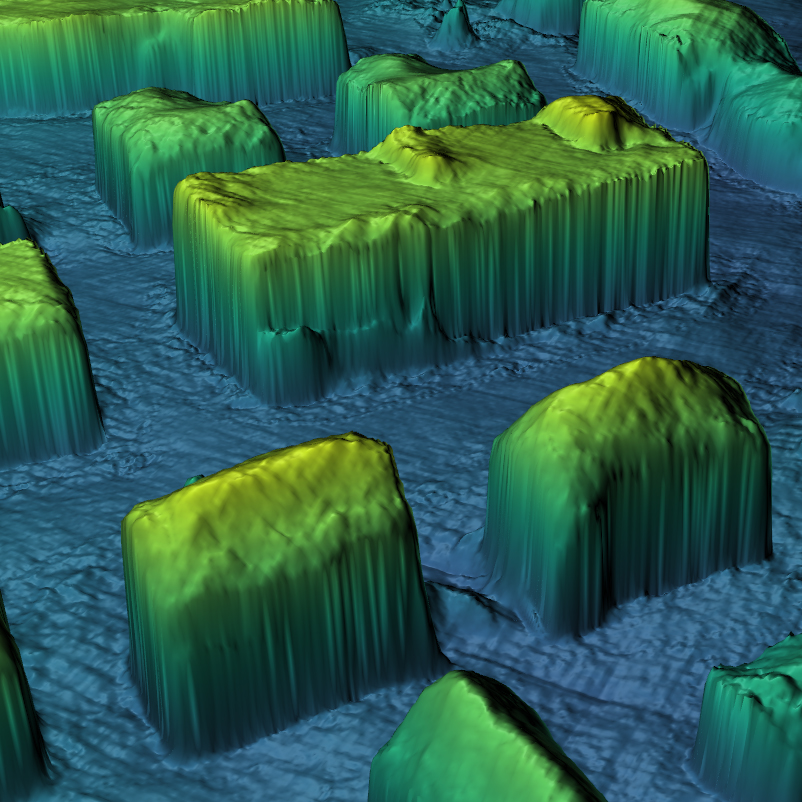} &
        \includegraphics[width=\mywidth,trim={0 0 0 0},clip]{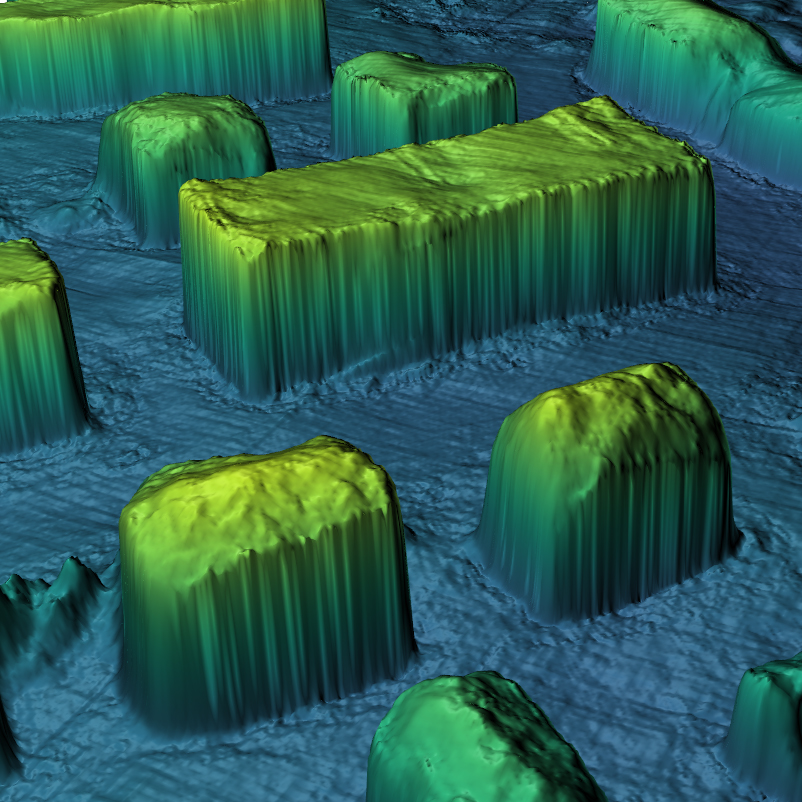} &
        \includegraphics[width=\mywidth,trim={0 0 0 0},clip]{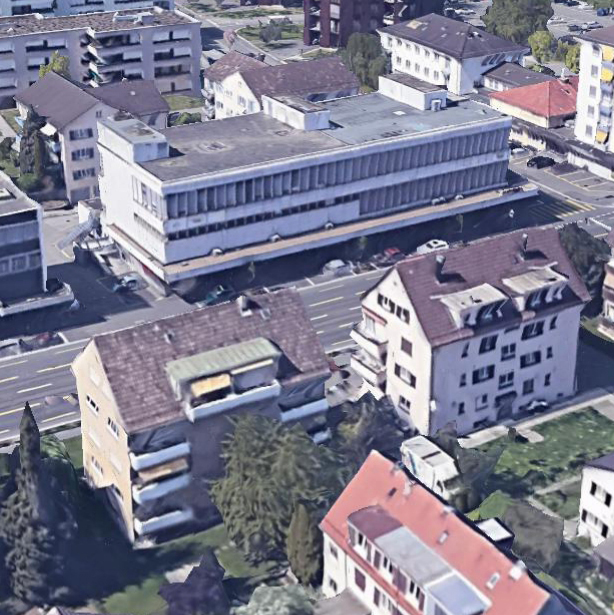}
        \smallskip
        \\
        \includegraphics[width=\mywidth,trim={0 0 0 0},clip]{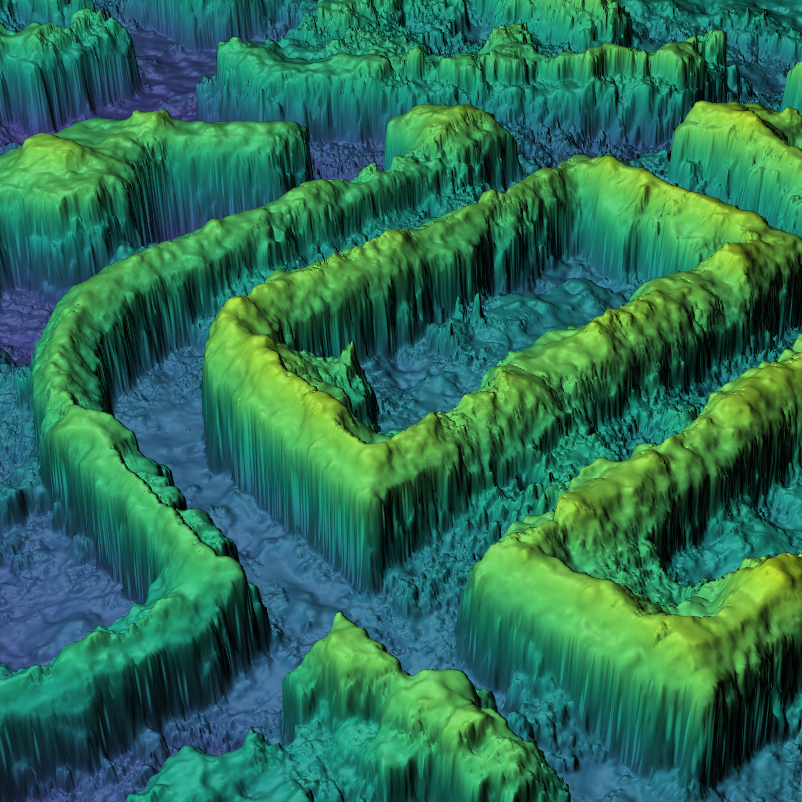} &
        \includegraphics[width=\mywidth,trim={0 0 0 0},clip]{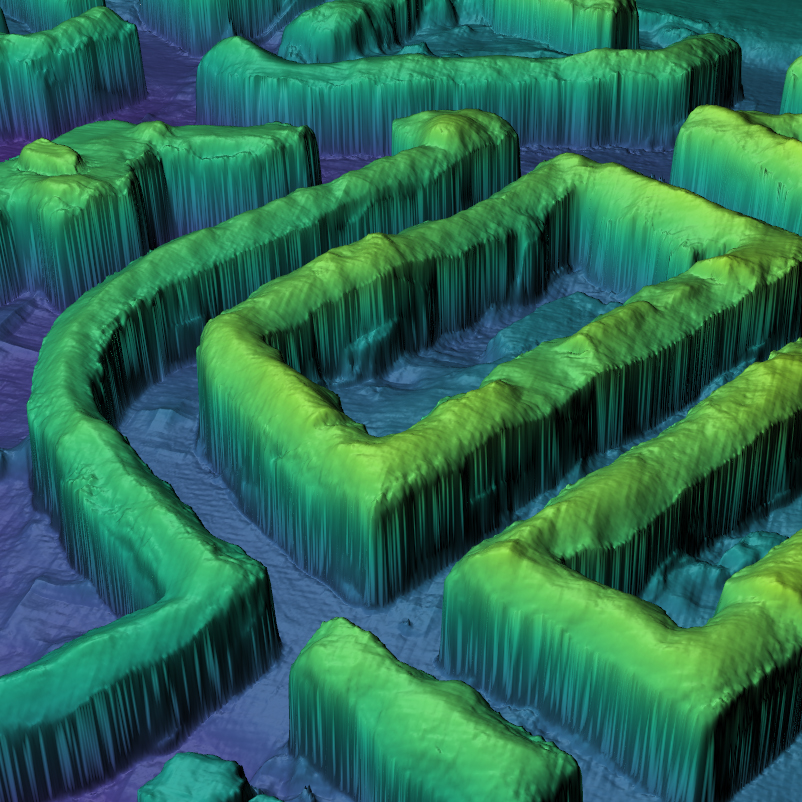} &
        \includegraphics[width=\mywidth,trim={0 0 0 0},clip]{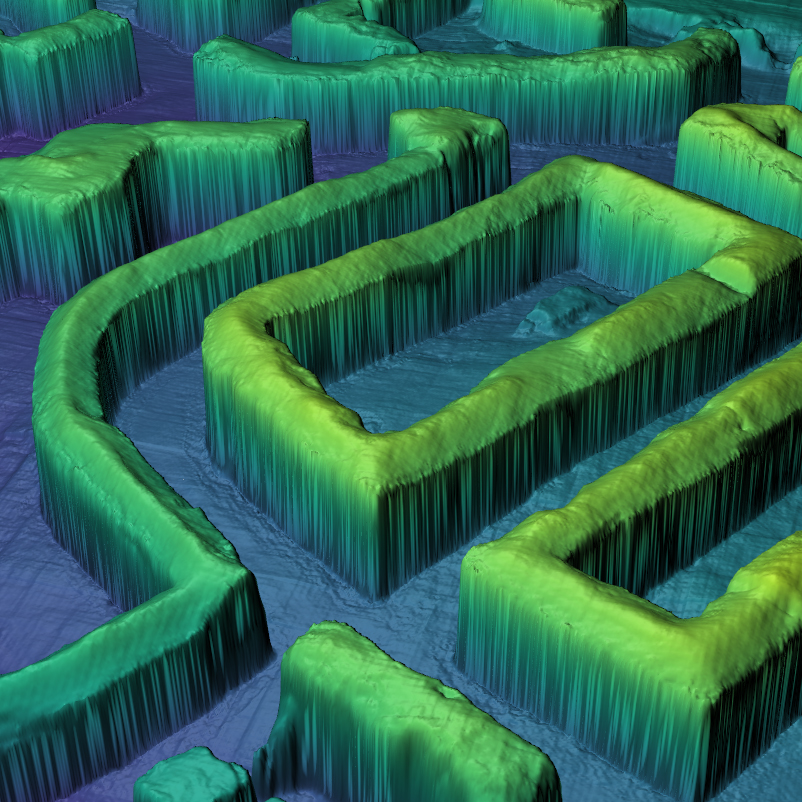} &
        \includegraphics[width=\mywidth,trim={0 0 0 0},clip]{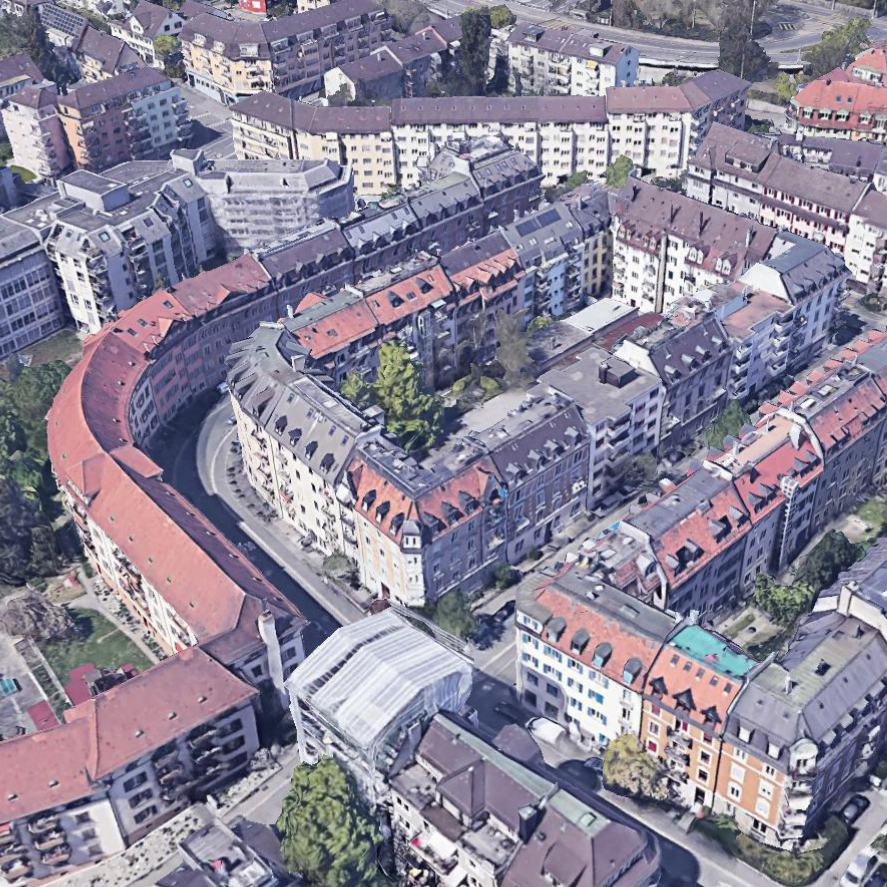}
        \\
    \end{tabular}
    \caption{Within-city and cross-city generalization. Row~1 shows a region in Berlin, rows~2--3 are from Zurich. For each example, \resdepth models were trained on a different part of the same city (\nth{2} column) or the respective other city (\nth{3} column). Heights are color-coded from blue to  yellow.}
    \label{fig:close-ups}
\end{figure*}

We provide additional visualizations of geographical generalization in Fig.~\ref{fig:close-ups}.

\bibliography{bib}

\end{document}